\documentclass{aa}  

\usepackage{graphicx}
\usepackage{txfonts}
\usepackage{lipsum}
\usepackage{subcaption}
\usepackage{rotating}
\usepackage{ulem}
\usepackage{pdflscape}
\usepackage{placeins}
\usepackage{gensymb}
\usepackage{etex}

\defcitealias{HensleyDraine2023ApJ...948...55H}{H\&D(2023)}
\defcitealias{Pontoppidan2024ApJ...963..158P}{KP5}
\defcitealias{McClure2009ApJ...693L..81M}{McClure(2009)}

\newcommand{\orcid}[1]{\href{https://orcid.org/#1}{\includegraphics[width=8pt]{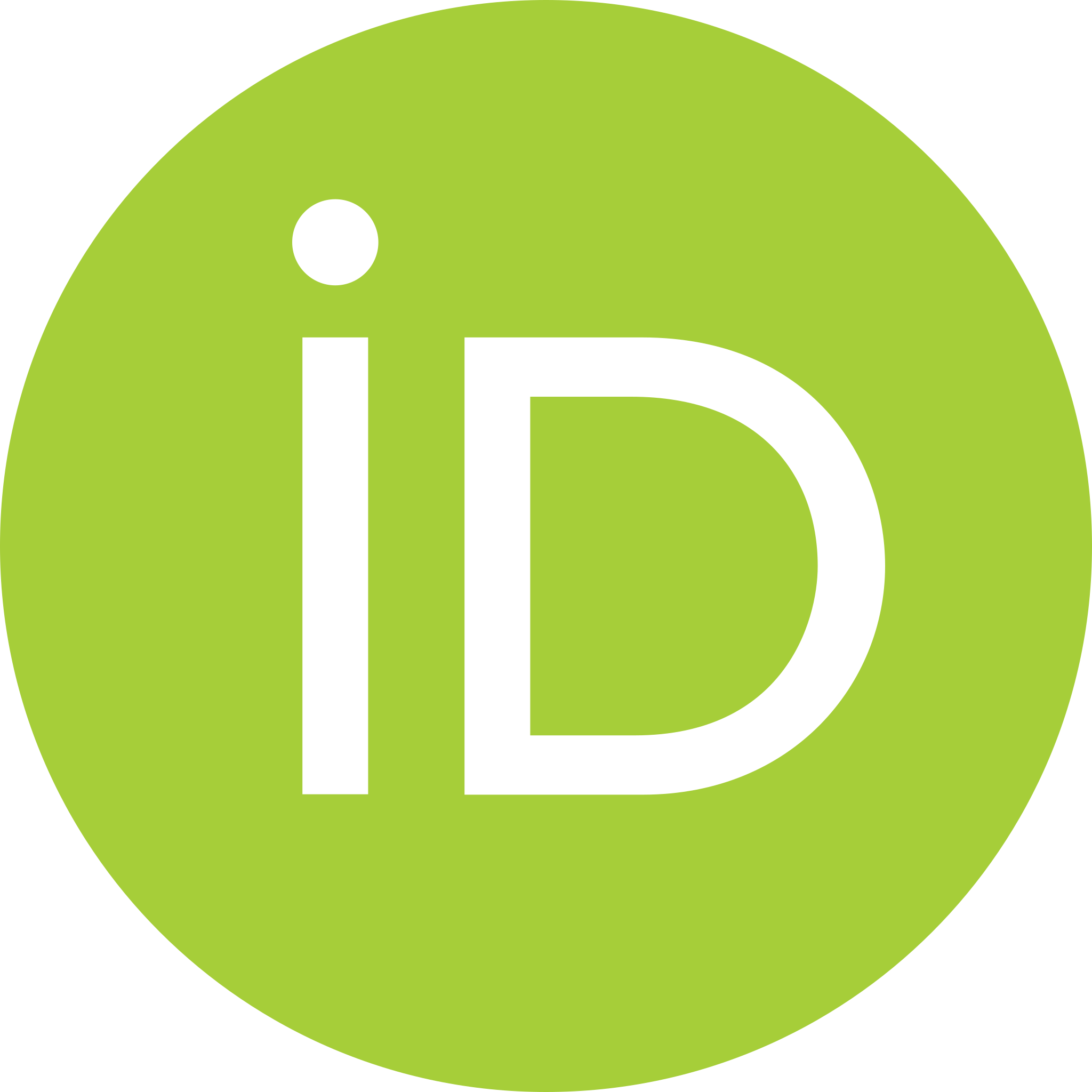}}}
\usepackage{hyperref}
\hypersetup{colorlinks = true, citecolor = blue, linkcolor = red, urlcolor = blue, filecolor = cyan, pdfstartview = FitH}

\begin{document}

\title{JWST/MIRI Detection of Molecular H$_2$ Winds from an Edge-on Class II Source HV Tau C}
\titlerunning{JWST detection of molecular H$_2$ winds from HV Tau C}
\authorrunning{Pathak, V. C., et al.}

\author{Vinod Chandra Pathak\orcid{0009-0006-3123-053X}\inst{1}
        \and P. Manoj\orcid{0000-0002-3530-304X}\inst{1}
        \and Himanshu Tyagi\orcid{0000-0002-9497-8856}\inst{1}
        \and B. Shridharan\orcid{0000-0002-2585-0111}\inst{1}
        \and Th. Henning\orcid{0000-0002-1493-300X}\inst{2}
        \and I. Kamp\orcid{0000-0001-7455-5349}\inst{3}
        \and A. Caratti o Garatti\orcid{0000-0001-8876-6614}\inst{4}
        \and B. Banerjee\orcid{0000-0001-8075-3819}\inst{1}
        \and Mayank Narang\orcid{0000-0002-0554-1151}\inst{5}
        \and E. F. van Dishoeck\orcid{0000-0001-7591-1907}\inst{6,7}
        \and Kamber Schwarz\orcid{0000-0002-6429-9457}\inst{2}
        \and Manuel G\"udel\orcid{0000-0001-9818-0588}\inst{8,9}
        \and Alice Somigliana\orcid{0000-0003-2090-2928}\inst{2}
        \and Giulia Perotti\orcid{0000-0002-8545-6175}\inst{11,2}
        \and Aditya M. Arabhavi\orcid{0000-0001-8407-4020}\inst{3}
        \and Sujay Vijay Jadhav\orcid{0009-0007-8087-5657}\inst{1}
        \and G\"oran Olofsson\orcid{0000-0003-3747-7120}\inst{10}
        }

\institute
{Department of Astronomy and Astrophysics, Tata Institute of Fundamental Research, Homi Bhabha Road, Colaba, Mumbai 400005, India\\
\email{vinodchandrapathak139@gmail.com, vinod.pathak@tifr.res.in}
\and Max-Planck-Institut für Astronomie (MPIA), Königstuhl 17, D-69117 Heidelberg, Germany
\and Kapteyn Astronomical Institute, Rijksuniversiteit Groningen, Postbus 800, 9700AV Groningen, The Netherlands
\and INAF – Osservatorio Astronomico di Capodimonte, Salita Moiariello 16, 80131 Napoli, Italy
\and Jet Propulsion Laboratory, California Institute of Technology, 4800 Oak Grove Drive, Pasadena, CA 91109, USA  
\and Leiden Observatory, Leiden University, 2300 RA Leiden, The Netherlands
\and Max-Planck-Institut f\"ur extraterrestrische Physik (MPE), Giessenbachstr.\ 1, 85748 Garching, Germany
\and Dept. of Astrophysics, University of Vienna, T\"urkenschanzstr. 17, A-1180 Vienna, Austria
\and ETH Z\"urich, Institute for Particle Physics and Astrophysics, Wolfgang-Pauli-Str. 27, 8093 Z\"urich, Switzerland
\and Department of Astronomy, Stockholm University, AlbaNova University Center, 10691 Stockholm, Sweden
\and Niels Bohr Institute, University of Copenhagen, NBB BA2, Jagtvej 155A, 2200 Copenhagen, Denmark
}

\abstract
{The evolution of protoplanetary disks is regulated by the interplay between accretion onto the central star and mass loss through jets and winds. While atomic and ionized outflows are commonly observed, direct detections of molecular winds in evolved Class~II disks remain rare, limiting our understanding of their role in disk evolution and dispersal.}
{We aim to characterize the spatial, thermal, kinematic, and dynamical properties of molecular hydrogen (H$_2$) emission associated with the nearly edge-on Class~II disk HV~Tau~C, and to assess the dynamical impact of its molecular wind. We also simultaneously constrain the accretion properties using H\,\textsc{i} recombination lines detected in the same mid-infrared (mid-IR) spectrum, enabling us to investigate the accretion--ejection connection in this system.}
{We analyze mid-IR \textit{JWST}/MIRI-MRS data from the MINDS Cycle~1 GTO program, extracting spatially resolved pure-rotational H$_2$ emission. Rotational diagrams are used to determine excitation temperatures and column densities, while position--velocity diagrams provide the kinematics of the extended emission. Combining excitation and kinematic information, we estimate the wind mass-loss rate, momentum rate, and mechanical luminosity. Accretion luminosities and mass-accretion rates are derived from H\,\textsc{i} lines, accounting for uncertainties in visual extinction due to the edge-on geometry.}
{We detect spatially extended pure-rotational H$_2$ emission from HV~Tau~C, revealing a wide-angled, biconical molecular wind extending well beyond the near-infrared scattered-light disk, the ALMA 887~$\mu$m dust continuum, and the compact $^{12}$CO ($J$=3--2) gas disk. The H$_2$ rotational diagram requires at least two temperature components: a warm component at $\sim$600~K and a hot component at $\sim$2000~K, similar to those found in much younger protostars. The kinematics indicate outward motions of a few tens of km\,s$^{-1}$ and dynamical timescales of tens to hundreds of years. The inferred mass-loss rate is about $10^{-8}\,M_\odot\,\mathrm{yr^{-1}}$, while accretion rates lie between $10^{-10}$ and $10^{-8}\,M_\odot\,\mathrm{yr^{-1}}$. The measured accretion rate may be underestimated due to geometrical effects, and the true accretion rate could be higher than that inferred from the MIR H\,\textsc{i} lines.}
{These results suggest that wide-angled molecular H$_2$ winds may persist into the Class~II phase, with inferred outflow rates that are comparable to those observed in some protostellar systems. Such outflows may play a significant role in angular momentum removal, disk evolution, and disk dispersal even in relatively evolved systems.}


\date{Submitted March 13, 2026 / Accepted July 26, 2026}

\maketitle
\nolinenumbers

\section{Introduction}\label{sec:introduction}

Protoplanetary disks are the natural outcome of angular momentum conservation during the gravitational collapse of molecular cloud cores and represent the sites where planetary systems form \citep[e.g.,][]{Shu1987ARA&A..25...23S, Stahler2004fost.book.....S, Hartmann2009apsf.book.....H, Williams2011ARA&A..49...67W, Armitage2020apfs.book.....A, Tabone2024A&A...691A..11T}. These disks supply the gas and dust reservoir required for planet formation over timescales of a few million years, while simultaneously regulating the mass and angular momentum evolution of young stars \citep[e.g.,][]{Andrews2020ARA&A..58..483A, Armitage2020apfs.book.....A}. Their evolution is governed by the coupled processes of mass accretion onto the central star and mass loss through collimated jets and wide-angled winds, which together control stellar mass assembly, disk lifetimes, and the eventual dispersal of circumstellar material \citep[e.g.,][]{Hartmann2009apsf.book.....H, Frank2014prpl.conf..451F, Bally2016ARA&A..54..491B, Pascucci2023ASPC..534..567P}.

Numerous surveys of young star-forming regions have demonstrated that protoplanetary disks are short-lived, with infrared excess emission tracing warm dust declining rapidly after $\sim$3--5~Myr and only a small fraction of disks surviving beyond $\sim$10~Myr \citep[e.g.,][]{Haisch2001ApJ...553L.153H, Hernandez2007ApJ...662.1067H, Hillenbrand2008PhST..130a4024H, Ribas2014A&A...561A..54R, Pfalzner2024ApJ...963..122P}. Viscous accretion alone cannot account for the observed rapid dispersal of disks, as the large angular momentum reservoir of disk material cannot be efficiently removed on a few Myr timescales. Moreover, studies indicate that disk dispersal proceeds in an inside-out, two-timescale manner, where the final clearing phase is much shorter than the overall disk lifetime \citep[i.e.,][]{Koepferl2013MNRAS.428.3327K, Ercolano2015MNRAS.452.3689E}. This indicates that additional mass-loss mechanisms must operate alongside viscous accretion to regulate disk evolution \citep[e.g.,][]{Alexander2014prpl.conf..475A, Ercolano2017RSOS....470114E}. Disk clearing, therefore, sets the timescale available for giant planet formation and ultimately shapes the architecture of emerging planetary systems \citep[e.g.,][]{Williams2011ARA&A..49...67W, Armitage2020apfs.book.....A}.

Mass loss from protoplanetary disks is now understood to arise from a combination of interconnected processes. High angular and spectral resolution observations of young stellar objects (YSOs) have revealed that disk-driven outflows are a ubiquitous feature of star formation. These outflows generally manifest in two components: highly collimated jets with opening angles $\lesssim15^{\circ}$ and velocities of several hundred km~s$^{-1}$, and slower, wide-angle winds with opening angles of $\sim$40-50$^{\circ}$ and velocities of a few tens of km~s$^{-1}$ \citep[e.g.,][]{Ray2007prpl.conf..231R, Frank2014prpl.conf..451F, Bally2016ARA&A..54..491B, Pascucci2023ASPC..534..567P, Narang2024ApJ...962L..16N, Tychoniec2024A&A...687A..36T, Schwarz2025ApJ...980..148S, Pascucci2025NatAs...9...81P, Bajaj2025AJ....169..296B}. These outflows play a central role in removing mass and angular momentum from the disk-star system.

Observationally, jets and winds are detected across all evolutionary stages of star formation. During the deeply embedded Class~0 and Class~I phases, powerful molecular outflows traced by CO, SiO, and H$_2$ dominate the mass-loss budget and have been extensively characterized through millimeter, infrared, and submillimeter observations \citep[e.g.,][]{Bontemps1996A&A...311..858B, Bachiller1996ARA&A..34..111B, Bally2016ARA&A..54..491B, Mottram2017A&A...600A..99M, Ray2021NewAR..9301615R, NarangMayank2025AJ....169..192N, ChinFeiLee2020A&ARv..28....1L}. High-resolution \textit{Atacama Large Millimeter Array} (\textit{ALMA}) observations have further revealed the detailed morphology and kinematics of these outflows, while recent \textit{James Webb Space Telescope} (\textit{JWST}) observations have begun to probe their warm molecular and atomic components with unprecedented sensitivity \citep[e.g.,][Tyagi et al. 2026, under review]{Vleugels2014A&A...568L...5C, Podio2015A&A...581A..85P, CarattioGaratti2024A&A...691A.134C, Federman2024ApJ...966...41F, Narang2024ApJ...962L..16N, Tychoniec2024A&A...687A..36T,  NarangMayank2025AJ....169..192N, vanDishoeck2025A&A...699A.361V, Vleugels2025A&A...695A.145V, Federman2026ApJ...998..282F, Narang2026ApJ..1004..188N, Francis2026A&A...711A.258F}.

In more evolved Class II systems, however, observational constraints on disk winds differ substantially. Most studies rely on high-resolution optical and ultraviolet (UV) spectroscopy, tracing atomic and ionized gas through forbidden emission lines such as [O\,\textsc{i}] at 6300~\AA\ and, in some cases, mid-IR diagnostics such as [Ne\,\textsc{ii}] at 12.81~$\mu$m (e.g., \citealt{Hartigan1995ApJ...452..736H, Frank2014prpl.conf..451F, Nisini2018A&A...609A..87N, Pascucci2023ASPC..534..567P}). While magnetohydrodynamic (MHD) winds are often invoked to explain these observations, radiation hydrodynamical simulations coupled with thermochemical models have shown that photoevaporative winds can reproduce several of these tracers under certain conditions (e.g., \citealt{Ercolano2016MNRAS.460.3472E, Weber2020MNRAS.496..223W, Weber2022MNRAS.517.3598W}). These observations demonstrate that jets and winds persist into the Class II phase, and in some cases reveal rotational signatures consistent with magneto-centrifugal launching scenarios (e.g., \citealt{Tabone2017A&A...607L...6T, Nazari2024A&A...686A.201N}). However, the molecular component of disk winds in evolved systems remains poorly constrained. This observational gap limits our understanding of how outflows transition from the embedded protostellar phase to optically visible disks and how long molecular material can survive under increasing stellar irradiation in disks and outflows.

From a theoretical perspective, several mechanisms have been proposed to drive disk winds during the Class~II phase, including MHD disk winds, and photoevaporative flows driven by high-energy irradiation, particularly X-rays (e.g., \citealt{Owen2010MNRAS.401.1415O, Picogna2019MNRAS.487..691P}), as well as UV radiation and hybrid models combining magnetic and thermal driving (e.g., \citealt{Bai2016ApJ...818..152B, Ercolano2017RSOS....470114E, Pascucci2023ASPC..534..567P, Kadam2025A&A...695A.167K, Nakatani2026A&A...706A.295N}). These models predict distinct wind geometries, kinematics, and thermochemical structures, as well as differing efficiencies in launching and sustaining molecular material. Molecular winds are therefore of particular interest, as they trace relatively dense gas lifted from the disk surface and can carry a significant fraction of the mass and angular momentum flux \citep[e.g.,][]{Nisini2018A&A...609A..87N, Trapman2022ApJ...926...61T, Tabone2022MNRAS.512.2290T, Kadam2025A&A...695A.167K}. Nevertheless, direct observational constraints on molecular winds from Class~II disks remain sparse \citep[but see][]{Arulanantham2024ApJ...965L..13A, Schwarz2025ApJ...980..148S, Pascucci2025NatAs...9...81P, Vlasblom2025A&A...693A.278V, Kurtovic2026A&A...705A..97K, Somigliana2026arXiv260623794S}.

Previous attempts to detect molecular winds in evolved disks have faced significant challenges. Near-infrared rovibrational lines of H$_2$ at 2.12 $\mu$m trace hot gas but are strongly affected by extinction and often lack the spatial resolution needed to disentangle disk, wind, and jet components \citep[e.g.,][]{Beck2008ApJ...676..472B}. Millimetre observations of CO with ALMA have revealed slow molecular winds in a few systems, but interpretation is complicated by optical depth effects, confusion with disk rotation, and limited sensitivity to warm gas \citep[e.g.,][]{Louvet2018A&A...618A.120L, Nisini2018A&A...609A..87N}. Mid-infrared spectroscopy with instruments like \textit{Spitzer} has enabled detections of pure-rotational H$_2$ lines, but the limited sensitivity and the lack of sufficient spatial and spectral resolution have prevented the localization of the emitting gas and detailed studies of its kinematics \citep[e.g.,][]{Lahuis2007ApJ...665..492L, Pontoppidan2010ApJ...720..887P, Gudel2010A&A...519A.113G}. Consequently, spatially resolved detections of warm molecular winds, particularly from H$_2$, in Class~II disks have remained rare. Detections of molecular winds have also been reported using other molecular lines, such as CO and H$_2$O \citep[e.g.,][]{Pontoppidan2011ApJ...733...84P, Bast2011A&A...527A.119B, Brown2013ApJ...770...94B, Banzatti2022AJ....163..174B}.

\begin{table}[htbp!]
\caption{Source properties of HV~Tau~C}
\label{tab:HV_Tau_C_source_Properties}
\centering
\resizebox{\columnwidth}{!}{
\begin{tabular}{lll}
\hline\hline
Parameter & Value & Refs \\
\hline
Spectral Type & K6 & 1 \\
Distance & $138.0 \pm 0.9$~pc & 2 \\
887~$\mu$m cont. RA & 04$^\mathrm{h}$38$^\mathrm{m}$35.50894$^\mathrm{s}$ & 3 \\
887~$\mu$m cont. Dec & +26$^\circ$10$'$41.1526$''$ & 3 \\
Stellar Mass ($M_*$) & $1.43 \pm 0.42$~$M_{\odot}$ & 3 \\
Stellar Luminosity ($L_*$) & $0.03 \pm 0.02$~$L_{\odot}$ & 1 \\
$T_{\mathrm{eff}}$ & $4200 \pm 200$~K & 1 \\
Inclination ($i$) & $79.2 \pm 0.5\degree$ (nearly edge-on) & 3 \\
PA (Disk Major Axis) & $108.8 \pm 0.5\degree$ & 3 \\
Dust Disk Diameter ($R_{\mathrm{out}}$) & $1.35 \pm 0.04\arcsec$ & 3 \\
Gas Disk Diameter ($R_{\mathrm{out}}$) & $3.96\arcsec$ & 3 \\
Dust Disk Mass ($M_{\mathrm{disk}}^{\mathrm{dust}}$) & $(5.3 \pm 0.6) \times 10^{-5}$~$M_{\odot}$ & 3 \\
$V_{\mathrm{LSR}}$ & $6.45 \pm 0.06$~km~s$^{-1}$ & 3 \\
Accretion Luminosity ($L_{\mathrm{acc}}$) & $< 0.06$~$L_{\odot}$ & 3 \\
Bolometric Luminosity ($L_{\mathrm{bol}}$) & $0.071$~$L_{\odot}$ & 3 \\
Mass Accretion Rate ($\dot{M}_{\mathrm{acc}}$) & $10^{-8}$--$10^{-10}$~$M_{\odot}$~yr$^{-1}$ & 3 \\
\hline
\end{tabular}
}
\tablefoot{References: (1) \citet{Simon2019ApJ...884...42S}; \\ 
(2) \citet{gaia2023}; (3) this work.}
\end{table}

The \textit{JWST} has opened a new observational window onto molecular H$_2$ gas in protoplanetary disks and their associated outflows. In particular, the Mid-Infrared Instrument (MIRI) Medium Resolution Spectrometer (MRS) \citep{Argyriou2023AA...675A.111A, Wright2023PASP..135d8003W} provides unprecedented sensitivity and spatial resolution for detecting pure rotational H$_2$ emission, enabling direct constraints on the spatial distribution, excitation conditions, and kinematics of warm molecular gas. These capabilities make \textit{JWST}/MIRI uniquely suited to probing the molecular component of disk winds that has remained largely inaccessible to previous facilities \citep[e.g.,][]{Schwarz2025ApJ...980..148S, Pascucci2025NatAs...9...81P, Vlasblom2025A&A...693A.278V, Bajaj2025AJ....169..296B, Somigliana2026arXiv260623794S}. As part of the \textit{JWST}/MIRI Mid-Infrared Disk Survey (MINDS) Guaranteed Time Observation (GTO) program \citep{Henning2024PASP..136e4302H}, we investigate disk structure, chemistry, and gas dispersal processes during the Class~II phase. In this context, the present study focuses on the Class~II source HV~Tau~C, for which \textit{JWST}/MIRI observations provide a unique opportunity to investigate molecular winds during the planet-forming phase of disk evolution.

HV~Tau~C is a young Classical T~Tauri star located in the Taurus star-forming region at a distance of $\sim$138~pc and constitutes the widely separated ($\sim4^{\prime\prime}$, $\sim$550~AU) component of the hierarchical HV~Tau triple system, whose close A-B pair is diskless and weak-lined \citep{Simon1992ApJ...384..212S, Reipurth1993A&A...278...81R, White2001ApJ...556..265W}. Initially misidentified as a Herbig-Haro object, HV~Tau~C is now firmly established as a young stellar object surrounded by a nearly edge-on protoplanetary disk \citep{Woitas1998A&A...338..122W, Monin2000A&A...356L..75M}. The central star has a spectral type of K6-K7, an effective temperature of $\sim$4200~K, and an estimated age of $\sim$1-2~Myr \citep{White2004ApJ...616..998W, Simon2019ApJ...884...42S}. Dynamical measurements indicate a stellar mass of $1.3$-$1.6~M_{\odot}$, significantly higher than expected from its spectral type \citep{Guilloteau2014A&A...567A.117G, Simon2019ApJ...884...42S}. Scattered-light and millimetre observations reveal a compact dust disk embedded within a more extended gaseous disk in Keplerian rotation, viewed at a high inclination \citep{Stapelfeldt2003ApJ...589..410S, Simon2019ApJ...884...42S}. The system is actively accreting and drives a bipolar outflow consisting of a fast atomic jet and a slower, wide-angle molecular component traced by forbidden atomic lines and near-infrared H$_2$ emission \citep{Stapelfeldt2003ApJ...589..410S, White2004ApJ...616..998W, Beck2008ApJ...676..472B}. The key source properties are listed in Table \ref{tab:HV_Tau_C_source_Properties}. The nearly edge-on geometry of HV~Tau~C provides particularly favourable conditions for studying wide-angled disk winds. The strong attenuation of the stellar continuum enhances the contrast of extended emission above and below the disk midplane, while the viewing geometry allows kinematic signatures of outflowing material to be disentangled from rotational motion.

In this work, we present a detailed study of molecular hydrogen emission from HV~Tau~C using \textit{JWST}/MIRI-MRS observations, complemented by archival ALMA interferometric data. We detect spatially extended pure rotational H$_2$ emission tracing a wide-angle molecular wind emerging from the disk. We characterize its thermal structure using H$_2$ rotational diagrams and its kinematics through position-velocity analysis. From these diagnostics, we derive key dynamical properties of the wind, including mass-loss rates, momentum rates, and mechanical luminosities. We also estimate accretion luminosities and mass accretion rates from H~\textsc{i} recombination lines detected in the same MIRI spectra, accounting for the nearly edge-on geometry. This combined analysis allows us to assess the dynamical role of molecular winds in the evolution of the HV~Tau~C disk during the Class~II phase.

The paper is organized as follows: Section~\ref{sec:obs_and_data_reduction} describes the observations and data reduction; Section~\ref{sec:hv_tau_c_alma_characteristics} presents the stellar properties and disk structure derived from ALMA data; Section~\ref{sec:overall_1D_spectra} summarizes the spectral features detected in the MIRI/MRS spectrum of HV~Tau~C; Section~\ref{sec:extended_H2_emission} presents the analysis of the extended H$_2$ emission, including its morphology (Section~\ref{subsec:extended_H2_morphology}), semi-opening angle measurements (Section~\ref{subsec:winds_semi_opening_angle}), excitation conditions (Section~\ref{subsec:excitation_conditions_winds}), kinematics (Section~\ref{subsec:winds_kinematics}), and dynamical properties (Section~\ref{subsec:h2_wind_dynamics}); Sections~\ref{sec:extinction_Av_text} and~\ref{sec:HI_accretion_measurement} present the extinction and accretion analysis; and, Sections~\ref{sec:discussion} and~\ref{sec:conclusions} discuss the implications and summarize our findings.

\section{Observations and Data Reduction} \label{sec:obs_and_data_reduction}

\subsection{JWST MIRI/MRS} \label{sec:miri_mrs}

HV~Tau~C was observed with the MIRI (\citealt{Wells2015PASP..127..646W, Labiano2021AA...656A..57L, Wright2015PASP..127..595W, Wright2023PASP..135d8003W})/MRS (\citealt{Wells2015PASP..127..646W, Argyriou2023AA...675A.111A}) instrument onboard the \textit{JWST}, as part of the Cycle~1 GTO program (Program ID:~1282; PI:~Thomas Henning). These observations were carried out on 2023~September~27, with a total exposure time of 1236.116~s ($\sim$20.6~min), employing a four-point dither pattern optimized for single-mosaic point-source spectroscopy.

The MIRI/MRS instrument provides spectral coverage with a resolving power ranging from approximately 1500~to~4000, depending on the wavelength \citep{Jones2023MNRAS.523.2519J, Argyriou2023AA...675A.111A, Pontoppidan2024ApJ...963..158P}. HV~Tau~C was observed across all four MIRI/MRS channels: 4.9--7.65~$\mu$m (Channel~1), 7.51--11.7~$\mu$m (Channel~2), 11.55--17.98~$\mu$m (Channel~3), and 17.7--27.9~$\mu$m (Channel~4), thus covering the 4.9--27.9~$\mu$m wavelength range. The spatial resolution of MIRI/MRS ranges from approximately $0.27\arcsec$ (at 5~$\mu$m) to $1.03\arcsec$ (at 28~$\mu$m) \citep{Law2023AJ....166...45L}. The spatial field of view varies by channel, ranging from approximately $3.2\arcsec\times3.7\arcsec$ for Channel~1~Short to $6.6\arcsec\times7.7\arcsec$ for Channel~4~Long.

The data were retrieved from the Mikulski Archive for Space Telescopes (MAST). The data reduction was performed using the \textit{JWST} Science Data Processing Pipeline (v2025.1; \citealt{Bushouse2024zndo..14153298B}), together with calibration reference files version~1.18.0 and the Calibration Reference Data System (CRDS) context \texttt{jwst\_1364.pmap} (CRDS version~12.1.1). The reduction followed the standard three-stage workflow, similar to the procedures described in 
\citet{narang_et_al_2024, Federman_et_al_2024, Neufeld_et_al_2024ApJ...966L..22N, tyagi_et_al_2025}. The standard three-stage pipeline was applied: detector-level calibration (\texttt{Stage~1}), spectral calibration (\texttt{Stage~2}), and cube building with spectral extraction (\texttt{Stage~3}). The same dataset was used as in \citet{Shridharan2026A&A...708A..22S}. This resulted in fully calibrated three-dimensional spectral data cubes and one-dimensional extracted spectra across all 12 MIRI/MRS sub-bands.

To mitigate the effects of bad pixels, we implemented a pixel replacement algorithm using the minimum gradient (\texttt{mingrad}) method during Stage~2. A residual fringe correction was also applied at the cube level. In Stage~3, spectral cubes were generated using the \texttt{output\_type=`band'} setting, which automatically organizes the data by spectral band. Outlier detection was enabled (\texttt{outlier\_detection.skip=False}) with an $11\times11$~pixel kernel (\texttt{kernel\_size=`11 11'}) and a 99.5\% detection threshold (\texttt{threshold\_percent=99.5}), alongside a secondary artifact check within the IFU data cubes (\texttt{ifu\_second\_check=True}). For the 1D spectral extraction, automatic source centering (\texttt{ifu\_autocen=True}) and residual fringe correction (\texttt{ifu\_rfcorr=True}) were enabled to optimize the accuracy of the extracted spectra. The final data products consist of flux-calibrated 3D spectral datacubes and corresponding 1D extracted spectra for all 12~MIRI/MRS sub-bands.

\subsection{Near-InfraRed Camera (NIRCam) Imaging} \label{sec:nircam_imaging}

We used publicly available imaging data from the JWST/NIRCam instrument \citep{Rieke2023PASP..135b8001R}. The HV~Tau system was observed on 2024 March 02 as part of the JWST Cycle 2 General Observer (GO) Program (Program ID: 4290; PI: François Ménard). The third component of the system, HV~Tau~C, lies within the field of view (FOV). We utilized the NIRCam imaging to trace scattered light and thermal emission from the nearly edge-on disk around HV~Tau~C.

NIRCam observations were obtained in FULL subarray mode using a 4-point standard dither pattern to optimize spatial coverage and mitigate detector artifacts. Imaging was performed with four filters: F115W and F200W in the short-wavelength (SW) channel, and F300M and F460M in the long-wavelength (LW) channel. Each integration simultaneously employed one SW and one LW filter, corresponding to the F115W/F300M and F200W/F460M pairs. Although acquired in dual-channel mode, the data products are provided as separate FITS files for each filter. The F115W and F200W mosaics cover a FOV of approximately $133\arcsec \times 133\arcsec$, sampled at a pixel scale of $0.031\arcsec$, with image dimensions of $4346 \times 4325$ and $4347 \times 4324$, respectively. The reference sky position is $(\alpha,\, \delta) = (04^{\mathrm{h}}38^{\mathrm{m}}32.77^{\mathrm{s}},\, +26^{\circ}10'13.8\arcsec)$ for both filters. The F300M and F460M filters were observed at a coarser pixel scale of $0.063\arcsec$, with image dimensions of $2079 \times 2070$ and $2079 \times 2071$, respectively, and centered at $(\alpha,\, \delta) \approx (04^{\mathrm{h}}38^{\mathrm{m}}32.78^{\mathrm{s}},\, +26^{\circ}10'14.8\arcsec)$. HV~Tau was imaged in all four bands with a total exposure time of 515.364~s per filter.
We used the publicly available, pipeline-reduced Level~3 science data products downloaded from the MAST.

\subsection{ALMA} \label{sec:alma}

We use publicly available ALMA observations of HV~Tau~C from Cycles 4 and 5 (Project~ID:~2016.1.00460.S; PI:~Fran\c{c}ois M\'{e}nard,\citealt{Villenave2020A&A...642A.164V}).
Our analysis focuses on high-resolution Band 7 continuum data of HV~Tau~C centered at 338.2632~GHz (887~$\mu$m). The dataset consists of a four-dimensional FITS cube with dimensions $1024 \times 1024 \times 1 \times 1$, corresponding to Right Ascension, Declination, frequency, and Stokes~I. The pixel scale is 0.01$''$ in both spatial directions, resulting in a 1024 × 1024 pixel image covering $\sim$10.2$'' \times 10.2''$. The restoring beam is $0.11\arcsec \times 0.08\arcsec$ with a position angle of $3.92^{\circ}$ east of north. The celestial coordinate system is FK5 (J2000), and the image intensity is expressed in Jy\,beam$^{-1}$.

We also analyzed ALMA Band~7 observations targeting the $^{12}$CO~($J$=3--2) pure rotational transition at a rest frequency of 345.796~GHz. The dataset comprises a three-dimensional spectral cube with dimensions 576 × 576 × 80, corresponding to Right Ascension, Declination, and frequency, and contains only Stokes~I emission. The spectral axis includes 80 channels with frequency decreasing across channels, spanning 345.8075--345.7711~GHz. This corresponds to a velocity range of $-10$ to $+21.6$~km~s$^{-1}$ in the Local Standard of Rest (kinematic; LSRK) reference frame using the radio velocity definition. The pixel scale is $0.06\arcsec$ in both spatial directions, corresponding to an image size of $\sim$34.6$\arcsec \times 34.6\arcsec$ for the $576 \times 576$ pixel map. The synthesized beam is $0.48\arcsec \times 0.34\arcsec$ with a position angle of $13.25^{\circ}$ east of north. The data are expressed in units of Jy\,beam$^{-1}$. The celestial coordinate system follows the ICRS, and the spectral frame is referenced to the LSRK standard.

\section{Stellar and Disk Properties of HV Tau C from ALMA Observations} \label{sec:hv_tau_c_alma_characteristics}

\begin{figure*}[htbp!]
    \centering
    \begin{minipage}[t]{0.350\textwidth}
        \centering
        \includegraphics[width=\linewidth]{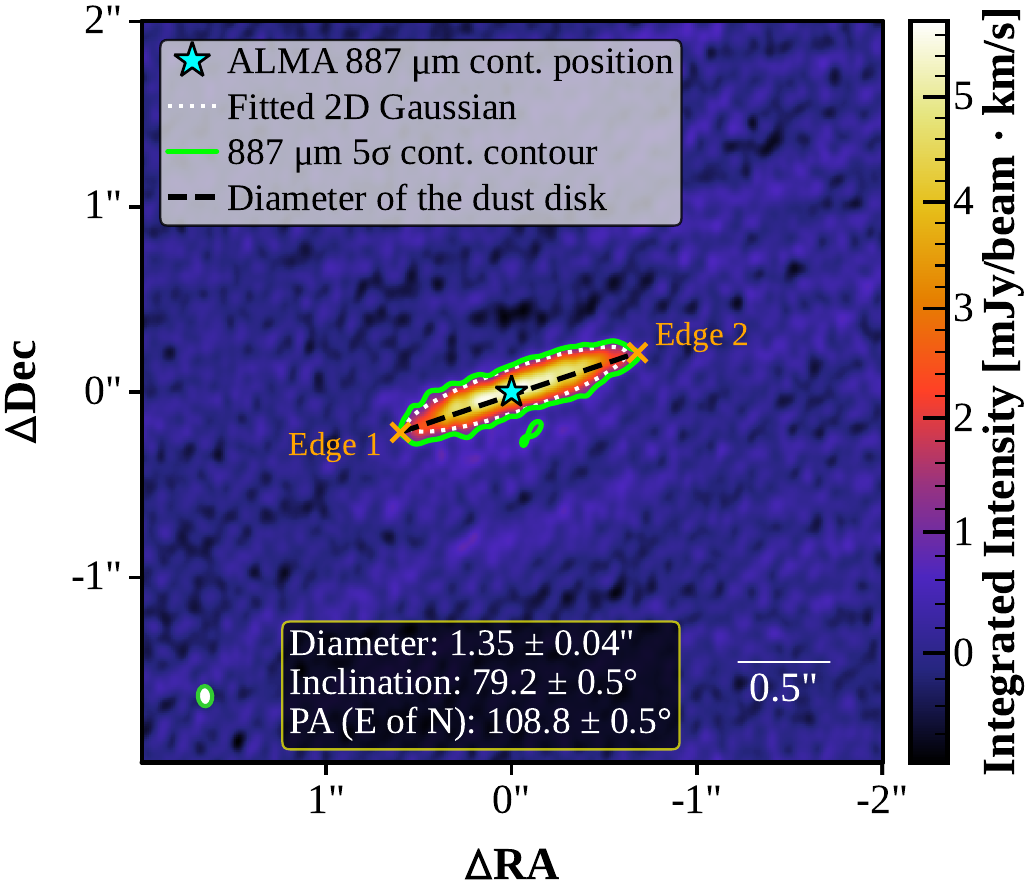}
        \subcaption{ALMA Band~7 continuum image.}\label{fig:HV_Tau_C_ALMA_sub1}
    \end{minipage}
    \hfill
    \begin{minipage}[t]{0.640\textwidth}
        \centering
        \includegraphics[width=\linewidth]{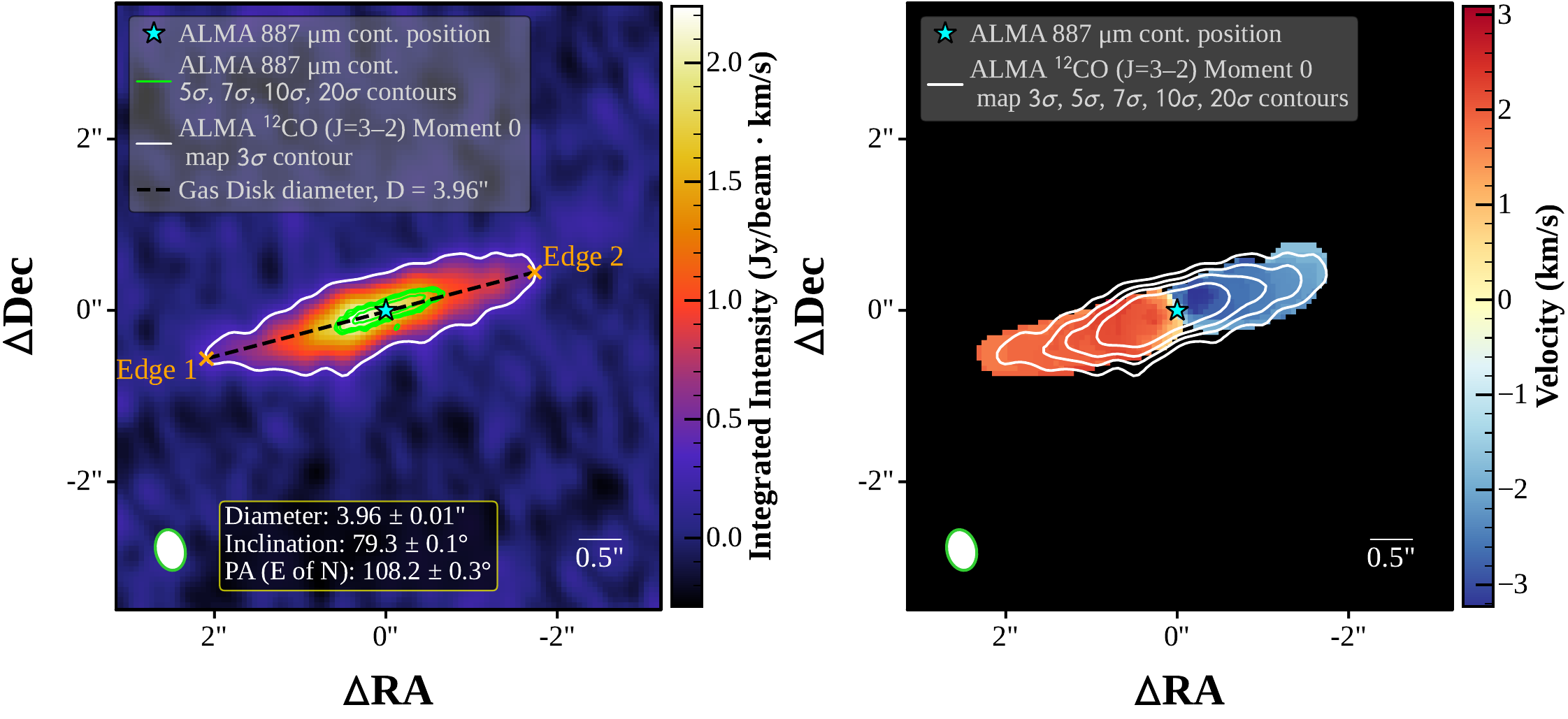}
        \subcaption{$^{12}$CO ($J$=3--2) moment~0 and moment~1 maps.}\label{fig:HV_Tau_C_ALMA_sub2}
    \end{minipage}
    
    \vspace{0.02\textwidth}
    
    \begin{minipage}[t]{0.605\textwidth}
        \centering
        \includegraphics[width=\linewidth]{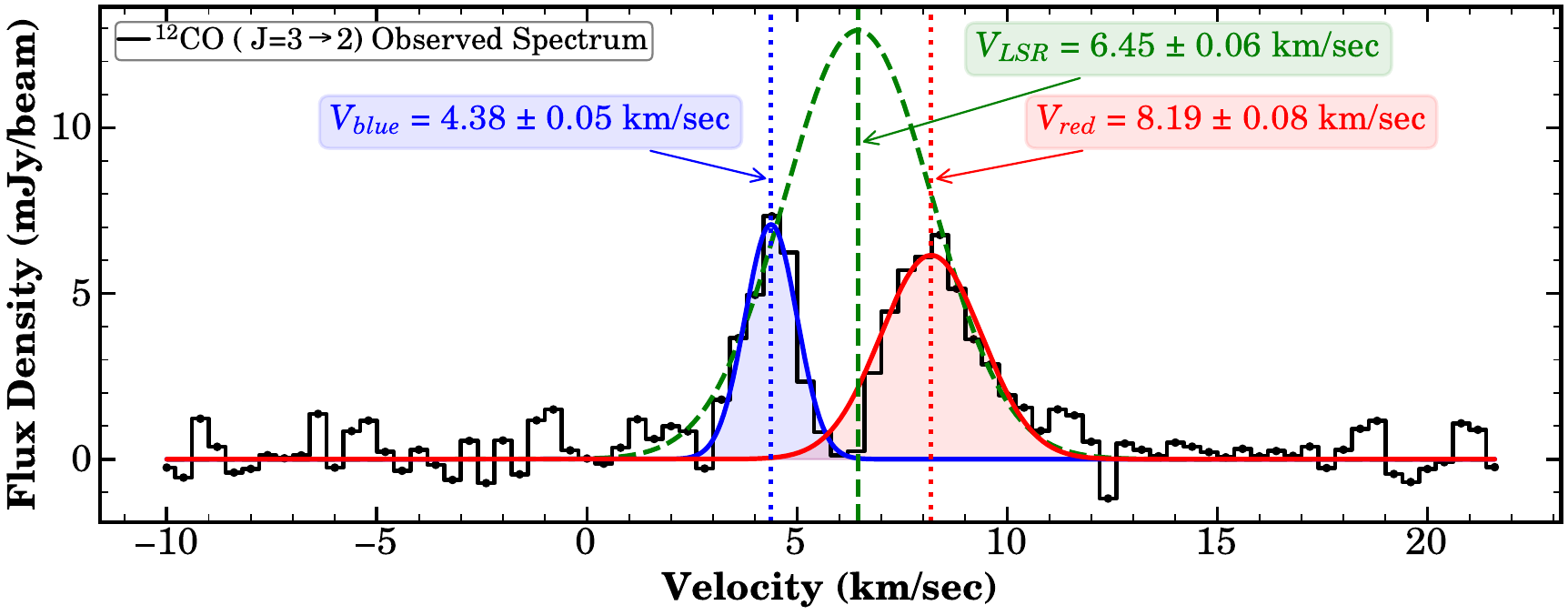}
        \subcaption{$^{12}$CO ($J$=3--2) Spectral Profile.}\label{fig:HV_Tau_C_ALMA_sub3}
    \end{minipage}
    \hfill
    \begin{minipage}[t]{0.385\textwidth}
        \centering
        \includegraphics[width=\linewidth]{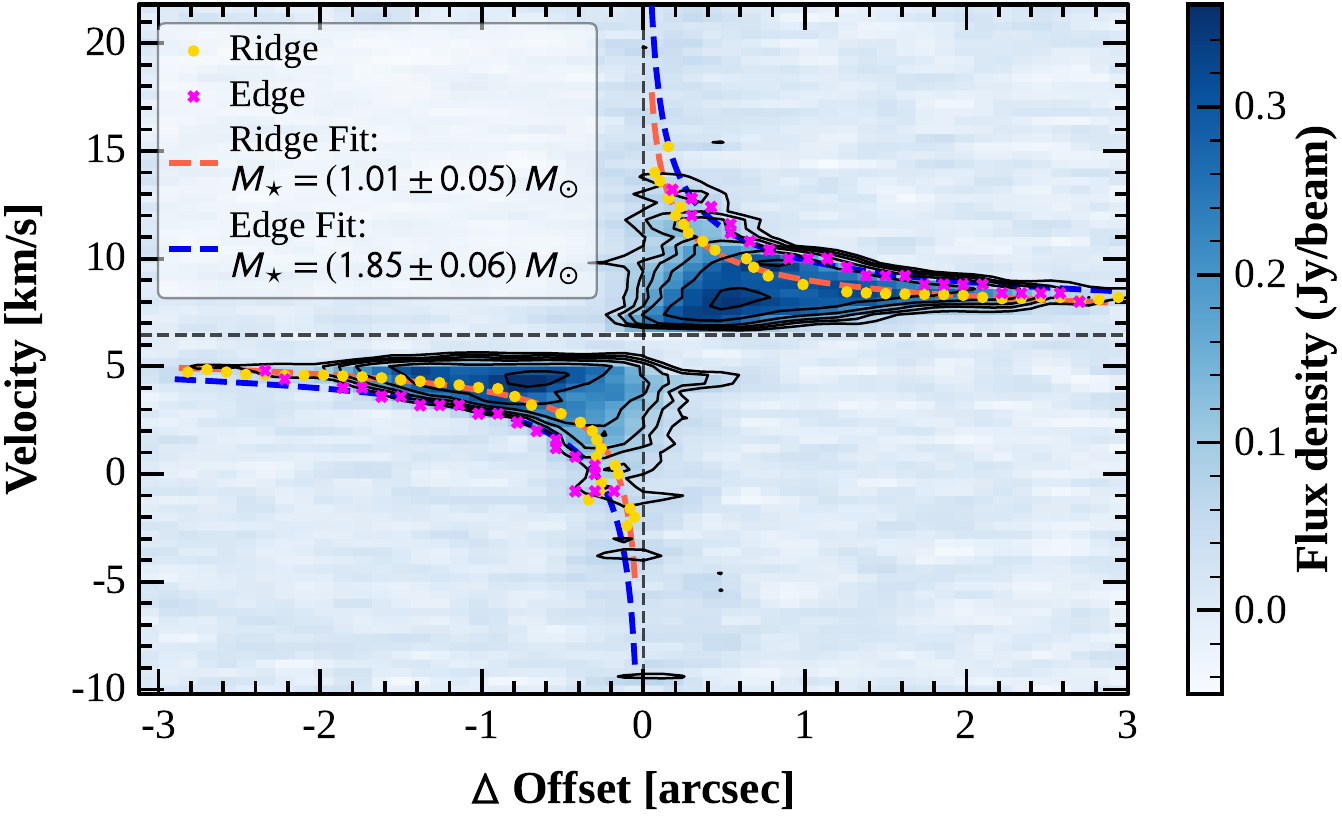}
        \subcaption{PV diagram with fitted ridge and edge profiles.}\label{fig:HV_Tau_C_ALMA_sub4}
    \end{minipage}
    
    \caption{\small ALMA Band~7 and $^{12}$CO ($J$=3--2) observations of HV~Tau~C.
    (a) Continuum image at 887~$\mu$m (color scale). The lime-green contour marks the 5$\sigma$ level, outlining the dust disk. The star symbol indicates the continuum peak, and the black dashed line traces the disk diameter connecting the two extreme points (``Edge1'' and ``Edge2'') on the 5$\sigma$ contour along the major axis. 
    (b) Moment~0 (left) and moment~1 (right) maps of $^{12}$CO ($J$=3--2). Lime-colored contours denote the 887~$\mu$m continuum; white contours denote the 3$\sigma$ CO emission. The gas disk diameter is measured by connecting edge points along the 3$\sigma$ contour.
    (c) $^{12}$CO ($J$=3--2) observed spectral profile (black curve) with Gaussian fits: blue- and red-shifted components (blue and red), and the total fit (green) centered at the systemic velocity $v_{\rm LSR}$. 
    (d) Position--velocity diagram of $^{12}$CO ($J$=3--2) emission extracted along the disk major axis. Black contours represent intensities above 3$\sigma$. Ridge and edge profiles are shown in yellow and magenta, respectively, with their Keplerian fits in red and blue.}
    \label{fig:HV_Tau_C_ALMA_data}
\end{figure*}

In this section, we present a comprehensive analysis of the ALMA Band 7 continuum and $^{12}$CO ($J$=3--2) line cube observations of HV Tau C, aiming to characterize both the dust and gas disk geometries. We also derive the dust mass and the stellar dynamical mass of the central pre-main-sequence star, which will be used in subsequent analyses.

\subsection{Disk Geometry: Size, Inclination, and Position Angle}

Figure~\ref{fig:HV_Tau_C_ALMA_sub1} shows the ALMA Band~7 continuum image of HV~Tau~C at an observed frequency of 338.263~GHz (corresponding to a wavelength of 887~$\mu$m). A two-dimensional elliptical Gaussian fit was performed to determine the central position and morphological parameters of the dust emission. The fit yields a peak flux of ($6.0 \pm 0.1$) mJy~beam$^{-1}$, with full width at half-maximum (FWHM) sizes of $(0.881 \pm 0.021)\arcsec \times (0.165 \pm 0.003)\arcsec$, at a position angle (PA) of $(108.8 \pm 0.3)^\circ$ east of north. The centroid of the emission is located at (RA, Dec) = (04$^{\mathrm h}$38$^{\mathrm m}$35.5089$^{\mathrm s}$, +26$^\circ$10$'$41.1526$''$).

To estimate the disk size and inclination ($i$), we measured the spatial extent of emission above the $5\sigma$ level. The RMS noise was estimated to be $\sigma \simeq 0.18$~mJy~beam$^{-1}$ from emission-free regions of the image. After rotating the image by the measured PA to align the disk horizontally, the major and minor axis extents were determined by counting the number of pixels above the $5\sigma$ threshold along the horizontal and vertical directions, respectively. This yields projected dimensions of 1.35\arcsec $\times$ 0.25\arcsec.

Assuming the disk is geometrically thin and intrinsically axisymmetric -- a standard approximation for estimating disk inclinations from projected axial ratios -- the apparent major-axis corresponds to the physical diameter of the disk. The inclination of the disk, measured from a face-on orientation, can be calculated from the axial ratio using the relation $(\text{minor axis})/(\text{major axis}) = \cos i$. The physical radius $r$ of the disk is calculated using the relation $\tan\theta = r / D$, where $\theta$ is the angular radius and $D$ is the distance to the object. Since the source HV Tau C is at a distance of ~138~pc (see Table \ref{tab:HV_Tau_C_source_Properties}), we estimate the dust disk radius and inclination to be approximately 92.8~AU and $79.2^\circ$, respectively.

From the $^{12}$CO ($J$=3--2) spectral cube, we generated the integrated intensity map (moment~0). Figure~\ref{fig:HV_Tau_C_ALMA_sub2} (left) presents the $^{12}$CO ($J$=3--2) moment~0 map (in color scale) overlaid with continuum contours at 5$\sigma$, 7$\sigma$, 10$\sigma$, and 20$\sigma$ levels. To estimate the RMS noise level ($\sigma$) of the CO emission, we defined a circular aperture of radius 6\arcsec centered on the continuum peak and selected velocity channels devoid of detectable CO emission (i.e. line-free channels). The RMS noise, computed from this region, is 27.40~mJy\,beam$^{-1}$.

Following the same procedure used for the continuum image analysis, we fitted a two-dimensional Gaussian model to the CO moment-0 map to derive the geometrical properties of the gas disk. The best-fit position angle (PA) is $(108.2 \pm 0.3)^\circ$, with a major axis of $(2.65 \pm 0.03)\arcsec$ and a minor axis of $(0.49 \pm 0.01)\arcsec$. The corresponding inclination, derived from the axial ratio assuming a geometrically thin disk (i.e., $\cos i = b/a$), is $(79.3 \pm 0.5)^\circ$. We note that this estimate neglects the finite vertical scale height of the CO-emitting layer and therefore represents a first-order approximation of the disk inclination. The $^{12}$CO emission appears significantly more extended than the dust continuum emission, as expected due to the higher optical depth of the CO line and the radial drift of large dust grains. The spatial extent of the CO-emitting region, measured above the $5\sigma$ level, is approximately $3.96~\arcsec$ in diameter, about three times larger than the dust disk -- a well-known observational trend in protoplanetary disks \citep[e.g.,][]{Andrews2012ApJ...744..162A, Ansdell2018ApJ...859...21A, Trapman2019A&A...629A..79T}. The gas disk is well aligned with the dust continuum emission, indicating a coplanar configuration, with the derived PA and inclination consistent with those of the dust disk.

\subsection{Dust Mass Estimation}

To estimate the dust mass of the HV Tau C disk, we calculated the total flux density enclosed within the $5\sigma$ contour of the ALMA Band 7 continuum (at 887 $\mu$m) image. Assuming that this emission originates entirely from thermal radiation of dust grains, the dust mass ($M_{\rm dust}$) can be derived using the relation \citep[e.g.,][]{Hildebrand1983QJRAS..24..267H, Ansdell2016ApJ...828...46A}:
$$
 M_{\rm dust} = \frac{F_{\nu}\,d^2}{\kappa_{\nu}\,B_{\nu}(T_{\rm dust})},
$$
where $F_{\nu}$ is the total flux density, $d$ is the distance to the source, $\kappa_{\nu}$ is the dust opacity, and $B_{\nu}(T_{\rm dust})$ is the Planck function at the characteristic dust temperature $T_{\rm dust}$.

For the dust opacity, we adopt the commonly used expression from \citet{Beckwith1990AJ.....99..924B},
$ \kappa_{\rm \nu}=10\,(\nu/1000\,{\rm GHz})\;{\rm cm^2\,g^{-1}} $. This assumes a power-law dependence of the form $\kappa_{\nu} \propto \nu^{\beta}$ with an emissivity index $\beta = 1$, which is typical for protoplanetary disks. However, we note that the value of $\kappa_{\nu}$, and hence the estimated dust mass, carries significant uncertainty due to variations in grain properties and the assumed $\beta$.

Here, we assume a dust temperature of $T_{\rm dust} = 20$~K, corresponding to the median dust temperature of Taurus disks \citep{Andrews2005ApJ...631.1134A}. Adopting a distance of $d = 138$~pc and an integrated continuum flux of 89.8~mJy (measured within the $5\sigma$ contour), we estimate the total dust mass in the HV~Tau~C disk to be approximately $(5.3 \pm 0.6) \times 10^{-5}\,M_\odot$. This estimate is consistent with the value of $M_{\rm dust} = 3.84 \times 10^{-5}\,M_\odot$ reported for HV~Tau~C by \citet{Miotello2023ASPC..534..501M}. Assuming a canonical gas-to-dust ratio of 100, the corresponding total disk mass would be $(5.3 \pm 0.6) \times 10^{-3}\,M_\odot$.

\subsection{Gas Kinematics and Dynamical Stellar Mass}\label{subsec:dynamical_mass_using_keplerian_rotation}

Figure~\ref{fig:HV_Tau_C_ALMA_sub3} shows the $^{12}$CO ($J$=3--2) spectrum of HV~Tau~C, extracted within a circular aperture of 6\arcsec\ radius centered on the ALMA 887~$\mu$m continuum peak. The spectrum exhibits a clear double-peaked profile, a characteristic signature of Keplerian rotation in a circumstellar disk. We fitted Gaussian functions to both the blue- and red-shifted peaks to determine their centroid velocities. The systemic velocity ($V_{\mathrm{LSR}}$) of the star--disk system was obtained by fitting a single Gaussian to the entire CO profile.

Similar to the moment 0 map, we also generated an intensity-weighted velocity (moment~1) map, as shown in Figure \ref{fig:HV_Tau_C_ALMA_sub2} (right). To minimize noise contributions, the moment~1 map is constructed using only pixels with emission above the 5$\sigma$ threshold.  The moment~1 map reveals a clear velocity gradient along the disk’s major axis, consistent with rotation. The velocity structure further confirms that disk rotation occurs about an axis closely aligned with the projected minor axis.

To estimate the dynamical stellar mass of the central pre-main sequence star in HV~Tau~C, we construct a position-velocity (PV) diagram along the major axis of the gas disk. The PV diagram is generated by extracting slices perpendicular to the disk’s major axis, ensuring that the slices encompass the full extent of the $^{12}$CO ($J$=3--2) emission. Figure~\ref{fig:HV_Tau_C_ALMA_sub4} presents the resulting PV diagram, with overlaid intensity contours at 5$\sigma$, 7$\sigma$, 10$\sigma$, 20$\sigma$, and 100$\sigma$ levels. To derive the stellar mass, we fit Keplerian rotation curves separately to the red-shifted and blue-shifted components of the emission.

Assuming a geometrically thin, axisymmetric disk in Keplerian rotation, the rotational velocity at a radial distance $r$ from the central star is given by
$$
 V_{\rm rot} = \sqrt{\frac{G M_\star}{r}},
$$
where $G$ is the gravitational constant and $M_\star$ is the stellar mass. For an inclined disk, the observable component is the line-of-sight velocity, $V_{\rm los} = V_{\rm rot} \sin i$, where $i$ is the inclination angle, measured from face-on ($i = 0^\circ$) to edge-on ($i = 90^\circ$). This $V_{\rm los}$ corresponds to the velocity distribution observed in the PV diagram.

To derive the rotational velocity distribution from the PV diagram, we employ two complementary approaches -- the \textit{ridge} and \textit{edge} methods \citep[e.g.,][]{Ohashi2014ApJ...796..131O, Seifried2016MNRAS.459.1892S, Alves2017A&A...603L...3A, YusukeAso2017ApJ...849...56A, Yen2017ApJ...834..178Y, Sai2020ApJ...893...51S, Maret2020A&A...635A..15M} -- as described in detail by \citet{YusukeAso2024PKAS...39...27A}. The \textit{ridge} method traces the intensity-weighted centroid or Gaussian-fitted peak of the emission in one-dimensional (1D) cuts taken along the positional axis (x-cuts) at each velocity channel, and along the velocity axis (v-cuts) at each spatial position, thereby characterizing the bulk of the rotating gas. In contrast, the \textit{edge} method identifies the outermost position or velocity where the emission exceeds a fixed threshold (typically 5$\sigma$), and thus traces the maximum velocity extent of the disk. After extraction, a filtering procedure removes spurious or non-Keplerian points, ensuring that only data consistent with rotational motion are retained \citep[e.g.,][]{YusukeAso2024PKAS...39...27A}. The filtered ridge and edge data are then fitted with a Keplerian power-law model to estimate the dynamical stellar mass, adopting the disk inclination obtained from the continuum observations.

We derive a stellar mass of $M_\star = 1.85 \pm 0.06\,M_\odot$ from the edge fit and $M_\star = 1.01 \pm 0.05\,M_\odot$ from the ridge fit. The quoted uncertainties reflect only the statistical errors of the individual fits and do not include the known systematic effects of the two methods. The substantial difference between the two estimates likely reflects these systematics rather than a genuine ambiguity in the disk kinematics. The edge method is sensitive to the highest-velocity emission and can be biased by optical depth effects, emission from elevated disk layers, or local deviations from a simple emitting surface, all of which may lead to an overestimate of the rotational velocity and hence the stellar mass. In contrast, the ridge method traces the intensity-weighted bulk emission and can be biased toward lower apparent velocities because of beam smearing, projection effects, and the weighting of the line-emitting surface, which may cause an underestimate of the true Keplerian speed. Similar discrepancies between ridge- and edge-based masses have been reported in previous studies \citep[e.g.,][]{YusukeAso2024PKAS...39...27A}.

Given these systematics, we adopt $M_\star = 1.43 \pm 0.42\,M_\odot$ as a conservative estimate, where the central value is the midpoint of the ridge and edge determinations and the uncertainty is half of their difference. This choice is intended to capture the dominant method-dependent uncertainty rather than the much smaller formal fitting errors. A formal inverse-variance weighted mean would yield $M_\star = 1.35 \pm 0.04\,M_\odot$, but this would significantly underrepresent the true uncertainty because the two estimates are not free of systematic biases. The resulting stellar mass remains consistent, within uncertainties, with previous dynamical estimates for HV~Tau~C based on disk kinematics.

Previous studies have reported comparable stellar mass estimates for HV~Tau~C based on disk kinematics. From IRAM $^{12}$CN~(2--1) observations, \citet{Guilloteau2014A&A...567A.117G} derived $M_\star = 1.59 \pm 0.08~M_\odot$, while ALMA $^{12}$CO~(3--2) data analyzed by \citet{Simon2019ApJ...884...42S} yielded $M_\star = 1.33 \pm 0.04~M_\odot$. Our adopted value, $M_\star \approx 1.43~M_\odot$, is therefore in good agreement with these measurements.

\section{JWST/MIRI-MRS Spectrum of HV Tau C: Overview and Emission Line Inventory} \label{sec:overall_1D_spectra}

\begin{figure*}[htbp!]
\centering
\includegraphics[width=\textwidth]{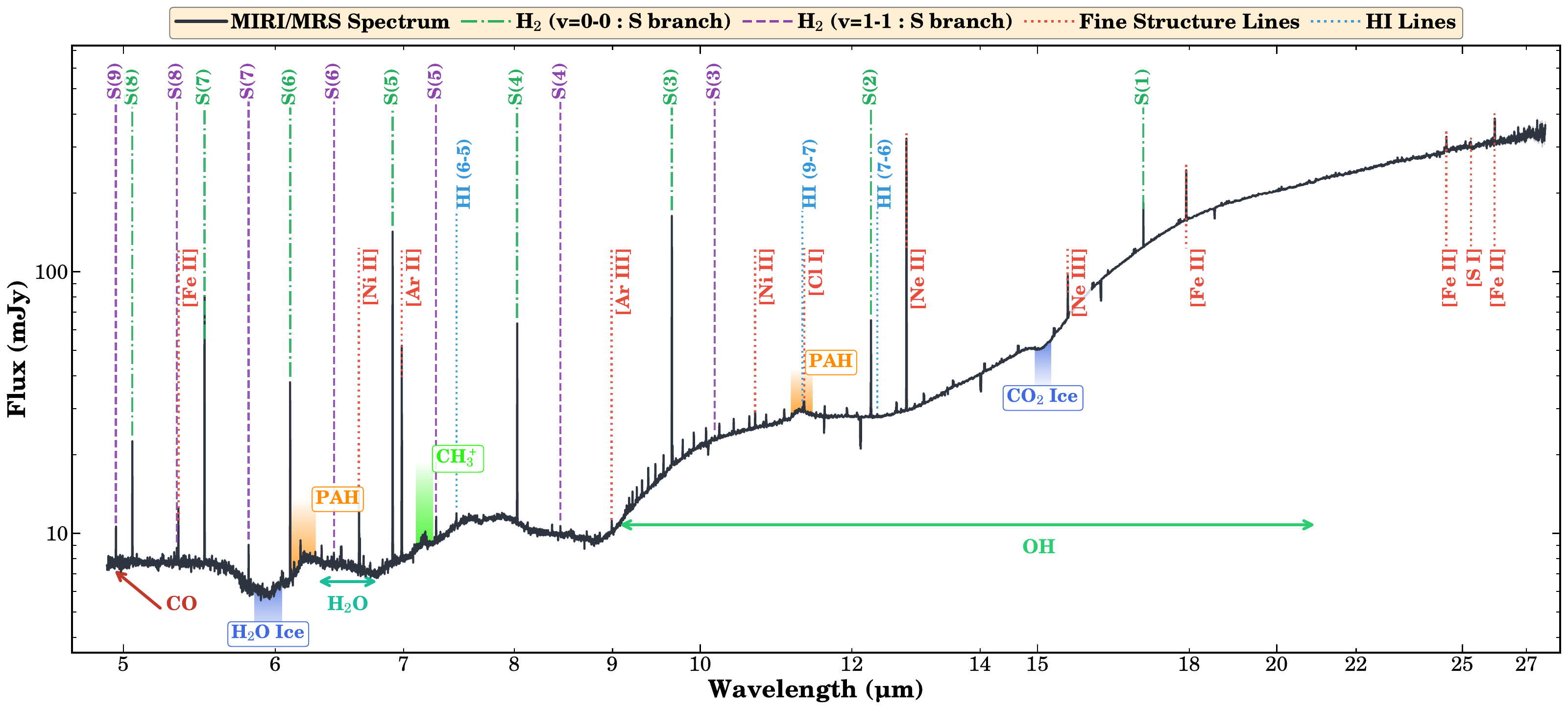}
\caption{Full \textit{JWST} MIRI/MRS spectrum of the Class II source HV Tau C, covering 5--27.5\,$\mu$m. Prominent features include: H$_2$ transitions (S(1)--S(8) in the $v=0$--0 ground state and S(3)--S(9) in the $v=1$--1 state), CO ($v=1$--0) around 5\,$\mu$m, ro-vibrational H$_2$O lines at 6--7\,$\mu$m, fine-structure lines ([Fe\,{\sc ii}], [Ni\,{\sc ii}], [Ne\,{\sc ii}], [Ar\,{\sc ii}], [Ar\,{\sc iii}], [S\,{\sc i}], [S\,{\sc iii}]) etc., H\,{\sc i} recombination lines (6--5, 7--6, 9--7), and OH lines starting at 9.1\,$\mu$m in the ground vibrational state ($v=0$--0).}
\label{fig:hv_tau_c_full_spectrum}
\end{figure*}

The full {\it JWST}-MIRI/MRS 1D spectrum of the source HV Tau C, covering the wavelength range 4.9-27.9 $\mu$m, is presented in Figure \ref{fig:hv_tau_c_full_spectrum}. The spectrum was extracted using a wavelength-dependent circular aperture with radius equal to twice the Point Spread Function (PSF) FWHM, where the MIRI/MRS PSF FWHM is given by $\theta(\lambda) = 0.033\times\lambda + 0.106'' \quad (\lambda\ {\rm in}\ \mu{\rm m})$ following \citet{Law2023AJ....166...45L}.

The MIRI/MRS spectrum of protoplanetary disks is known to be rich in molecular and atomic features, including pure-rotational H$_2$ lines, ro-vibrational lines of CO, pure-rotational and ro-vibrational lines of H$_2$O, atomic fine-structure transitions, H\,{\sc i} recombination lines, and PAH emission and ice absorption bands \citep[e.g.,][]{Banzatti2023AJ....165...72B, Kospal2023ApJ...945L...7K, Grant2023ApJ...947L...6G, vanDishoeck2023FaDi..245...52V, Henning2024PASP..136e4302H, Krijt2025ApJ...990L..72K, Arulanantham2025AJ....170...67A}. Nearly face-on disks are typically rich in molecular emission tracing the inner disk regions \citep[e.g.,][]{Gasman2023A&A...679A.117G, Xie2023ApJ...959L..25X, Pontoppidan2024ApJ...963..158P, RomeroMirza2024ApJ...964...36R}. In multiple companion Class~II systems, primaries are often line-rich while secondary components tend to show comparatively weaker molecular emission \citep[e.g.,][]{Kurtovic2026A&A...705A..97K, Somigliana2026arXiv260623794S}. In contrast to this general trend, the tertiary component HV~Tau~C is the line-rich source in the system, whereas the primary pair HV~Tau~AB shows comparatively weaker line emission. The HV~Tau~C spectrum is dominated by tracers of outflowing material (winds and jets) rather than inner disk molecular emission, likely due to its nearly edge-on geometry.

In the MIRI spectrum of HV~Tau~C, we detect several molecular H$_2$ emission lines above the continuum. In particular, we identify pure rotational H$_2$ transitions in the ground vibrational state ($v=0$–0), ranging from S(1) at 17.03~$\mu$m to S(8) at 5.05~$\mu$m \citep[as detected in other Class~II sources also, e.g., see][]{Franceschi2024A&A...687A..96F, Henning2024PASP..136e4302H}. These transitions trace wide-angle H$_2$ emission. In addition, we detect weak, low signal-to-noise (S/N) ratio H$_2$ transitions in the first excited vibrational state ($v =$ 1--1), from S(3) at 10.18~$\mu$m to S(9) at 4.95~$\mu$m.

Several atomic fine-structure lines are also present, such as [Fe\,\textsc{ii}] at 5.34\,$\mu$m, [Ni\,\textsc{ii}] at 6.64\,$\mu$m, [Ar\,\textsc{ii}] at 6.98\,$\mu$m, [Ar\,\textsc{iii}] at 8.99\,$\mu$m, [Ni\,\textsc{ii}] at 10.68\,$\mu$m, [Cl\,\textsc{i}] at 11.33\,$\mu$m, [Ne\,\textsc{ii}] at 12.81\,$\mu$m, [Ne\,\textsc{iii}] at 15.56\,$\mu$m, [Fe\,\textsc{ii}] at 17.94\,$\mu$m, [S\,\textsc{iii}] at 18.71\,$\mu$m, [Fe\,\textsc{ii}] at 24.52\,$\mu$m, [S\,\textsc{i}] at 25.25\,$\mu$m and [Fe\,\textsc{ii}] at 25.98\,$\mu$m \citep[as detected in other Class~II sources, e.g., see][Shridharan et al. in prep]{Bajaj2024AJ....167..127B, Pascucci2025NatAs...9...81P, Schwarz2025ApJ...980..148S}. Some of the H\,\textsc{i} lines, i.e., 6-5, 7-6, 9-7 are also detected in this source \citep[as detected in other Class~II sources, e.g., see][]{Franceschi2024A&A...687A..96F, Tofflemire2025ApJ...985..224T, Shridharan2026A&A...708A..22S}.

Given that HV~Tau~C hosts an edge-on disk, emission from molecular lines arising in the inner disk may be attenuated or obscured; transitions originating within the inner few au may therefore appear weaker. Nevertheless, we detect ro-vibrational CO emission near 5~$\mu$m and H$_2$O ro-vibrational lines around 6--7~$\mu$m, similar to those observed in other Class~II disks such as DF~Tau \citep[e.g.,][]{Grant2024A&A...689A..85G}. Furthermore, we identify several pure-rotational OH lines (v = 0--0) starting at $\sim$9.1~$\mu$m, which may indicate UV irradiation of the disk surface \citep[detected in other Class~0/I, and Class~II sources, e.g., see][]{Tabone2024A&A...691A..11T, Zannese2024NatAs...8..577Z, Neufeld_et_al_2024ApJ...966L..22N, Watson2026ApJ...999..264W}. The spectrum also further reveals emission from methyl cation CH$_3^+$ at 7.15~$\mu$m \citep[e.g., see in other Class~II sources,][]{Henning2024PASP..136e4302H, Zannese2025A&A...696A..99Z}. As seen in other disk environments, CH$_3^+$ excitation may be driven either by collisions in dense, warm gas or by FUV-pumped H$_2$ in lower-density, highly irradiated regions \citep[e.g.,][]{Zannese2025A&A...696A..99Z}.

The spectrum also exhibits relatively faint PAH emission features at 6.2~$\mu$m and 11.3~$\mu$m, detected at the source position and confined within $\sim$2 PSF FWHM. In addition, an H$_2$O ice absorption feature is present near $\sim$6~$\mu$m, along with a CO$_2$ ice absorption band around $\sim$15.1~$\mu$m \citep[see, e.g., detected in other YSOs,][]{Sturm2024A&A...689A..92S, Tychoniec2024A&A...687A..36T, Vlasblom2025A&A...693A.278V, Dartois2025A&A...698A...8D}.

In this paper, we present a detailed analysis of the molecular H$_2$ lines and the H~\textsc{i} recombination lines. The rich set of fine-structure lines and other molecular species, particularly OH, will be discussed in detail in an upcoming paper (Pathak et al., in prep.).

\section{\texorpdfstring{Extended H$_2$ Emission}{Extended H2 Emission}} \label{sec:extended_H2_emission}

\begin{figure*}[htbp!]
\centering
\includegraphics[width=\textwidth]{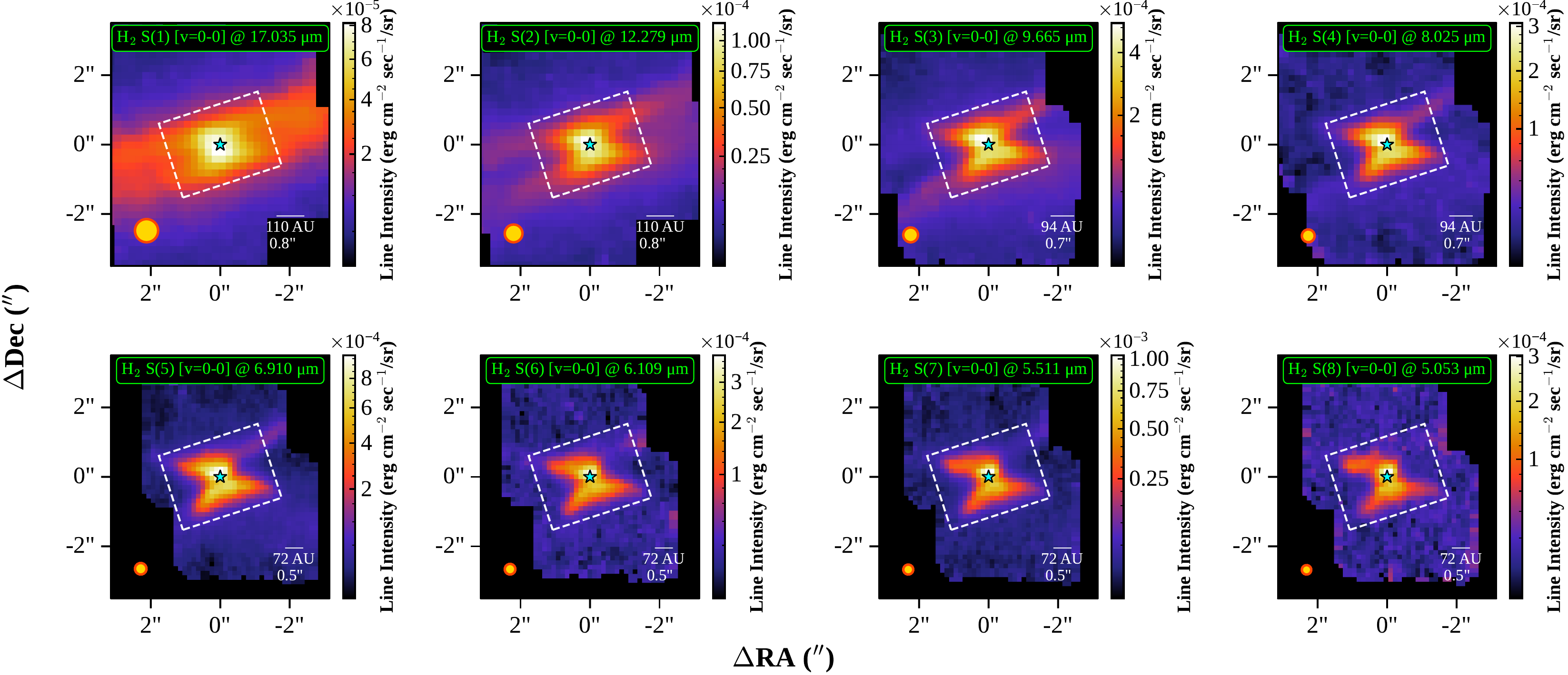}
\caption{Continuum-subtracted line-intensity maps of all pure-rotational H$_2$ lines S(1)--S(8) in the ground vibrational state ($v=0$--$0$), showing wide-angled H$_2$ emission. The star marks the JWST/MIRI 14.0\,$\mu$m continuum photocenter. The dashed rectangular region that encloses the H$_2$~S(8) transition marks the area from which the spectra were extracted to derive the line fluxes used in the rotational diagram analysis.}
\label{fig:hv_tau_c_H2_v00_extended_linemaps}
\end{figure*}

\begin{figure*}[htbp!]
\centering
\includegraphics[width=0.85\textwidth]{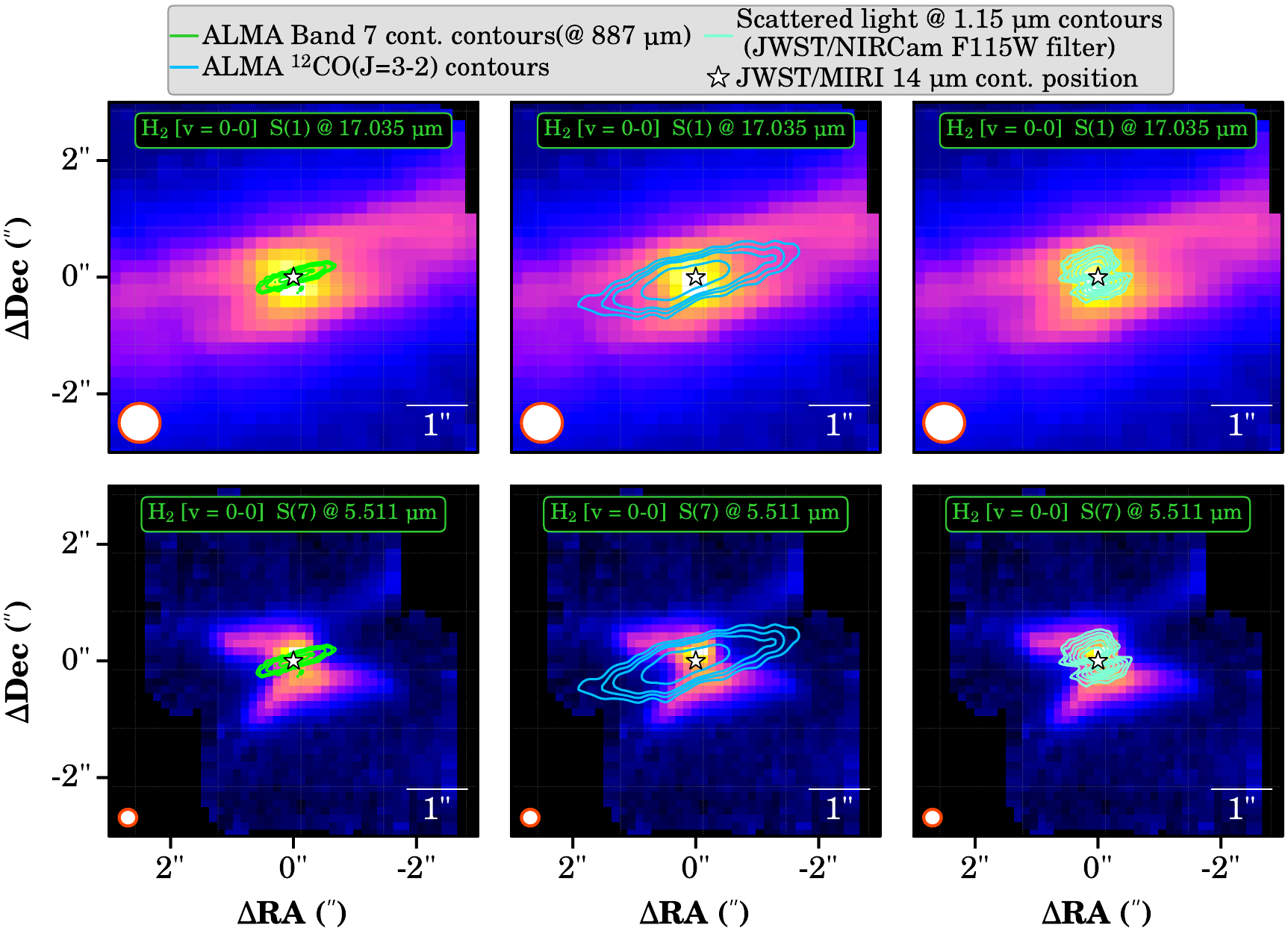}
\caption{Comparison of molecular hydrogen emission with tracers of the disk and its surroundings in HV Tau C. \textit{Top row:} Continuum-subtracted intensity map of the H$_2$ S(1) line at 17.035\,$\mu$m. \textit{Bottom row:} Same for the H$_2$ S(7) line at 5.511\,$\mu$m. The star symbol marks the JWST/MIRI 14.0\,$\mu$m continuum photocenter. \textit{First column:} H$_2$ emission overlaid with ALMA 887\,$\mu$m dust continuum contours (lime green). \textit{Second column:} H$_2$ emission overlaid with ALMA $^{12}$CO ($J$=3--2) integrated intensity contours (cyan). \textit{Third column:} H$_2$ emission overlaid with JWST/NIRCam F115W (1.15\,$\mu$m) scattered light contours (light green).}
\label{fig:JWST_MIRI_NIRCAM_ALMA_comparison_images_of_V-HV-TAU-C}
\end{figure*}

\begin{figure*}[htbp!]
\centering
\includegraphics[width=\textwidth]{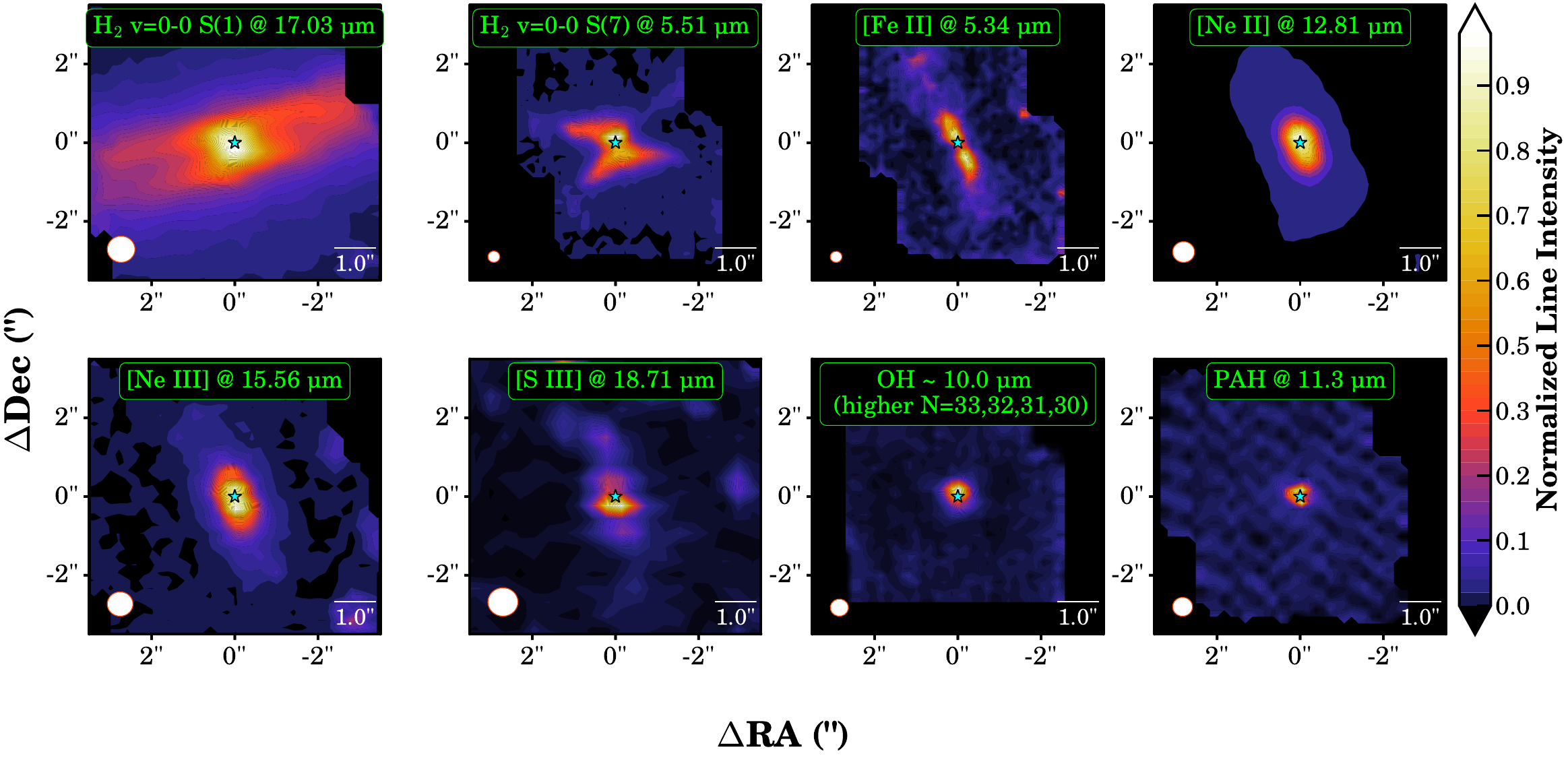}
\caption{Comparison of molecular hydrogen emission with potential UV-related tracers in the \textit{JWST}/MIRI-MRS wavelength range. The figure shows normalized continuum-subtracted line-intensity maps (displayed as colormaps) of the extended H$_2$ $v=0$--0 S(1) line at 17.03~$\mu$m (a lower-excitation transition) and the H$_2$ $v=0$--0 S(7) line at 5.51~$\mu$m (a higher-excitation transition). These are compared with the spatial morphologies of several potential UV tracers, including [Ne\,{\sc ii}] at 12.81~$\mu$m, [Ne\,{\sc iii}] at 15.56~$\mu$m, [S\,{\sc iii}] at 18.71~$\mu$m, OH emission near 10~$\mu$m (constructed by combining line maps with $N_{\mathrm{low}} = 33, 32, 31,$ and $30$ to increase the S/N ratio), and broad PAH emission at 11.3~$\mu$m. The map of [Fe\,{\sc ii}] at 5.34~$\mu$m, which traces the collimated jet, is also shown for comparison. The green star marks the $\sim$14.0~$\mu$m continuum centroid.}
\label{fig:HV_Tau_C_comparison_of_H2_morphology_with_UV_tracers}
\end{figure*}

Recent {\it JWST} observations of Class~II sources with the MIRI/MRS and NIRSpec instruments have revealed a variety of extended molecular H$_2$ outflow structures, owing to the high spatial resolution and sensitivity of these instruments. These include the ring-like cold H$_2$ emission in FZ~Tau \citep{Pontoppidan2024ApJ...963..158P}, the bipolar X-shaped H$_2$ winds in the edge-on disk of TAU~042021 observed with MIRI \citep{Arulanantham2024ApJ...965L..13A}, and similar extended H$_2$ structures detected with NIRSpec in TAU~042021, HH~30, and FS~Tau~B \citep{Pascucci2025NatAs...9...81P}. Additional extended H$_2$ emission has been reported in CX~Tau \citep{Anderson2024ApJ...977..213A}, SY~Cha \citep{Schwarz2025ApJ...980..148S}, and binary Class~II systems \citep{Kurtovic2026A&A...705A..97K}. More recently, similar extended H$_2$ structures have also been identified in HK~Tau~B \citep{Somigliana2026arXiv260623794S} and in the JDISC survey of extended H$_2$ winds \citep{Narang2026ApJ..1004..188N}.

To search for extended H$_2$ emission in HV~Tau~C, we constructed continuum-subtracted line-intensity maps for all detected H$_2$ transitions. For each transition, spectra were extracted from individual spatial pixels across a wavelength interval of $\pm 10\Delta\lambda$ centered on the line, where $\Delta\lambda = \lambda / R(\lambda)$ and the spectral resolution is given by $R(\lambda) = 4603 - 128\lambda + 10^{-7.4\lambda}$ \citep{Law2023AJ....166...45L}. The local continuum was subtracted from each spectrum, and a Gaussian profile centered on the rest wavelength was fitted to the residual line emission. The integrated flux from each fit was then used to generate two-dimensional maps tracing the spatial distribution of the H$_2$ emission.

\subsection{\texorpdfstring{Morphology of the Extended H$_2$ Emission}{Morphology of the Extended H2 Emission}}
\label{subsec:extended_H2_morphology}

The two-dimensional continuum-subtracted line-intensity maps of all detected H$_2$ transitions from S(1) to S(8) in the ground vibrational state ($v = 0$--$0$) are presented in Figure~\ref{fig:hv_tau_c_H2_v00_extended_linemaps}. The maps reveal a clear bi-conical morphology. From lower- to higher-excitation transitions (i.e., from S(1) to S(8)), the emission becomes progressively narrower, forming a nested structure. For the first time, extended H$_2$ emission in HV~Tau~C is revealed through mid-IR pure-rotational H$_2$ lines in the ground vibrational state, enabled by the high spatial resolution and sensitivity of \textit{JWST}/MIRI. The upper-right arm becomes fainter in the S(8) map, likely due to the lower S/N ratio at shorter wavelengths. Previous near-infrared observations by \citet{Beck2008ApJ...676..472B} also reported spatially extended H$_2$ emission in the $v = 1$--$0$ S(1) line at 2.12~$\mu$m, extending to a projected distance of $\sim$200~au from the central star.

A key question is whether the observed spatially extended H$_2$ emission originates from a disk wind, a UV-excited photoevaporative wind, or the UV-excited surface of the disk. To examine this, we compared the spatial extent of the H$_2$ emission with that of the near-infrared scattered-light emission at 1.15~$\mu$m obtained with \textit{JWST}/NIRCam F115W filter, the 887~$\mu$m dust continuum observed with ALMA Band~7, and the $^{12}$CO ($J$ = 3--2) emission from ALMA. As shown in Figure~\ref{fig:JWST_MIRI_NIRCAM_ALMA_comparison_images_of_V-HV-TAU-C}, the H$_2$ emission is significantly more extended than all of these tracers. In addition, we examined the spatial distribution of other JWST-accessible UV-related tracers, including PAH emission, OH lines, and the ionic lines [Ne\,{\sc ii}], [Ne\,{\sc iii}], and [S\,{\sc iii}], shown in Figure~\ref{fig:HV_Tau_C_comparison_of_H2_morphology_with_UV_tracers}. We also compare the morphologies of these UV tracers with the collimated jet traced by [Fe\,{\sc ii}] at 5.34~$\mu$m. The H$_2$ emission appears to trace the wider component of the outflow (especially in the higher transitions; in the lower transition, it might also be affected by the beam, as it is extended in both directions, i.e., radially along the disk surface and vertically as well). The fine-structure lines ([Ne\,{\sc ii}], [Ne\,{\sc iii}], and [S\,{\sc iii}]) lie along the jet but are wider compared to the [Fe\,{\sc ii}] jet and narrower than H$_2$ emission, whereas the OH emission (at ~10~$\mu$m) and broader PAH emission (at ~11.3~$\mu$m) is unresolved, i.e., it lies within the 2 PSF FWHM region at the center. This suggests that the extended H$_2$ emission does not completely trace a UV-driven component that is spatially distinct from the disk surface, but rather follows a wide-angled wind launched from the disk surface.

\begin{figure*}[htbp!]
\centering
\includegraphics[width=\textwidth]{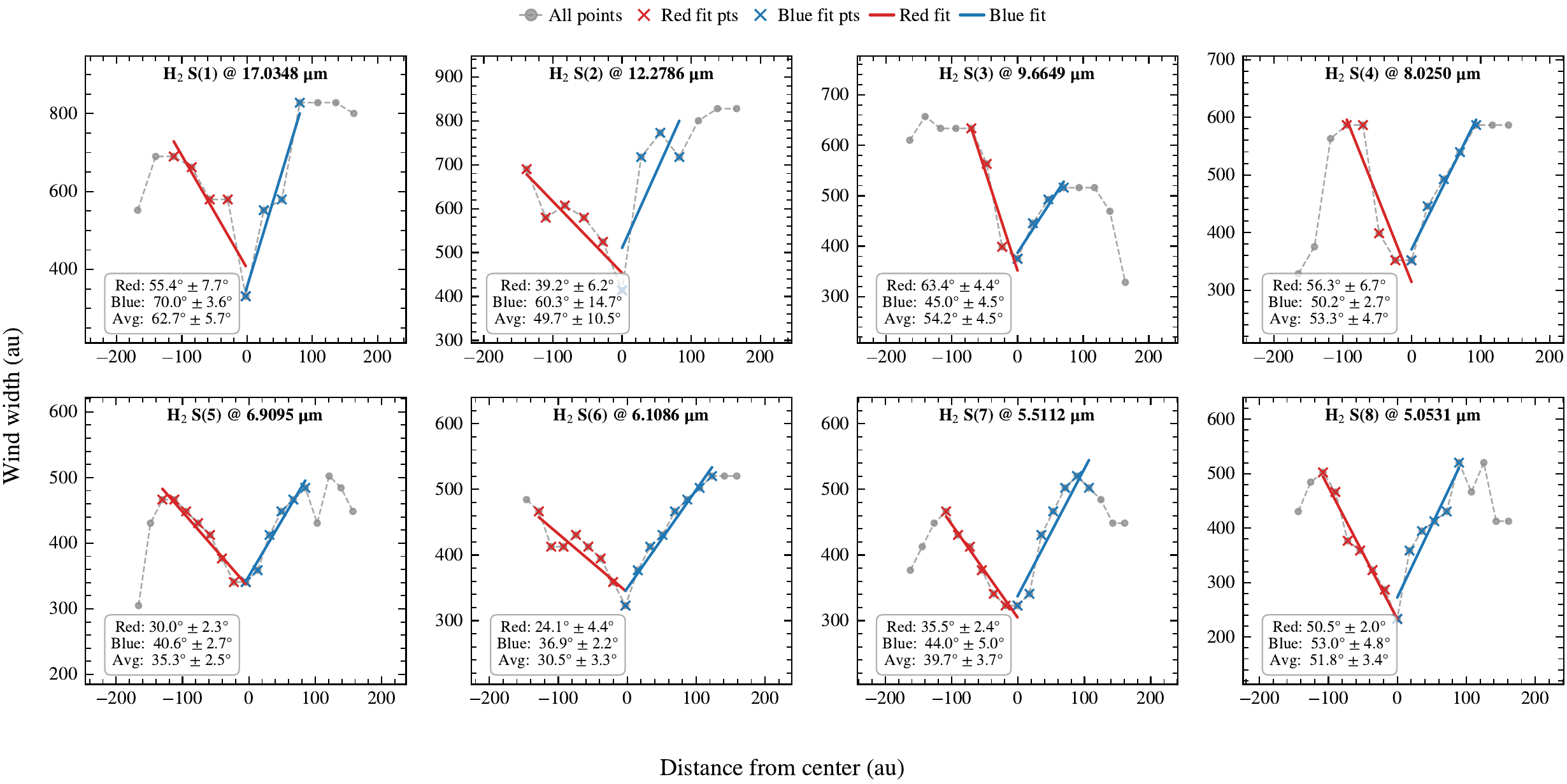}
\caption{Measurement of the outflow opening angle in HV~Tau~C. The edges of the red- and blue-shifted molecular wind lobes are identified (gray points connected by dashed lines). A linear fit is applied to the edge points of each lobe separately to determine the wind opening angle.}
\label{fig:hv_tau_c_H2_v=0-0_opening_angle}
\end{figure*}

\begin{table*}[htbp]
\caption{Measured semi-opening angles of the H$_2$ emission in HV~Tau~C}
\label{tab:semi_opening_angle}
\centering
\begin{tabular}{lcccc}
\hline\hline
Transition & Wavelength ($\mu$m) & Red-shifted Lobe ($^\circ$) & Blue-shifted Lobe ($^\circ$) & Average ($^\circ$) \\
\hline
H$_2$ $\nu$=0--0 S(1) & 17.035 & $55.4 \pm 7.7$ & $70.0 \pm 3.6$ & $62.7 \pm 5.7$ \\
H$_2$ $\nu$=0--0 S(2) & 12.279 & $39.2 \pm 6.2$ & $60.3 \pm 14.7$ & $49.7 \pm 10.5$ \\
H$_2$ $\nu$=0--0 S(3) & 9.665  & $63.4 \pm 4.4$ & $45.0 \pm 4.5$ & $54.2 \pm 4.5$ \\
H$_2$ $\nu$=0--0 S(4) & 8.026  & $56.3 \pm 6.7$ & $50.2 \pm 2.7$ & $53.3 \pm 4.7$ \\
H$_2$ $\nu$=0--0 S(5) & 6.909  & $30.0 \pm 2.3$ & $40.6 \pm 2.7$ & $35.3 \pm 2.5$ \\
H$_2$ $\nu$=0--0 S(6) & 6.109  & $24.1 \pm 4.4$ & $36.9 \pm 2.2$ & $30.5 \pm 3.3$ \\
H$_2$ $\nu$=0--0 S(7) & 5.511  & $35.5 \pm 2.4$ & $44.0 \pm 5.0$ & $39.7 \pm 3.7$ \\
H$_2$ $\nu$=0--0 S(8) & 5.053  & $50.5 \pm 2.0$ & $53.0 \pm 4.8$ & $51.8 \pm 3.4$ \\
\hline
\end{tabular}
\end{table*}

Potential alternatives to a disk wind interpretation include emission from the disk surface layers or from a remnant envelope. However, the disk surface is expected to have a spatial extent comparable to that of the molecular gas traced by $^{12}$CO, which is clearly not the case here. A remnant envelope would likely produce a more symmetric and radially broader emission morphology, potentially accompanied by signatures of infall. The observed bi-conical H$_2$ morphology and its substantial extension beyond the well-defined Keplerian disk traced by CO therefore argue against both the disk surface and a simple envelope scenario. Taken together, these considerations indicate that the extended H$_2$ emission most likely traces a molecular disk wind launched from the disk surface. Furthermore, the velocity structure of the H$_2$ emission (see Section~\ref{subsec:winds_kinematics}) provides additional support for a wind interpretation.

As discussed in Section~\ref{sec:overall_1D_spectra}, we also detect H$_2$ lines from S(3) to S(9) in the first excited vibrational state ($v = 1$--$1$). The H$_2$ $v = 1$--$1$ S(3) transition at 10.178~$\mu$m is blended with the [Fe\,{\sc ii}] $^4\!P_{3/2}$--$^4\!P_{5/2}$ line\footnote{\url{https://physics.nist.gov/PhysRefData/ASD/lines_form.html}}. However, the continuum-subtracted line-intensity maps of these transitions show no clear evidence for extended emission (see Appendix~\ref{appsec:v=1-1_transitions_H2_linemaps}), likely due to the lower S/N ratio of these higher-excitation lines.

\subsection{Measuring the Semi-Opening Angle} \label{subsec:winds_semi_opening_angle}

The continuum-subtracted line intensity maps of the pure rotational H$_2$ transitions in the ground vibrational state ($v=0$--0) show a nested morphology, in which the emission becomes progressively narrower from S(1) through S(8). To quantify this behaviour, we measure the semi-opening angle of the wind outflow traced by each transition.

To perform this measurement, the wind outflow axis is first aligned vertically (north-south) by rotating the 2D maps by an angle corresponding to the disk position angle minus 90$^\circ$, such that the disk major axis lies horizontally along the X-axis. We subtracted the ambient H$_2$ emission from the two-dimensional continuum-subtracted line maps using small background apertures placed far from the prominent H$_2$ emission. The maps were then cropped to include only the regions showing wind--outflow emission, reducing the likelihood of spurious edge identification. Edge positions are determined using the \texttt{Sobel} filter from the \texttt{skimage.filters} package, which enhances spatial gradients, in combination with a \texttt{threshold\_mean} mask that suppresses background noise. The detected horizontal edge traces for all eight H$_2$ transitions are shown in Appendix~\ref{appfig:hv_tau_c_H2_v_0_0_lines_detected_edges}.

For each horizontal pixel row, we measure the distance between the detected edge and the central vertical axis, corresponding to the outflow symmetry axis. This yields the semi-width of the flow as a function of distance from the source along the Y-axis, measured independently for the northern and southern sides. A linear function is then fitted to these profiles to obtain the geometric opening behaviour of the flow, as illustrated in Figure~\ref{fig:hv_tau_c_H2_v=0-0_opening_angle}. The semi-opening angle, $\theta_{\mathrm{wind}}$, is defined as
$$
\theta_{\mathrm{wind}} = \cos^{-1}\left(\frac{d/\sin{i}}{l_{\mathrm{edge}}}\right),
$$
where $~i$ is the inclination of the HV~Tau~C disk, $d/\sin{i}$ represents the de-projected wind outflow extent from the source center, and $l_{\mathrm{edge}}$ is the fitted edge distance on each side of the axis. The resulting semi-opening angles for all detected $v=0$--0 transitions are listed in Table~\ref{tab:semi_opening_angle}.

It is difficult to robustly quantify the semi-opening angle of the wind in this source. While a visual inspection of the H$_2$ line maps suggests that the emission becomes progressively narrower from the lower S(1) transition to the higher S(8) transition, our derived measurements do not show a clear decreasing trend, with average values spanning $\sim$30\degree to 62\degree. This discrepancy can be attributed to several observational factors: (i) the relatively poor S/N ratio of the H$_2$ emission in this Class II source, making precise identification of the wind edges challenging; (ii) for the lowest transitions (notably S(1) and S(2)), significant ambient or spatially extended H$_2$ emission fills the IFU field, complicating the isolation of the wide-angled wind component; and (iii) for the higher transitions (notably S(7) and S(8)), one of the wind arms suffers from particularly low S/N ratio, preventing an accurate measurement of its lateral extent. These effects collectively limit our ability to trace a clear trend in wind collimation with excitation level.

\subsection{\texorpdfstring{Excitation Mechanism of H$_2$: Rotational Diagram}{Excitation Mechanism of H2: Rotational Diagram}} 
\label{subsec:excitation_conditions_winds}

To investigate the excitation conditions of the H$_2$ emission associated with the disk wind in HV~Tau~C, we constructed a rotational diagram using fluxes from eight pure rotational H$_2$ transitions (S(1)--S(8)) in the ground vibrational state ($v=0$--0) and five additional pure rotational transitions (S(5)--S(9)) in the first vibrational state ($v=1$--1). In the optically thin limit -- appropriate for these H$_2$ lines given their small Einstein $A$ coefficients -- the level populations provide direct constraints on the rotational temperature ($T_{\rm rot}$) and the total column density ($N_{\rm tot}$) of the emitting gas \citep[e.g.,][]{Goldsmith1999ApJ...517..209G, Manoj2013ApJ...763...83M, Franceschi2024A&A...687A..96F, Schwarz2025ApJ...980..148S}.

The spectrum extraction was performed using a constant rectangular aperture encompassing the full spatial extent of the H$_2$ $v=0$--0 S(8) line emission (area $= 3.00\arcsec \times 2.24\arcsec$; see Figure~\ref{fig:hv_tau_c_H2_v00_extended_linemaps}). For each transition, within a wavelength window extending $8\,\times\,\Delta\lambda$ on either side of the line center, the local continuum was subtracted, and a Gaussian profile was fitted using the \texttt{LMFIT} Python package \citep{Newville2025zndo..16175987N}. The integrated line fluxes and their associated uncertainties, propagated from the fitted parameters, are listed in Table~\ref{apptab:hv_tau_c_all_H2_transitions_rotational_diagram_params_for_rectangular_S8_aperture}, and the corresponding line profiles are shown in Figure~\ref{appfig:hv_tau_c_all_H2_lines_gaussian_fit_profiles_for_rectangular_S8_aperture}.

\begin{figure*}[htbp!]
\centering
\begin{subfigure}[t]{0.48\textwidth}
\centering
\includegraphics[width=\textwidth]{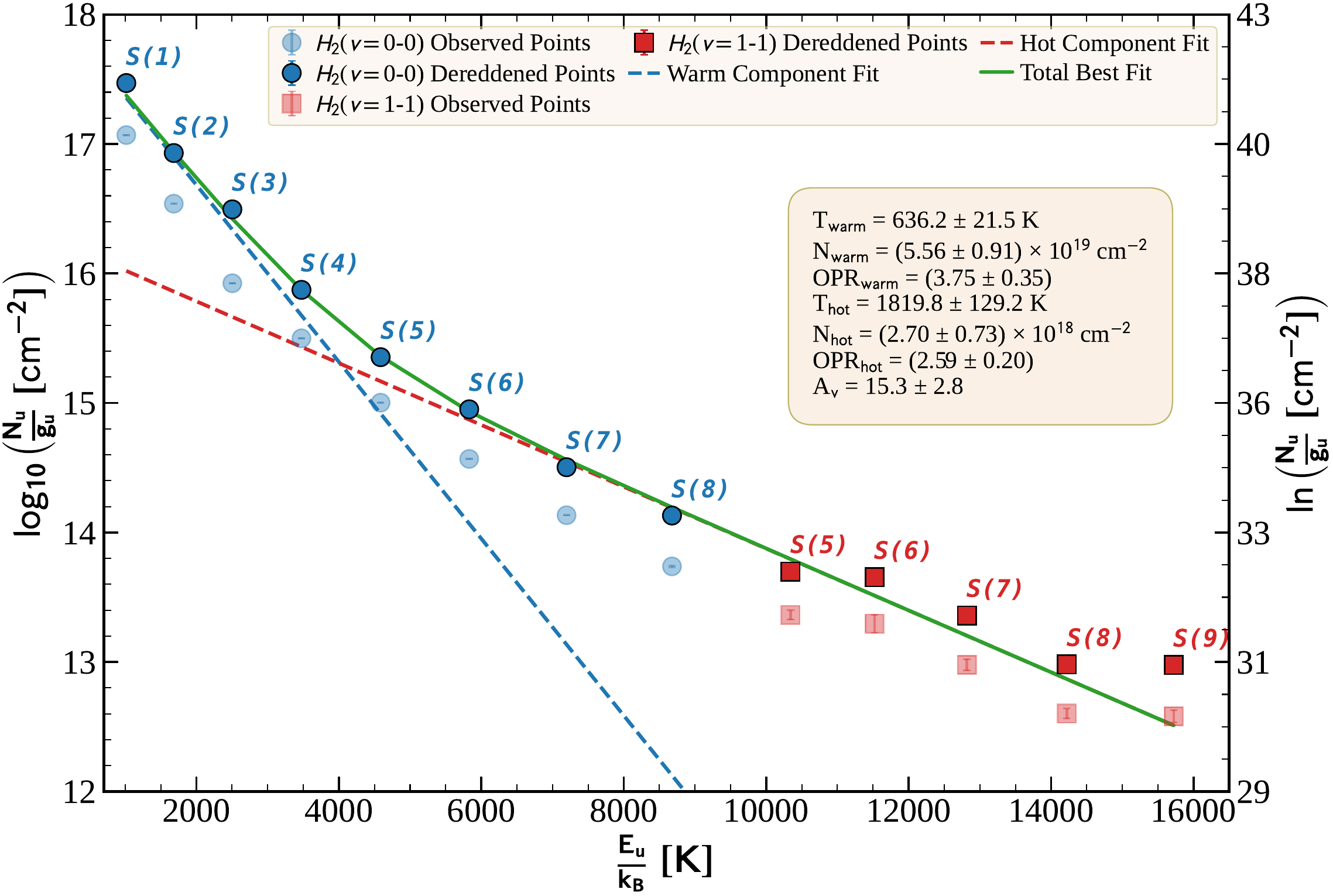}
\caption{Two-component model fit.}
\label{subfig:hv_tau_c_H2_rot_diag_two_comp}
\end{subfigure}
\hfill
\begin{subfigure}[t]{0.48\textwidth}
\centering
\includegraphics[width=\textwidth]{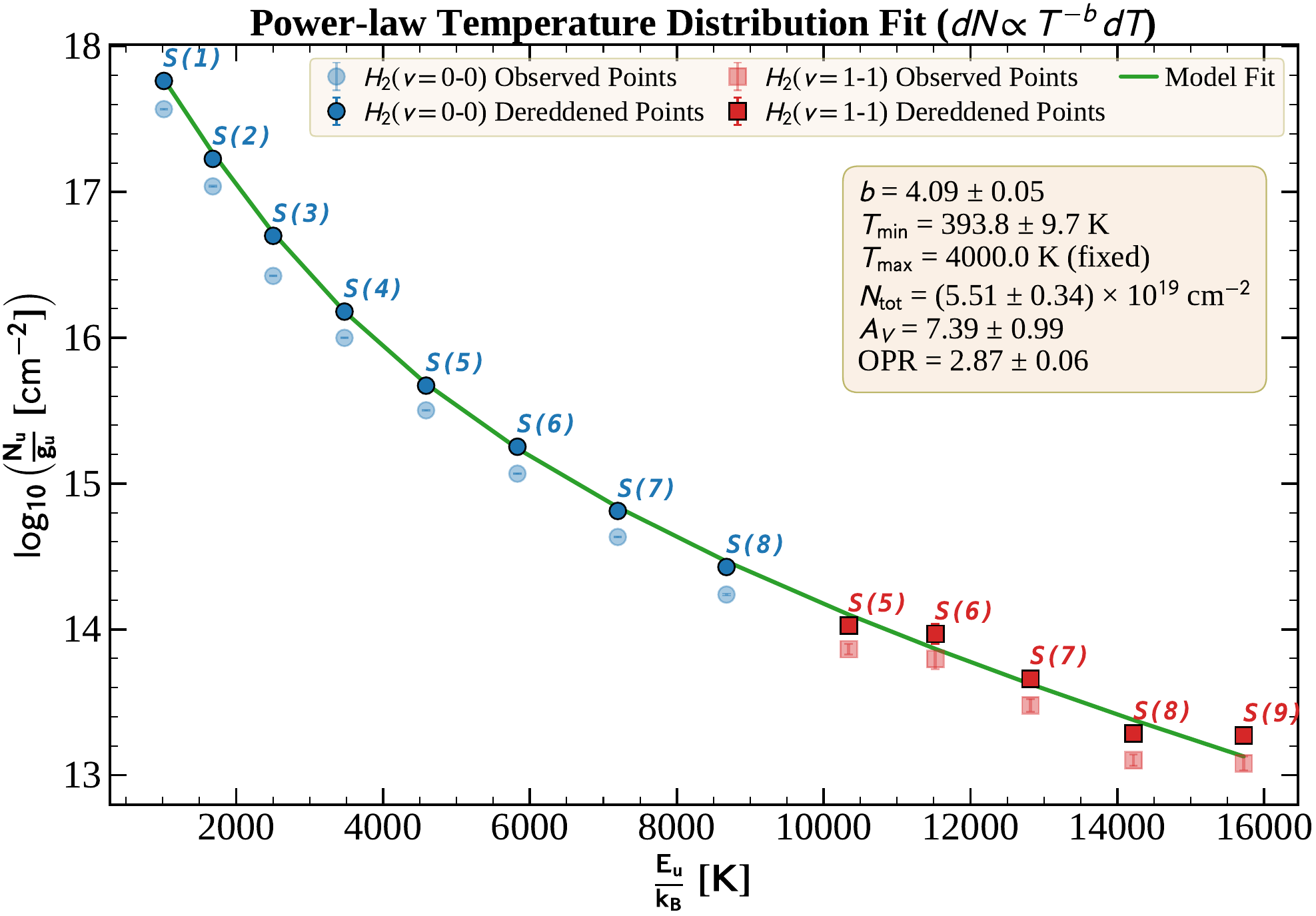}
\caption{Power-law model fit.}
\label{subfig:hv_tau_c_H2_rot_diag_powerlaw}
\end{subfigure}
\caption{Rotational diagrams of the detected H$_2$ pure rotational emission toward HV~Tau~C, showing $\log(N_u/g_u)$ as a function of upper-level energy $E_u/k_{\rm B}$. Both observed and extinction-corrected values are displayed; the latter are derived using the extinction law of \citetalias{McClure2009ApJ...693L..81M}. Spectra were extracted from the rectangular aperture enclosing the H$_2$ $v=0$--0 S(8) emission extent (see the white dashed rectangular aperture in Figure~\ref{fig:hv_tau_c_H2_v00_extended_linemaps}). (\textit{a}) Two-component fit comprising warm and hot temperature components; dashed lines show the individual components and the solid line the combined best-fit model. (\textit{b}) Power-law temperature distribution fit assuming $dN \propto T^{-b}\,dT$ between $T_{\rm min}$ and $T_{\rm max}$; the solid line indicates the best-fit model. Blue circles correspond to $v=0$--0 transitions and red squares to $v=1$--1 transitions. Best-fit parameters for both models are listed in the insets.}

\label{fig:hv_tau_c_H2_rotational_diagrams_comparison}
\end{figure*}

The resulting rotational diagram is shown in Figure~\ref{subfig:hv_tau_c_H2_rot_diag_two_comp}. The observed data points -- with $v=0$--0 pure rotational transitions shown in blue and $v=1$--1 transitions in red -- deviate systematically from a single straight-line fit (dashed blue line) and exhibit clear positive curvature. This curvature indicates that a single-temperature model cannot adequately describe the excitation conditions of the H$_2$ gas. The change in slope across the diagram suggests the presence of at least two distinct thermal components, motivating a simultaneous two-component fit to reproduce the dereddened line fluxes, consistent with previous studies \citep[e.g.,][]{Narang2024ApJ...962L..16N, Schwarz2025ApJ...980..148S, Navarro2025ApJ...995..199N}.

To model the excitation conditions, we simultaneously fit for the temperatures, column densities, and ortho-to-para ratios (OPR) of both the warm and hot components, along with the line-of-sight visual extinction ($A_V$), as described in Tyagi et al. 2026 (under review). To account for extinction effects, we adopt three independent extinction laws: the \citet{HensleyDraine2023ApJ...948...55H} extinction law (hereafter \citetalias{HensleyDraine2023ApJ...948...55H}), the KP5 extinction law from \citet{Pontoppidan2024ApJ...963..158P} (hereafter \citetalias{Pontoppidan2024ApJ...963..158P}), and the \citet{McClure2009ApJ...693L..81M} extinction law (hereafter \citetalias{McClure2009ApJ...693L..81M}). 

For each extinction prescription, we fit a seven-parameter model consisting of two temperatures ($T_{\rm warm}$ and $T_{\rm hot}$), two column densities ($N_{\rm warm}$ and $N_{\rm hot}$), two ortho-to-para ratios (OPR$_{\rm warm}$ and OPR$_{\rm hot}$), and the visual extinction ($A_V$). The fitting procedure is carried out using the \texttt{LMFIT} Python package  \citep{Newville2025zndo..16175987N}. The complete model formulation and fitting methodology are presented in Appendix~\ref{appendix:two_component_model}. The best-fit model is shown in Figure~\ref{subfig:hv_tau_c_H2_rot_diag_two_comp}, and the corresponding best-fit parameters, together with their associated uncertainties, are listed in Table~\ref{tab:two_component_H2_parameters}.

From the fitted column densities, we estimate the total number of H$_2$ molecules in the wind as

$$
N_{\mathrm{H}_2} = N_{\mathrm{tot}} \times A,
$$

where $A = 3.00\arcsec \times 2.24\arcsec$ ($A = 1.58 \times 10^{-10}$\, sr) represents the solid angle subtended by the emitting region.

The total molecular mass is then given by

$$
M_{\mathrm{H}_2} = \mu_{\mathrm{H}_2}\,m_{\mathrm{H}}\,N_{\mathrm{H}_2},
$$

where $\mu_{\mathrm{H}_2} = 2$ is the mean molecular weight and $m_{\mathrm{H}} = 1.67 \times 10^{-27}$\,kg is the mass of a hydrogen atom.

To examine whether the H$_2$ emission arises from a continuous temperature distribution rather than discrete thermal components, we additionally explore a power-law temperature distribution model. Although a two-temperature fit provides a useful phenomenological description, a more realistic physical scenario -- particularly in shocked or UV-irradiated gas -- may involve a continuous distribution of temperatures. To test this hypothesis and to facilitate comparison with previous studies of warm H$_2$ emission in protostellar environments \citep[e.g.,][]{Neufeld2008ApJ...678..974N, Neufeld2012ApJ...749..125N, Manoj2013ApJ...763...83M}, we model the gas using a power-law temperature distribution. The differential column density is parameterized as
$$
\frac{dN}{dT} = a\,T^{-b}, \qquad T \in [T_{\min},T_{\max}].
$$

where $dN$ represents the column density within the temperature interval $[T, T+dT]$, $a$ is a normalization constant, and $b > 0$ is the power-law index that regulates the relative contribution of hot versus cold gas. The full derivation and complete model equations are provided in Appendix~\ref{appendix:powerlaw_model}.

In our modeling, we fix $T_{\rm max} = 4000$~K, corresponding to the typical dissociation energy of H$_2$ gas \citep[see,][]{Neufeld2008ApJ...678..974N}, while allowing $T_{\rm min}$ to vary from 100~K upward in order to probe the presence of colder gas components. The average OPR ratio adopts ${\rm OPR}_{\rm th} = 3$ as the theoretical high-temperature equilibrium value. The model therefore includes five free parameters: the power-law index $b$, the minimum temperature $T_{\rm min}$, the total column density $N_T$, the visual extinction $A_V$, and the average OPR of the H$_2$ gas. As in the two-component model, we account for extinction using three different extinction laws: \citetalias{HensleyDraine2023ApJ...948...55H}, \citetalias{ Pontoppidan2024ApJ...963..158P}, and \citetalias{McClure2009ApJ...693L..81M}.

The best-fit power-law model to the full set of H$_2$ transitions is shown in Figure~\ref{subfig:hv_tau_c_H2_rot_diag_powerlaw}, and the corresponding best-fit parameters are reported in Table~\ref{tab:two_component_H2_parameters}. The fit reveals a gas temperature distribution characterized by a power-law index $b \sim 4$ and a minimum temperature $T_{\rm min} = 394$~K, indicating the presence of a substantial reservoir of warm molecular gas within the wind. However, the JWST/MIRI wavelength coverage begins with the S(1) transition and does not include the S(0) line; consequently, the coldest H$_2$ gas component is not directly constrained by the present data. The derived minimum temperature should therefore be regarded as an upper limit to the true lowest gas temperature, and the inferred column density of the cold component -- and hence the total H$_2$ column density -- likely represents lower limits.

Overall, the agreement between the total molecular mass inferred from the power-law model and the warm-component mass derived from the two-component fit suggests that the bulk of the observed H$_2$ emission originates from warm molecular gas associated with the disk wind. These results indicate that the excitation of H$_2$ in HV~Tau~C is consistent with a multi-temperature gas distribution, as expected in shock-heated molecular winds.

\begin{table}[htbp]
\caption{Best-fit parameters from two-component and power-law fits using different extinction laws}
\label{tab:two_component_H2_parameters}
\centering
\footnotesize
\resizebox{\columnwidth}{!}{
\begin{tabular}{llll}
\hline\hline
Parameter & \citetalias{Pontoppidan2024ApJ...963..158P} & \citetalias{HensleyDraine2023ApJ...948...55H} & \citetalias{McClure2009ApJ...693L..81M} \\
\hline
\multicolumn{4}{c}{Two Component Fit} \\
\hline
$T_{\rm rot}^{\rm warm}$ (K)                         & $621.4 \pm 21.3$   & $650.6 \pm 22.8$   & $636.2 \pm 21.5$ \\
$N_{\rm H_2}^{\rm warm}$ ($\times10^{19}$ cm$^{-2}$) & $3.87 \pm 0.54$    & $3.21 \pm 0.37$    & $5.56 \pm 0.91$ \\
OPR$^{\rm warm}$                                     & $3.27 \pm 0.29$    & $3.96 \pm 0.38$    & $3.75 \pm 0.35$ \\
$M_{\rm wind}^{\rm warm}$ ($\times10^{-6}$ M$_\odot$) & $2.63 \pm 0.37$   & $2.18 \pm 0.25$    & $3.78 \pm 0.62$ \\
$T_{\rm rot}^{\rm hot}$ (K)                          & $1777.5 \pm 128.8$ & $1826.2 \pm 130.3$ & $1819.8 \pm 129.2$ \\
$N_{\rm H_2}^{\rm hot}$ ($\times10^{18}$ cm$^{-2}$)  & $1.55 \pm 0.46$    & $1.66 \pm 0.46$    & $2.70 \pm 0.73$ \\
OPR$^{\rm hot}$                                      & $2.72 \pm 0.21$    & $2.68 \pm 0.20$    & $2.59 \pm 0.20$ \\
$A_V$ (mag)                                          & $4.9 \pm 1.0$      & $10.7 \pm 1.9$     & $15.3 \pm 2.8$ \\
$M_{\rm wind}^{\rm hot}$ ($\times10^{-7}$ M$_\odot$) & $1.05 \pm 0.31$    & $1.13 \pm 0.31$    & $1.84 \pm 0.50$ \\
\hline
\multicolumn{4}{c}{Power Law Fit} \\
\hline
$b$                                                  & $4.18 \pm 0.03$    & $4.11 \pm 0.05$    & $4.09 \pm 0.05$ \\
$T^{\rm min}$ (K)                                    & $398.9 \pm 6.0$    & $401.7 \pm 11.2$   & $393.8 \pm 9.7$ \\
$N_{\rm H_2}^{\rm tot}$ ($\times10^{19}$ cm$^{-2}$)  & $4.48 \pm 0.13$    & $4.20 \pm 0.21$    & $5.51 \pm 0.34$ \\
$A_V$ (mag)                                          & $2.79 \pm 0.22$    & $5.19 \pm 0.75$    & $7.39 \pm 0.34$ \\
Avg. OPR                                             & $2.86 \pm 0.04$    & $2.93 \pm 0.08$    & $2.87 \pm 0.06$ \\
$M_{\rm wind}^{\rm total}$ ($\times10^{-6}$ M$_\odot$) & $3.05 \pm 0.10$  & $2.86 \pm 0.15$    & $3.75 \pm 0.24$ \\
\hline
\end{tabular}
}
\tablefoot{The fitted quantities include rotational temperatures, H$_2$ column densities, OPR ratios, and extinction. Wind masses are estimated from the emitting area and associated column densities as described in Sect.~\ref{subsec:excitation_conditions_winds}.}
\end{table}

\subsection{\texorpdfstring{Kinematics of H$_2$ Winds}{Kinematics of H2 Winds}} \label{subsec:winds_kinematics}

We investigate the velocity structure of the H$_2$ winds in HV~Tau~C using PV diagrams constructed for warm and hot gas components, traced by the S(2) \& S(1) lines (warm component) and S(8) \& S(7) lines (hot component) in the ground vibrational state ($v=0$--0).

\begin{figure*}[htbp!]
\centering
\includegraphics[width=\textwidth]{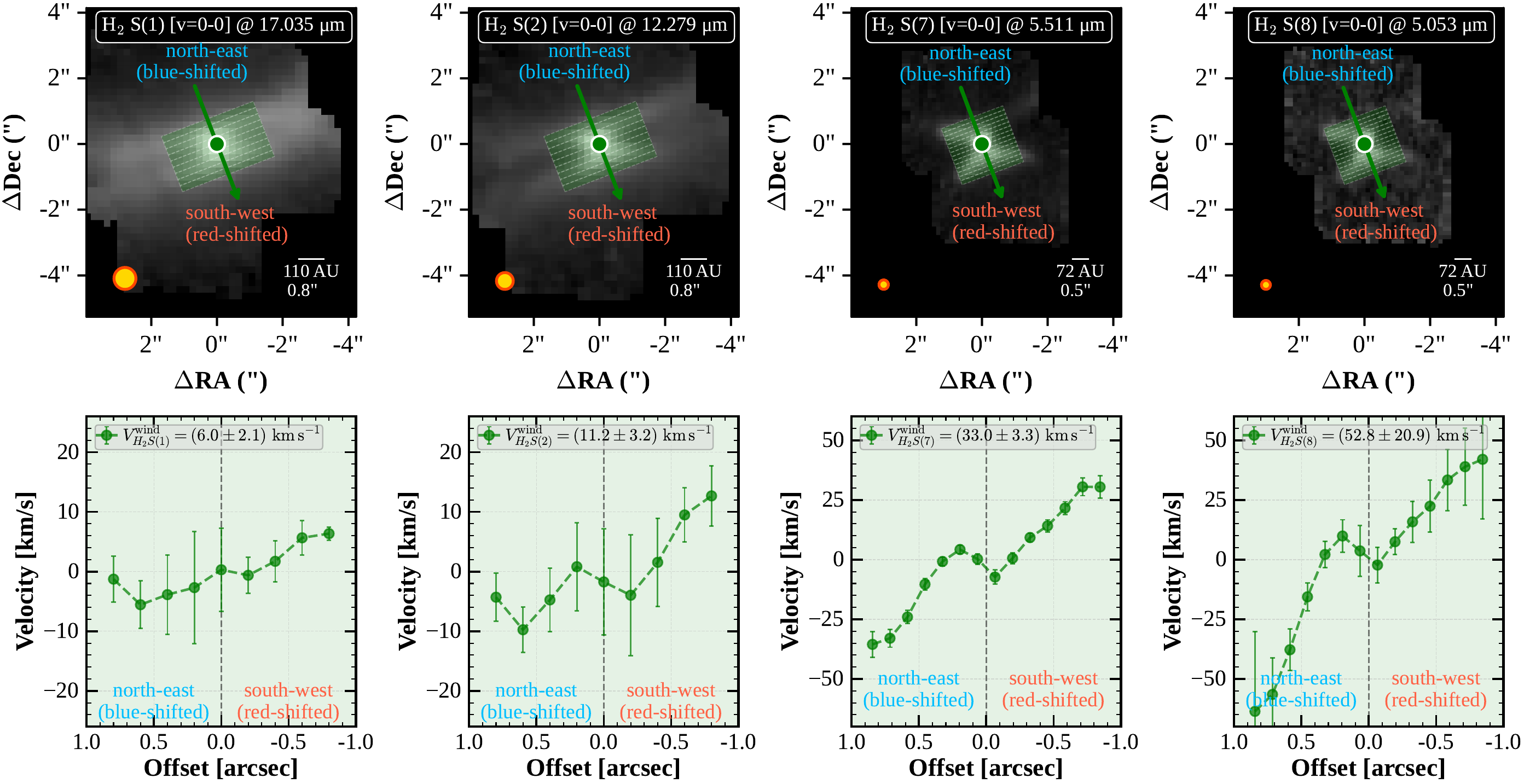}
\caption{PV diagrams for four H$_2$ transitions in the ground vibrational state ($v = 0$--0): S(1) and S(2), which trace the warm component, and S(7) and S(8), which trace the hot component. The upper panels show the PV paths (green rectangles with white PV slices) overlaid on the integrated-intensity maps of each transition, centered on the 14~$\mu$m JWST continuum position. The lower panels display the inclination- and systemic-velocity-corrected velocities as a function of positional offset from the source center. Note that the velocity (y-axis) scales differ between panels.}
\label{fig:PV_diagrams_along_H2_outflow_axis_for_S1_S2_S7_S8_lines}
\end{figure*}

\begin{table*}[htbp]
\caption{Velocity, extent, and dynamical timescale for molecular H$_2$ winds before and after inclination correction}
\label{tab:velocity_extent_dynamical_time_of_H2_winds}
\centering
\resizebox{\textwidth}{!}{
\begin{tabular}{c c | c c c | c c c}
\hline\hline
Component & Transition &
\multicolumn{3}{c|}{Before inclination correction} &
\multicolumn{3}{c}{After inclination correction} \\
 & &
Velocity $v$ (km s$^{-1}$) & Extent $d$ (AU) & $t_{\mathrm{dyn}}$ (year) &
Velocity $v_{\mathrm{corr}}$ (km s$^{-1}$) & Extent $d_{\mathrm{corr}}$ (AU) & $t_{\mathrm{dyn, corr}}$ (year) \\
\hline
Warm & S(1)    & $1.0 \pm 0.4$ & $155.1 \pm 1.0$ & $735.3 \pm 294.2$ & $6.0 \pm 2.1$  & $157.5 \pm 1.1$ & $129.7 \pm 52.3$ \\
     & S(2)    & $1.9 \pm 0.6$ & $155.1 \pm 1.0$ & $408.5 \pm 136.2$ & $11.2 \pm 3.2$ & $157.5 \pm 1.1$ & $72.0 \pm 24.3$ \\
     & Average & $1.4 \pm 0.4$ & $155.1 \pm 0.7$ & $561.2 \pm 159.3$ & $8.2 \pm 2.0$  & $157.5 \pm 0.8$ & $98.9 \pm 28.3$ \\
\hline
Hot  & S(7)    & $5.7 \pm 0.6$ & $155.1 \pm 1.0$ & $129.0 \pm 13.6$  & $33.0 \pm 3.3$ & $157.5 \pm 1.1$ & $22.8 \pm 2.6$ \\
     & S(8)    & $9.1 \pm 3.6$ & $155.1 \pm 1.0$ & $80.8 \pm 32.0$   & $52.8 \pm 20.9$ & $157.5 \pm 1.1$ & $14.2 \pm 5.8$ \\
     & Average & $7.4 \pm 1.9$ & $155.1 \pm 0.7$ & $104.9 \pm 17.8$  & $42.7 \pm 10.7$ & $157.5 \pm 0.8$ & $18.5 \pm 3.2$ \\
\hline
\end{tabular}
}
\end{table*}

\begin{figure*}[htbp!]
\centering
\includegraphics[width=\textwidth]{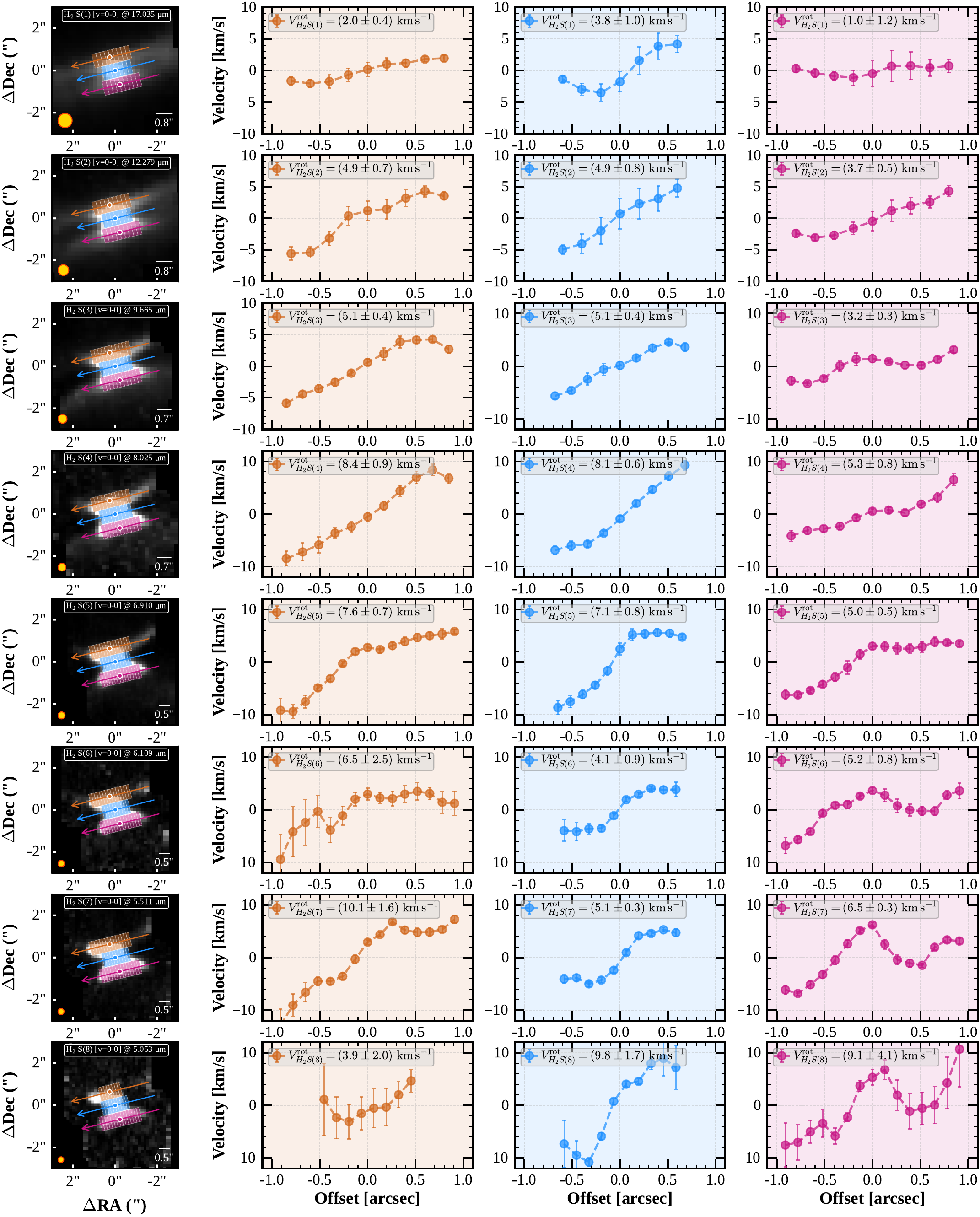}
\caption{PV diagrams for the eight H$_2$ pure rotational transitions S(1)--S(8) in the ground vibrational state $v=0$--0 toward HV~Tau~C. The left column displays three color-coded PV paths across the outflow axis, centered at the marked position (circle). The remaining columns show the corresponding PV diagrams for each transition along these paths, illustrating the rotational kinematic structure of the H$_2$ wind.}
\label{fig:hv_tau_c_all_H2_v=0-0_lines_PV_diagrams_across_the_H2_outflow_axis_for_three_PV_paths}
\end{figure*}

For each transition, we define a path along the outflow axis, perpendicular to the disk (indicated by the green arrow in the upper panels of Figure~\ref{fig:PV_diagrams_along_H2_outflow_axis_for_S1_S2_S7_S8_lines}), extending from the northeastern (positive offset) to the southwestern (negative offset) direction. The path length is chosen to fully encompass the H$_2$ emission detected in the S(8) $v=0$--0 transition (see the upper row of Figure~\ref{fig:PV_diagrams_along_H2_outflow_axis_for_S1_S2_S7_S8_lines}). The path width is set to one pixel per slice. For each positional slice, we extract the velocity-intensity profile and fit a Gaussian function to determine the mean velocity. The systemic velocity is established by computing the average of all slices' mean velocities and subtracting this offset from the entire velocity distribution. This correction addresses potential systematic uncertainties in the MIRI wavelength and velocity calibration. Although the \textit{JWST}/MIRI-MRS velocity resolution ranges from $\sim 80$ to $200~\text{km~s}^{-1}$ \citep{Labiano2021AA...656A..57L}, the absolute wavelength calibration accuracy is better than $9~\text{km~s}^{-1}$ at $5~\mu\text{m}$ and $27~\text{km~s}^{-1}$ at $28~\mu\text{m}$ \citep{Argyriou2023AA...675A.111A, Pontoppidan2024ApJ...963..158P}. These calibration uncertainties are propagated into our velocity measurements.

The systemic-velocity-corrected mean velocities (in $\text{km~s}^{-1}$) are plotted against positional offset (in arcseconds) along the wind path in the bottom row of Figure~\ref{fig:PV_diagrams_along_H2_outflow_axis_for_S1_S2_S7_S8_lines}. These data reveal a clear velocity gradient from the northern to southern regions of the wind outflow. We characterize the wind velocity as half the difference between the minimum and maximum velocities in the corrected distribution, representing the typical wind speed. All velocities are subsequently corrected for inclination effects, and Figure~\ref{fig:PV_diagrams_along_H2_outflow_axis_for_S1_S2_S7_S8_lines} displays the inclination-corrected velocities.

Given HV~Tau~C's nearly edge-on orientation (inclination angle $\sim 80\degree$), distinguishing between blueshifted and redshifted velocity components proves challenging. Nevertheless, our analysis clearly demonstrates that the lower transitions (S(1) and S(2)), which trace the warm gas component, exhibit systematically lower velocities compared to the higher transitions (S(7) and S(8)) that probe hotter gas. We have extended this analysis to include the remaining H$_2$ transitions (S(3) through S(6)), finding that all lines show consistent velocity gradients along the outflow axis (see Appendix~\ref{appfig:hv_tau_c_H2_v=0-0_PV_diagrams_along_H2_outflow_axis_for_S3_S4_S5_S6_lines} ). Furthermore, we observe a monotonic increase in wind velocity from the S(1) to S(8) transitions, indicating a temperature-dependent velocity structure within the molecular outflow.

The spatial extent of the wind is measured from the S(8) line emission using a rectangular aperture that encloses the entire emission region, as shown in Figure~\ref{fig:hv_tau_c_H2_v00_extended_linemaps}. Since the full measured length encompasses both outflow lobes, we adopt half of this value to represent the one-sided wind extent ($d$) from the central source. The dynamical timescale of the wind is calculated as, $ t_{\text{dyn}} = {d}/{\langle v \rangle},$ where $d$ is the one-sided wind extent, and $\langle v \rangle$ is the average wind velocity. To account for the wind outflow inclination angle $i$ relative to the line of sight, we apply inclination corrections following \citet{LiShanghuo2019ApJ...886..130L}:
$$
v_{\text{corr}} = \frac{\langle v \rangle}{\cos i}, \quad d_{\text{corr}} = \frac{d}{\sin i}, \quad t_{\text{dyn, corr}} = \frac{t_{\text{dyn}}}{\sin i / \cos i}.
$$
Using this methodology, we derive the kinematic properties of the wind outflow--including average velocity, spatial extent, and dynamical timescale--for both warm and hot components. Table~\ref{tab:velocity_extent_dynamical_time_of_H2_winds} presents these values both before and after inclination correction.

The dynamical timescale ($t_{\mathrm{dyn}}$) is estimated by adopting a constant average velocity, $\langle v \rangle$, as the characteristic velocity of the H$_2$ emission for the warm (tracing the outer, warmer part of the wind) and hot (tracing the inner, hotter component of the wind) gas separately (see Table~\ref{tab:velocity_extent_dynamical_time_of_H2_winds}). This approach implicitly assumes that the flow propagates at a constant velocity. However, our results, particularly the presence of a clear velocity gradient along the outflow axis (see Figure~\ref{fig:PV_diagrams_along_H2_outflow_axis_for_S1_S2_S7_S8_lines}), suggest that the flow is likely accelerating and stratified, as the velocity increases with offset in the PV diagrams, while the hotter H$_2$ lines consistently trace higher velocities than the warmer H$_2$ lines.
In such a scenario, the velocity near the launching region could be lower than the characteristic velocity measured at larger distances for both the warm and hot components. As a consequence, adopting a single average velocity likely overestimates the effective flow speed and therefore leads to an underestimation of the dynamical timescale. To account for this effect, we estimate lower and upper bounds on the dynamical timescale for each component by adopting the velocities measured at the smallest and largest offsets, respectively. In this approach, we define $v_{\min}$ as the velocity close to the source (i.e., at small offsets, just beyond the diffraction limit) and $v_{\max}$ as the velocity at the largest observed extent of the emission. 

For a fixed spatial extent $d$ for each component, the dynamical timescale is bounded as
$$
t_{\mathrm{dyn}}^{\min} = \frac{d}{v_{\max}}, \quad
t_{\mathrm{dyn}}^{\max} = \frac{d}{v_{\min}} \implies 
\frac{t_{\mathrm{dyn}}^{\max}}{t_{\mathrm{dyn}}^{\min}} = \frac{v_{\max}}{v_{\min}}.
$$

For the warm component, with velocities spanning $\sim 3$ to $\sim 8$ km s$^{-1}$, and for the hot component, with velocities from $\sim 10$ to $\sim 40$ km s$^{-1}$, this gives
$$
\left(\frac{t_{\mathrm{dyn}}^{\max}}{t_{\mathrm{dyn}}^{\min}}\right)_{\mathrm{warm}} \approx \frac{8}{3} \approx 2.7; \quad
\left(\frac{t_{\mathrm{dyn}}^{\max}}{t_{\mathrm{dyn}}^{\min}}\right)_{\mathrm{hot}} \approx \frac{40}{10} = 4.
$$

Thus, the dynamical timescale is uncertain by a factor of $\sim 2.7$--4 across the two components. In particular, if the flow is accelerating, the use of a single average velocity biases $t_{\mathrm{dyn}}$ toward lower values.

To investigate whether the H$_2$ winds exhibit rotational motion, we constructed PV diagrams by extracting three PV slices at different spatial locations across (i.e., orthogonal to) the outflow axis, as illustrated in Figure~\ref{fig:hv_tau_c_all_H2_v=0-0_lines_PV_diagrams_across_the_H2_outflow_axis_for_three_PV_paths}. These diagrams were generated for all pure rotational H$_2$ transitions from S(1) through S(8) in the ground vibrational state $v=0$--0, following the same methodology described above. In this case, we corrected the observed velocities using $ v_{\rm corr} = \langle v \rangle / \sin i, $
where the inclination angle $i$ accounts for projection effects. The resulting inclination-corrected velocities shown in Figure~\ref{fig:hv_tau_c_all_H2_v=0-0_lines_PV_diagrams_across_the_H2_outflow_axis_for_three_PV_paths} are additionally corrected for the systemic velocity. The central PV aperture (indicated in blue) aligns closely with the disk major axis. It therefore reveals the disk rotation signature in all eight lines, consistent with the ALMA disk rotation pattern discussed in Section~\ref{sec:hv_tau_c_alma_characteristics}. However, unlike ALMA, \textit{JWST}/MIRI-MRS does not provide sufficient velocity resolution to resolve the rotation at the disk center fully. Even so, a clear velocity gradient is visible across the right-to-left spatial offsets in all transitions.

\begin{table*}[htbp]
\caption{H$_2$ wind outflow dynamical quantities}
\label{tab:dynamical_quantities}
\centering
\footnotesize
\begin{tabular}{lccccc}
\hline\hline
Extinction Law & Component & 
$M_{\mathrm{wind}}$ ($M_\odot$) &
$\dot{M}_{\mathrm{loss}}$ ($M_\odot~\mathrm{yr}^{-1}$) &
$F_{\mathrm{H_2}}$ ($M_\odot~\mathrm{yr}^{-1}~\mathrm{km~s}^{-1}$) &
$L_{\mathrm{mech}}$ ($M_\odot~\mathrm{yr}^{-1}~\mathrm{km}^2~\mathrm{s}^{-2}$) \\
\hline
\citetalias{Pontoppidan2024ApJ...963..158P}
    & Warm  & $(2.63 \pm 0.37) \times 10^{-6}$ & $(2.66 \pm 0.85) \times 10^{-8}$ & $(2.19 \pm 0.88) \times 10^{-7}$ & $(9.05 \pm 5.22) \times 10^{-7}$ \\
    & Hot   & $(1.05 \pm 0.31) \times 10^{-7}$ & $(5.68 \pm 1.95) \times 10^{-9}$ & $(2.42 \pm 1.03) \times 10^{-7}$ & $(5.17 \pm 3.14) \times 10^{-6}$ \\
    & Total & $(3.05 \pm 0.97) \times 10^{-6}$ & $(5.19 \pm 1.27) \times 10^{-8}$ & $(1.32 \pm 0.43) \times 10^{-6}$ & $(1.69 \pm 0.83) \times 10^{-5}$ \\
\hline
\citetalias{HensleyDraine2023ApJ...948...55H}
    & Warm  & $(2.18 \pm 0.25) \times 10^{-6}$ & $(2.20 \pm 0.68) \times 10^{-8}$ & $(1.82 \pm 0.71) \times 10^{-7}$ & $(7.50 \pm 4.28) \times 10^{-7}$ \\
    & Hot   & $(1.13 \pm 0.31) \times 10^{-7}$ & $(6.11 \pm 1.99) \times 10^{-9}$ & $(2.61 \pm 1.07) \times 10^{-7}$ & $(5.57 \pm 3.33) \times 10^{-6}$ \\
    & Total & $(2.86 \pm 0.15) \times 10^{-6}$ & $(4.87 \pm 1.21) \times 10^{-8}$ & $(1.24 \pm 0.41) \times 10^{-6}$ & $(1.58 \pm 0.78) \times 10^{-5}$ \\
\hline
\citetalias{McClure2009ApJ...693L..81M}
    & Warm  & $(3.78 \pm 0.62) \times 10^{-6}$ & $(3.82 \pm 1.26) \times 10^{-8}$ & $(3.15 \pm 1.29) \times 10^{-7}$ & $(1.30 \pm 0.76) \times 10^{-6}$ \\
    & Hot   & $(1.84 \pm 0.50) \times 10^{-7}$ & $(9.95 \pm 3.18) \times 10^{-9}$ & $(4.25 \pm 1.73) \times 10^{-7}$ & $(9.07 \pm 5.39) \times 10^{-6}$ \\
    & Total & $(3.75 \pm 0.24) \times 10^{-6}$ & $(6.39 \pm 1.60) \times 10^{-8}$ & $(1.63 \pm 0.54) \times 10^{-6}$ & $(2.07 \pm 1.03) \times 10^{-5}$ \\
\hline
\end{tabular}

\tablefoot{The mass-loss rate ($\dot{M}_{\mathrm{loss}}$), momentum rate ($F_{\mathrm{H_2}}$), and mechanical luminosity ($L_{\mathrm{mech}}$) are shown for the warm and hot H$_2$ wind components derived using three different extinction laws. The ``Total'' values correspond to wind properties derived from the power-law temperature distribution.}
\end{table*}

To further examine rotation in the winds themselves, we produced PV diagrams for two distinct PV slices for both wind lobes (northern and southern lobes indicated in orange and magenta colors, respectively), centered on the outflow axis and oriented orthogonal to it. The inclination- and systemic-corrected PV diagrams for these wind-tracing apertures are presented in the second and fourth columns for all eight H$_2$ lines in Figure~\ref{fig:hv_tau_c_all_H2_v=0-0_lines_PV_diagrams_across_the_H2_outflow_axis_for_three_PV_paths} (each line shown in a separate row). These diagrams also display a clear velocity gradient from the left side to the right side along the offset axis. A rotational trend in the winds is particularly evident in at least the S(2), S(3), S(4), S(5), and S(7) transitions.

Additionally, we generated velocity-shift maps for all pure rotational H$_2$ lines in the ground vibrational state $v=0$--0, following the procedure described in Appendix~\ref{appsec:Remaining_PV_diagrams_and_Radial_Velocity_Shift_Maps}. The resulting maps are presented in Figure~\ref{appfig:hv_tau_c_v=0-0_H2_lines_radial_velocity_shift_maps}. These velocity maps also reveal a gradient in the velocity structure, particularly in the higher-excitation transitions from S(8) down to S(5), further supporting the presence of rotational motion within the winds. In contrast, the velocity structure in the lower-excitation transitions, such as S(1) and S(2), cannot be clearly resolved. This limitation is likely due to the relatively poorer velocity resolution of \textit{JWST}/MIRI--MRS at longer wavelengths, where these lower transitions are located.

\subsection{\texorpdfstring{H$_2$}{H2} wind dynamics}
\label{subsec:h2_wind_dynamics}

To quantify the dynamical role of the H$_2$ wind in HV~Tau~C, we estimate its mass-loss rate, momentum rate, and mechanical luminosity. These quantities are derived using the inclination-corrected velocities, projected extents, and dynamical timescales obtained in Section~\ref{subsec:winds_kinematics}, together with the column densities and wind masses estimated in Section~\ref{subsec:excitation_conditions_winds}. 

The mass-loss rate of the wind is given by
$
\dot{M}_{\mathrm{loss}} = M_{\mathrm{wind}}/t_{\mathrm{dyn}},
$
where $M_{\mathrm{wind}}$ is the total H$_2$ mass in the wind and $t_{\mathrm{dyn}}$ is the dynamical timescale of the outflow. This quantity represents the rate at which molecular gas is being removed from the circumstellar environment. The momentum rate carried by the wind is
$
F_{\mathrm{H_2}} = \dot{M}_{\mathrm{loss}} \, \langle v \rangle,
$
where $\langle v \rangle$ is the inclination-corrected characteristic velocity of the outflow. This quantity indicates the rate at which linear momentum is transferred to the surrounding gas, governing the wind’s ability to sweep up or displace circumstellar material. The mechanical luminosity, representing the kinetic energy injection rate, is computed as
$
L_{\mathrm{mech}} = 0.5 \, \dot{M}_{\mathrm{loss}} \, \langle v \rangle^{2},
$
which estimates the power available to heat, accelerate, or disperse H$_2$ gas. Together, $\dot{M}_{\mathrm{loss}}$, $F_{\mathrm{H_2}}$, and $L_{\mathrm{mech}}$ characterize the dynamical impact of the molecular wind on the disk environment and its potential role in extracting angular momentum. The derived values for the warm, hot, and total H$_2$ components are listed in Table~\ref{tab:dynamical_quantities}.

As established in Section~\ref{subsec:winds_kinematics}, the observed velocity gradient along the outflow axis indicates that the flow is stratified and likely accelerating. To account for this, we estimate lower and upper bounds on the dynamical timescale using the minimum ($v_{\min}$) and maximum ($v_{\max}$) observed velocities. Because the derived quantities scale as $\dot{M}_{\mathrm{loss}} \propto v$, $F_{\mathrm{H_2}} \propto v^{2}$, and $L_{\mathrm{mech}} \propto v^{3}$, this velocity spread introduces systematic uncertainties of factors of $\sim$2.5--4, $\sim$6--16, and $\sim$15--64, respectively. Crucially, if the flow is accelerating, adopting a single average velocity biases the dynamical timescale $t_{\mathrm{dyn}}$ toward lower values. Consequently, the derived $\dot{M}_{\mathrm{loss}}$, $F_{\mathrm{H_2}}$, and $L_{\mathrm{mech}}$ values should be regarded as upper limits.

\section{Estimating the line-of-sight extinction ($A_V$)}
\label{sec:extinction_Av_text}

HV~Tau~C is viewed at a nearly edge-on inclination, such that the line-of-sight toward the central star is strongly affected by circumstellar material in the disk. In this geometry, the observed emission at optical and near-infrared wavelengths is dominated by scattered light originating from the disk surface layers rather than by direct stellar radiation \citep[e.g.,][]{Wolff2017ApJ...851...56W, Villenave2020A&A...642A.164V, Sturm2024A&A...689A..92S}. As a result, extinction estimates derived from standard color-excess methods, which assume a direct view of the stellar photosphere, are not reliable for this system.

\begin{figure*}[htbp!]
\centering
\begin{subfigure}[t]{0.48\textwidth}
    \centering
    \includegraphics[width=\textwidth]{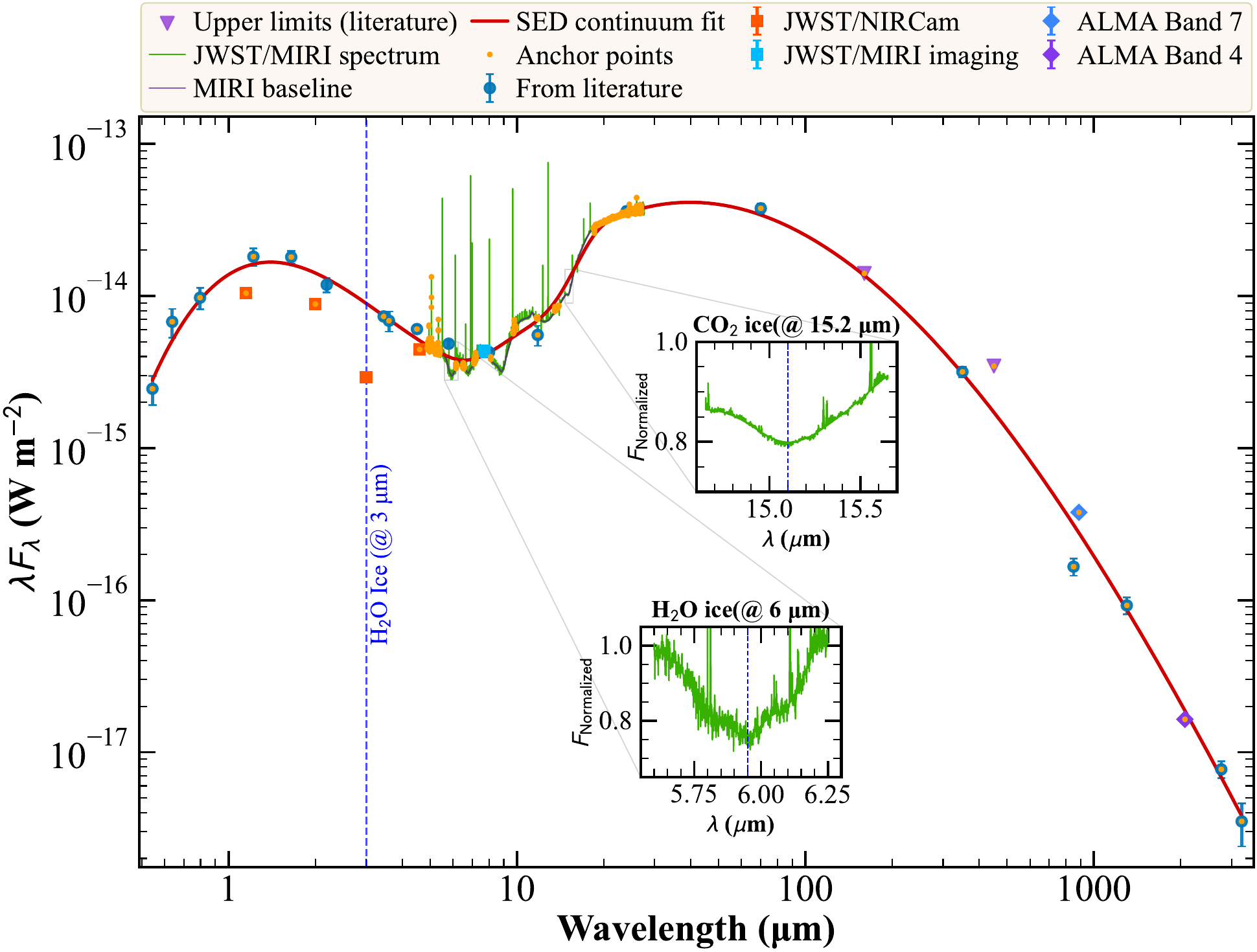}
    \caption{SED continuum fitting and ice feature analysis for extinction calculation. The solid red line shows the continuum fit, with small orange circles indicating anchor points. Vertical dashed lines mark the locations of H$_2$O and CO$_2$ ice features used for extinction measurements.}
    \label{fig:hv_tau_c_Av_calculation_using_ice}
\end{subfigure}
\hfill
\begin{subfigure}[t]{0.48\textwidth}
    \centering
    \includegraphics[width=\textwidth]{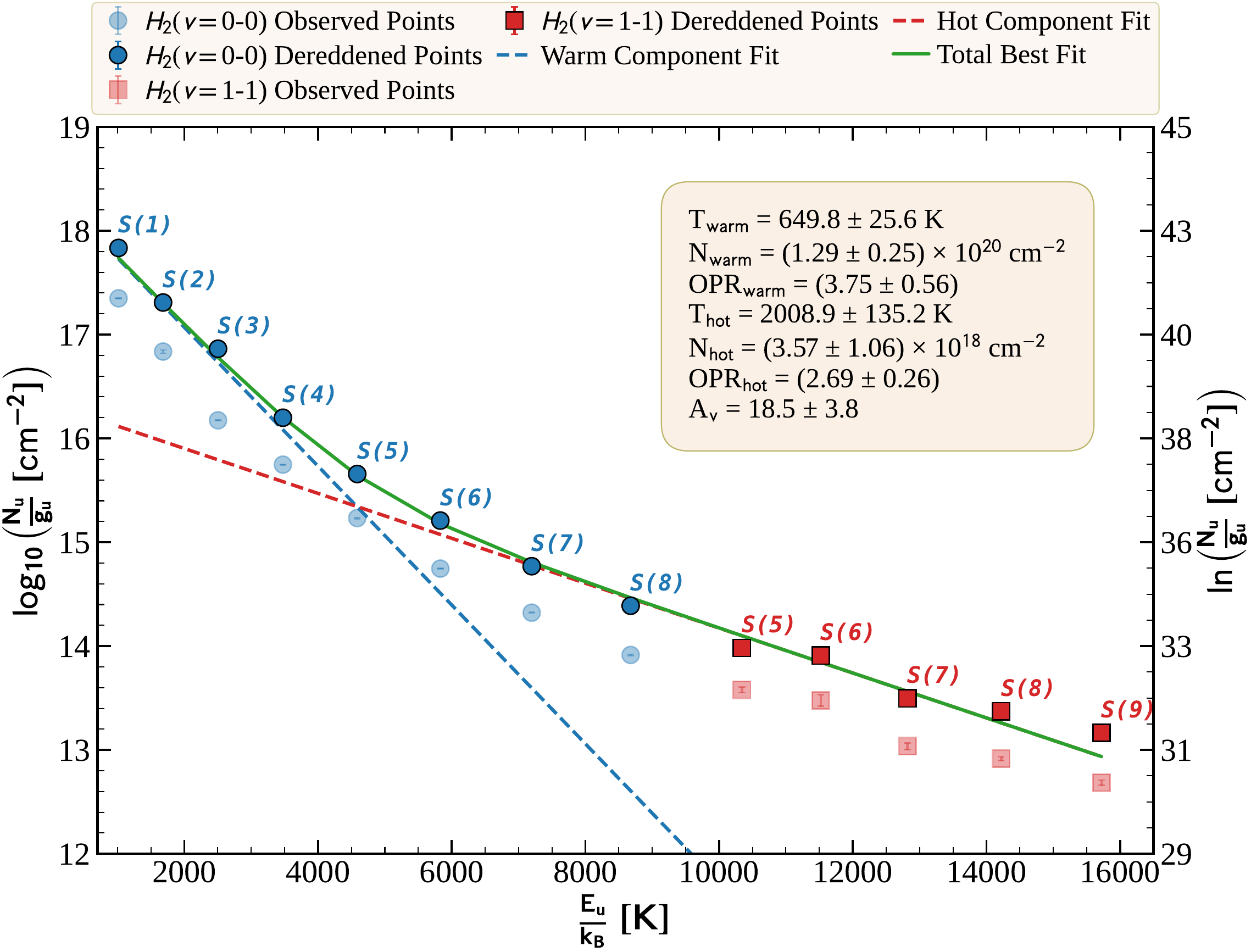}
    \caption{Two-component fit to the H$_2$ rotational diagram within a circular aperture of radius 2$\times$PSF FWHM, using the \texttt{LMFIT}. The OPR was treated as a free parameter. This particular fit employs the \citetalias{McClure2009ApJ...693L..81M} extinction law.}
    \label{fig:hv_tau_c_Av_calculation_using_h2_rotational_diagram}
\end{subfigure}
\caption{Extinction calculations for HV Tau C using two independent methods. (\subref{fig:hv_tau_c_Av_calculation_using_ice}) Spectral energy distribution analysis utilizing ice absorption features at 3.0, 6.0, and 15.0 $\mu$m. (\subref{fig:hv_tau_c_Av_calculation_using_h2_rotational_diagram}) H$_2$ rotational diagram analysis with two-temperature component fitting, providing complementary extinction estimates through molecular line diagnostics.}
\label{fig:hv_tau_c_Av_calculation_combined}
\end{figure*}

A more robust approach to estimating $A_V$ in such systems is provided by infrared ice absorption features (see also \citet{vanDishoeck2025A&A...699A.361V}, Appendix~G). In this work, we estimate the line-of-sight A$_V$ toward HV~Tau~C using two independent methods: (i) ice absorption features in the near-IR and mid-IR spectral energy distribution (SED), and (ii) molecular H$_2$ lines rotational diagram diagnostics constructed from continuum-subtracted line fluxes. These two approaches probe different physical components of the disk -- solid-phase ices and warm molecular gas. Given that HV~Tau~C is viewed nearly edge-on, it provides an excellent opportunity to investigate the ice inventory in a Class~II protoplanetary disk.

Previous studies have reported a wide range of extinction values toward HV~Tau~C using both ice absorption features and near-infrared H$_2$ line ratios. These measurements are summarized in Table~\ref{tab:Av_measurements_using_ices_and_H2}. \citet{Terada2007ApJ...667..303T} derived $A_V$ from the optical depth of the 3.0~$\mu$m H$_2$O ice feature observed with Subaru/IRCS-LRS. Their observations, obtained at two epochs (2002 September 25 UT and 2005 January 18 UT), revealed a significant variation of approximately 15~mag in $A_V$, suggesting temporal changes in the line-of-sight column density or disk structure. An independent gas-based extinction estimate was presented by \citet{Beck2008ApJ...676..472B}, who used near-infrared H$_2$ line ratios measured with the Gemini Near-infrared Integral Field Spectrograph (NIFS). By comparing pairs of transitions originating from the same upper energy level, specifically the 1--0~Q(3)/S(1) and 1--0~Q(2)/S(0) ratios, they constrained the extinction affecting the warm molecular gas component.

\subsection{SED fitting and \texorpdfstring{$A_V$}{AV} from ice features}
\label{subsec:Av_from_ice}

The SED of HV~Tau~C was constructed by combining photometric measurements available in the literature with new JWST photometric and spectroscopic observations, together with ALMA continuum data from Bands~4 and~7. The compiled dataset includes optical, near-infrared, far-infrared, and radio fluxes from \citet{Duchene2010ApJ...712..112D}, JWST/NIRCam photometry at 1.15~$\mu$m (F115W), 2.00~$\mu$m (F200W), 3.00~$\mu$m (F300M), and 4.60~$\mu$m (F460M), as well as JWST/MIRI photometry at 7.70~$\mu$m (F770W). In addition, JWST/MIRI--MRS spectroscopic observations covering the 4.9--27.9~$\mu$m wavelength range were incorporated. We further included ALMA Band~7 (887~$\mu$m) and Band~4 (2068~$\mu$m) continuum fluxes to extend the SED into the (sub-)millimeter regime. The MIRI--MRS spectra were extracted from the \texttt{x1d.fits} products using apertures corresponding to twice the wavelength-dependent PSF FWHM. Since ice absorption features are expected to originate predominantly from material concentrated near the disk midplane, we restricted the extraction to a compact region centered on the disk. The full SED is shown in Fig.~\ref{fig:hv_tau_c_Av_calculation_using_ice}, and details of the photometric data extraction and flux measurements of JWST/NIRCam \& MIRI are provided in Appendix~\ref{appsec:HV_Tau_C_JWST_NIRCam_and_MIRI_photometric_data_analysis}. The 1.15~$\mu$m and 2.00~$\mu$m NIRCam photometric points trace the continuum emission, while the 3.00~$\mu$m (F300M) filter captures the H$_2$O ice absorption feature, the 4.60~$\mu$m (F460M) filter includes the CO ice absorption band, and the 7.70~$\mu$m (F770W) filter is dominated by polycyclic aromatic hydrocarbon (PAH) emission. A detailed description of the JWST/NIRCam filter set is available online\footnote{\url{https://jwst-docs.stsci.edu/jwst-near-infrared-camera/nircam-instrumentation/nircam-filters}}.

To reliably measure the optical depths of the ice absorption features, it is essential to determine the underlying continuum baseline across a broad wavelength range. Instead of performing local continuum fits around individual absorption bands, we adopted a global SED-fitting approach to characterize the intrinsic continuum emission of the system. This method provides a physically motivated and self-consistent baseline, minimizing systematic uncertainties associated with local polynomial fitting and enabling robust derivation of ice-band optical depths.

\begin{table}[htbp]
\caption{$A_V$ estimates from different features and extinction laws.}
\label{tab:Av_measurements_using_ices_and_H2}
\centering
\footnotesize
\resizebox{\columnwidth}{!}{
\begin{tabular}{lll}
\hline\hline
Extinction Law & Feature & $A_V$ (mag) \\
\hline
\citetalias{Pontoppidan2024ApJ...963..158P} 
    & 3\,$\mu$m H$_2$O ice & $5.36 \pm 0.38$ \\
    & 6\,$\mu$m H$_2$O ice & $4.18 \pm 0.21$ \\
    & 15\,$\mu$m CO$_2$ ice & $2.91 \pm 0.03$ \\
    & H$_2$ rotational diagram & $6.2 \pm 1.4$ \\
\hline
\citetalias{HensleyDraine2023ApJ...948...55H} 
    & 3\,$\mu$m H$_2$O ice & -- \\
    & 6\,$\mu$m H$_2$O ice & -- \\
    & 15\,$\mu$m CO$_2$ ice & -- \\
    & H$_2$ rotational diagram & $12.8 \pm 2.5$ \\
\hline
\citetalias{McClure2009ApJ...693L..81M}
    & 3\,$\mu$m H$_2$O ice & -- \\
    & 6\,$\mu$m H$_2$O ice & $4.44 \pm 0.22$ \\
    & 15\,$\mu$m CO$_2$ ice & $4.63 \pm 0.55$ \\
    & H$_2$ rotational diagram & $18.5 \pm 3.8$ \\
\hline
Literature 
    & 3\,$\mu$m H$_2$O ice (1st epoch) & $44.1^{\mathrm{a}}$ \\
    & 3\,$\mu$m H$_2$O ice (2nd epoch) & $29.0^{\mathrm{a}}$ \\
    & 1--0 Q(3)/S(1) ratio & $20.3 \pm 0.5^{\mathrm{b}}$ \\
    & 1--0 Q(2)/S(0) ratio & $7 \pm 1^{\mathrm{b}}$ \\
\hline
\end{tabular}
}
\tablefoot{$A_V$ estimates derived from ice features and H$_2$ rotational diagnostics assuming various extinction laws. Literature values are from: $^{\mathrm{a}}$~\citet{Terada2007ApJ...667..303T}; $^{\mathrm{b}}$~\citet{Beck2008ApJ...676..472B}.  }
\end{table}

\begin{table*}[htbp]
\caption{Best-fit parameters of all physical quantities from two-component H$_2$ rotational diagrams for all six apertures}
\label{tab:all_6_apertures_best_fit_params_from_H2_rotational_diagrams}
\centering
\footnotesize
\resizebox{\textwidth}{!}{
\begin{tabular}{lcccccccc}
\hline\hline
Extinction Law & Aperture &
$T_{\mathrm{warm}}$ (K) & $N_{\mathrm{warm}}$ ($\times10^{18}$ cm$^{-2}$) &
OPR$_{\mathrm{warm}}$ & $T_{\mathrm{hot}}$ (K) &
$N_{\mathrm{hot}}$ ($\times10^{18}$ cm$^{-2}$) & OPR$_{\mathrm{hot}}$ &
$A_V$ (mag) \\
\hline
\citetalias{Pontoppidan2024ApJ...963..158P} 
    & AP1 & $674.9 \pm 18.9$ & $46.8 \pm 5.1$ & $3.32 \pm 0.21$ & $1819.8 \pm 125.6$ & $2.28 \pm 0.67$ & $2.68 \pm 0.20$ & $5.2 \pm 0.7$ \\
    & AP2 & $681.4 \pm 38.8$ & $52.1 \pm 15.8$ & $2.97 \pm 0.44$ & $1843.5 \pm 102.9$ & $3.16 \pm 0.87$ & $2.96 \pm 0.16$ & $3.8 \pm 1.7$ \\
    & AP3 & $632.9 \pm 19.0$ & $41.8 \pm 4.8$ & $3.12 \pm 0.22$ & $1768.1 \pm 152.1$ & $1.47 \pm 0.51$ & $2.74 \pm 0.20$ & $4.2 \pm 0.9$ \\
    & AP4 & $684.5 \pm 11.6$ & $41.0 \pm 4.6$ & $3.24 \pm 0.20$ & $2132.8 \pm 56.4$ & $1.15 \pm 0.15$ & $3.00 \pm 0.24$ & $4.8 \pm 0.7$ \\
    & AP5 & $734.3 \pm 48.6$ & $59.4 \pm 22.4$ & $2.99 \pm 0.51$ & $2097.5 \pm 160.9$ & $2.29 \pm 0.83$ & $3.06 \pm 0.27$ & $4.1 \pm 2.2$ \\
    & AP6 & $655.9 \pm 31.8$ & $46.2 \pm 12.7$ & $3.26 \pm 0.42$ & $1823.9 \pm 125.0$ & $2.66 \pm 0.77$ & $2.64 \pm 0.19$ & $4.6 \pm 1.6$ \\
\hline
\citetalias{HensleyDraine2023ApJ...948...55H} 
    & AP1 & $699.1 \pm 20.3$ & $37.3 \pm 3.4$ & $4.03 \pm 0.29$ & $1832.3 \pm 121.5$ & $2.59 \pm 0.71$ & $2.60 \pm 0.19$ & $10.5 \pm 1.2$ \\
    & AP2 & $711.1 \pm 39.7$ & $45.1 \pm 10.9$ & $3.43 \pm 0.58$ & $1876.6 \pm 106.4$ & $3.36 \pm 0.94$ & $2.97 \pm 0.15$ & $8.4 \pm 3.3$ \\
    & AP3 & $652.0 \pm 20.8$ & $35.7 \pm 3.5$ & $3.67 \pm 0.28$ & $1806.1 \pm 159.5$ & $1.56 \pm 0.53$ & $2.71 \pm 0.20$ & $8.9 \pm 1.7$ \\
    & AP4 & $700.2 \pm 13.0$ & $34.5 \pm 3.6$ & $3.72 \pm 0.27$ & $2116.3 \pm 60.0$ & $1.34 \pm 0.19$ & $2.93 \pm 0.25$ & $9.4 \pm 1.4$ \\
    & AP5 & $765.2 \pm 47.1$ & $52.7 \pm 15.6$ & $3.41 \pm 0.65$ & $2132.7 \pm 162.9$ & $2.47 \pm 0.90$ & $3.04 \pm 0.27$ & $9.1 \pm 4.0$ \\
    & AP6 & $696.4 \pm 30.2$ & $36.5 \pm 6.4$ & $3.91 \pm 0.56$ & $1871.9 \pm 122.2$ & $2.77 \pm 0.76$ & $2.64 \pm 0.18$ & $9.9 \pm 2.7$ \\
\hline
\citetalias{McClure2009ApJ...693L..81M}
    & AP1 & $682.2 \pm 20.1$ & $63.8 \pm 8.6$ & $3.83 \pm 0.27$ & $1819.6 \pm 121.2$ & $4.25 \pm 1.13$ & $2.52 \pm 0.19$ & $15.1 \pm 1.9$ \\
    & AP2 & $693.5 \pm 41.0$ & $66.1 \pm 25.3$ & $3.21 \pm 0.56$ & $1854.9 \pm 106.0$ & $4.82 \pm 1.83$ & $2.88 \pm 0.17$ & $11.2 \pm 5.1$ \\
    & AP3 & $640.5 \pm 20.0$ & $56.5 \pm 8.6$ & $3.51 \pm 0.27$ & $1803.9 \pm 161.0$ & $2.35 \pm 0.76$ & $2.62 \pm 0.19$ & $12.9 \pm 2.6$ \\
    & AP4 & $688.1 \pm 13.5$ & $56.0 \pm 9.4$ & $3.56 \pm 0.27$ & $2126.0 \pm 63.1$ & $2.04 \pm 0.37$ & $2.83 \pm 0.25$ & $13.7 \pm 2.2$ \\
    & AP5 & $751.1 \pm 50.6$ & $76.6 \pm 37.3$ & $3.16 \pm 0.63$ & $2114.7 \pm 169.0$ & $3.53 \pm 1.76$ & $2.97 \pm 0.29$ & $11.7 \pm 6.3$ \\
    & AP6 & $679.8 \pm 30.2$ & $60.2 \pm 17.8$ & $3.70 \pm 0.52$ & $1863.1 \pm 122.2$ & $4.34 \pm 1.41$ & $2.56 \pm 0.18$ & $14.0 \pm 4.0$ \\
\hline
\end{tabular}
}
\tablefoot{Best-fit physical parameters (temperatures, column densities, OPRs, and $A_V$) from two-component H$_2$ rotational diagram fits for six apertures under three different extinction laws.}
\end{table*}

To construct a smooth continuum model, we first removed spectral line emission from the MIRI--MRS spectra using the \texttt{pybaselines} Python package \citep{pybaselines2022zndo...5608581E}, which estimates the continuum by fitting a baseline to the observed spectrum. The baseline-corrected spectra were then combined with the multi-wavelength photometric data. We selected anchor points in the SED that were free of both narrow (e.g., emission lines) and broad (e.g., ice, PAH etc.) spectral features and used the \texttt{UnivariateSpline} function from the \texttt{scipy.interpolate} module to obtain a smooth interpolation of the continuum. We adopted a cubic spline ($k=3$) and a smoothing parameter $s=4$, which provides an optimal balance between fitting accuracy and smoothness. The resulting best-fit SED continuum is overplotted in Fig.~\ref{fig:hv_tau_c_Av_calculation_using_ice}.

The extinction due to ice was estimated by measuring the optical depths of the ice absorption bands centered at $\sim$3.0~$\mu$m (H$_2$O), $\sim$6.0~$\mu$m (H$_2$O), and $\sim$15.0~$\mu$m (CO$_2$) (following \citealt{tyagi_et_al_2025}). The optical depth at each wavelength, $\tau_\lambda$, was calculated as
$$
\tau_\lambda = \ln \left( \frac{F_{\mathrm{cont}}}{F_{\mathrm{obs}}} \right),
$$
where $F_{\mathrm{cont}}$ is the continuum flux derived from the fitted SED, and $F_{\mathrm{obs}}$ is the observed flux at the feature center. The corresponding extinction is given by $ A_\lambda = 1.086\, \tau_\lambda.$ 

To convert from $A_\lambda$ to $A_V$, we used empirical extinction laws that provide $A_\lambda / A_V$ at the relevant wavelengths. We adopted two extinction laws: (1) \citetalias{Pontoppidan2024ApJ...963..158P}, which includes all three ice features, and (2) \citetalias{McClure2009ApJ...693L..81M}, which covers the 6.0~$\mu$m H$_2$O and 15.0~$\mu$m CO$_2$ features but does not include the 3.0~$\mu$m H$_2$O ice band. We did not use the \citetalias{HensleyDraine2023ApJ...948...55H} law, as it does not include any ice absorption features. The resulting $A_V$ values derived from each ice feature are summarized in Table~\ref{tab:Av_measurements_using_ices_and_H2}.

\subsection{$A_V$ from H$_2$ Rotational Diagrams} \label{Av_from_H2_rotational_diagram}

A second, independent estimate of the $A_V$ was derived from a two-component fit to H$_2$ rotational diagrams using the \texttt{LMFIT} package~\citep[e.g., see][Tyagi et al. 2026, under review]{Narang2024ApJ...962L..16N}. We constructed rotational diagrams (following the same methodology detailed in Section~\ref{subsec:excitation_conditions_winds}) from continuum-subtracted H$_2$ line fluxes measured within a circular aperture of radius equal to 2~PSF~FWHM, as presented in Figure~\ref{fig:hv_tau_c_Av_calculation_using_h2_rotational_diagram}. The warm and hot H$_2$ components were fitted simultaneously, with the OPR for each component treated as a free parameter. $A_V$ is incorporated as a fitting parameter because it modulates the observed flux ratios across the wavelength range, thereby influencing the derived excitation temperatures and column densities. The best-fit $A_V$ values from the rotational diagram analysis are listed in Table~\ref{tab:Av_measurements_using_ices_and_H2} for all three considered extinction laws.

We further investigated the spatial variation of physical parameters -- temperature ($T$), column density ($N$), and OPR for both the warm and hot components, along with $A_V$ -- by extracting spectra from rectangular apertures aligned with the wind/jet and disk axes. Along the disk axis, three apertures (AP1, AP2, AP3) were defined, each with a width of 0.67\arcsec (corresponding to the PSF FWHM at the H$_2$ $v=0$--0 S(1) transition) and a height of 1.9\arcsec, as illustrated in Figure~\ref{appfig:ap123}. Along the wind/jet axis, three apertures (AP4, AP5, AP6) were defined; AP4 and AP6 have a width of 1.9\arcsec and a height of 0.67\arcsec, while AP5 has a width of 1.3\arcsec and a height of 0.67\arcsec, as shown in Figure~\ref{appfig:ap456}. For all apertures, the rotational diagram analysis was performed using three different extinction laws: \citetalias{Pontoppidan2024ApJ...963..158P, HensleyDraine2023ApJ...948...55H, McClure2009ApJ...693L..81M}. The measured H$_2$ line fluxes and associated parameters required to construct the rotational diagrams for all six apertures are provided in Table~\ref{apptab:H2_line_fluxes_for_various_rect_apertures}. The resulting best-fit parameters from the H$_2$ rotational diagram analysis are compiled in Table~\ref{tab:all_6_apertures_best_fit_params_from_H2_rotational_diagrams}.

Our analysis shows that the visual extinction toward HV~Tau~C varies depending on the diagnostic method and extinction law. The ice absorption features give lower values ($A_V \sim 3$--5~mag), tracing cold, dense material in the disk midplane, while the H$_2$ rotational diagrams yield higher values ($A_V \sim 6$--19~mag), probing warmer gas in the disk surface or outflow regions. Even though both our study and \citet{Terada2007ApJ...667..303T} measure nearly the same 3.0~$\mu$m H$_2$O ice optical depth, the extinction laws assumed in the two cases are not the same. The H$_2$-based $A_V$ remains relatively constant across different apertures, suggesting a widespread dust screen affecting the warm gas, whereas the ices are likely confined to localized, shielded zones. The choice of extinction law introduces the largest uncertainty, with $A_V$ differing by up to a factor of three between laws (e.g., \citetalias{Pontoppidan2024ApJ...963..158P} vs. \citetalias{McClure2009ApJ...693L..81M}). Compared to earlier measurements, our ice-based $A_V$ values are lower than those from \citet{Terada2007ApJ...667..303T}, while the H$_2$-based results are closer to those from \citet{Beck2008ApJ...676..472B}. Overall, the differences highlight that ice and H$_2$ diagnostics probe distinct regions of the disk and are sensitive to both geometry and dust properties.

\begin{table*}[htbp]
\caption{Accretion properties under different extinction laws}
\label{tab:accretion_rate_table}
\centering
\footnotesize
\resizebox{\textwidth}{!}{
\begin{tabular}{l l l c c c c c}
\hline\hline
Extinction Law & H\,{\sc i} Line & Property &
$A_V = 3.77\pm0.29$ &
$A_V = 4.31\pm0.22$ &
$A_V = 12.8\pm2.5$ &
$A_V = 20.3\pm0.5$ &
$A_V = 36.55\pm0.1$ \\
\hline
\citetalias{Pontoppidan2024ApJ...963..158P} 
    & H\,{\sc i}\,6--5 & $L_{\rm acc}$ &
    $0.006\pm0.004$ & $0.006\pm0.004$ & $0.010\pm0.006$ & $0.016\pm0.007$ & $0.041\pm0.014$ \\
    &  & $\log \dot{M}_{\rm acc}$ &
    $-9.58^{+0.23}_{-0.32}$ &
    $-9.57^{+0.22}_{-0.30}$ &
    $-9.35^{+0.20}_{-0.28}$ &
    $-9.16^{+0.17}_{-0.25}$ &
    $-8.75^{+0.15}_{-0.21}$ \\
    & H\,{\sc i}\,7--6 & $L_{\rm acc}$ &
    $0.009\pm0.005$ & $0.009\pm0.005$ & $0.025\pm0.013$ & $0.059\pm0.021$ & $0.389\pm0.080$ \\
    &  & $\log \dot{M}_{\rm acc}$ &
    $-9.43^{+0.21}_{-0.30}$ &
    $-9.40^{+0.21}_{-0.29}$ &
    $-8.97^{+0.19}_{-0.27}$ &
    $-8.59^{+0.16}_{-0.23}$ &
    $-7.77^{+0.09}_{-0.11}$ \\
\hline
\citetalias{HensleyDraine2023ApJ...948...55H}
    & H\,{\sc i}\,6--5 & $L_{\rm acc}$ &
    $0.006\pm0.003$ & $0.006\pm0.003$ & $0.008\pm0.005$ & $0.011\pm0.006$ & $0.021\pm0.009$ \\
    &  & $\log \dot{M}_{\rm acc}$ &
    $-9.61^{+0.24}_{-0.33}$ &
    $-9.60^{+0.24}_{-0.33}$ &
    $-9.45^{+0.22}_{-0.31}$ &
    $-9.32^{+0.20}_{-0.28}$ &
    $-9.03^{+0.18}_{-0.25}$ \\
    & H\,{\sc i}\,7--6 & $L_{\rm acc}$ &
    $0.006\pm0.004$ & $0.006\pm0.004$ & $0.009\pm0.005$ & $0.012\pm0.006$ & $0.022\pm0.010$ \\
    &  & $\log \dot{M}_{\rm acc}$ &
    $-9.56^{+0.23}_{-0.32}$ &
    $-9.55^{+0.23}_{-0.32}$ &
    $-9.41^{+0.21}_{-0.30}$ &
    $-9.28^{+0.20}_{-0.28}$ &
    $-9.02^{+0.18}_{-0.25}$ \\
\hline
\citetalias{McClure2009ApJ...693L..81M}
    & H\,{\sc i}\,6--5 & $L_{\rm acc}$ &
    $0.006\pm0.004$ & $0.006\pm0.004$ & $0.010\pm0.005$ & $0.015\pm0.007$ & $0.036\pm0.013$ \\
    &  & $\log \dot{M}_{\rm acc}$ &
    $-9.59^{+0.23}_{-0.32}$ &
    $-9.58^{+0.23}_{-0.31}$ &
    $-9.37^{+0.21}_{-0.29}$ &
    $-9.20^{+0.18}_{-0.26}$ &
    $-8.81^{+0.15}_{-0.22}$ \\
    & H\,{\sc i}\,7--6 & $L_{\rm acc}$ &
    $0.007\pm0.004$ & $0.007\pm0.004$ & $0.012\pm0.007$ & $0.020\pm0.009$ & $0.054\pm0.020$ \\
    &  & $\log \dot{M}_{\rm acc}$ &
    $-9.52^{+0.22}_{-0.31}$ &
    $-9.50^{+0.22}_{-0.31}$ &
    $-9.27^{+0.20}_{-0.28}$ &
    $-9.07^{+0.18}_{-0.26}$ &
    $-8.63^{+0.14}_{-0.20}$ \\
\hline
\end{tabular}
}
\tablefoot{$L_{\rm acc}$ is in units of L$_\odot$. $\dot{M}_{\rm acc}$ is in units of M$_\odot$~yr$^{-1}$.}
\end{table*}

\section{Mid-IR H\textsc{i} Lines: Accretion Diagnostics} 
\label{sec:HI_accretion_measurement}

Hydrogen recombination (H\,\textsc{i}) lines in the mid-IR wavelength regime provide valuable diagnostics of gas accretion in YSOs \citep[e.g.,][]{Tofflemire2025ApJ...985..224T, Shridharan2026A&A...708A..22S}. In particular, transitions from the H\,{\sc i} Humphreys and Pfund series trace ionized gas associated with accretion-related processes. Owing to their longer wavelengths, these lines suffer significantly less extinction than commonly used optical and near-infrared tracers such as H$\alpha$ or Br$\gamma$, making them particularly well suited for probing accretion in evolved YSO systems.

\begin{figure}[htbp!]
\centering
\includegraphics[width=\columnwidth]{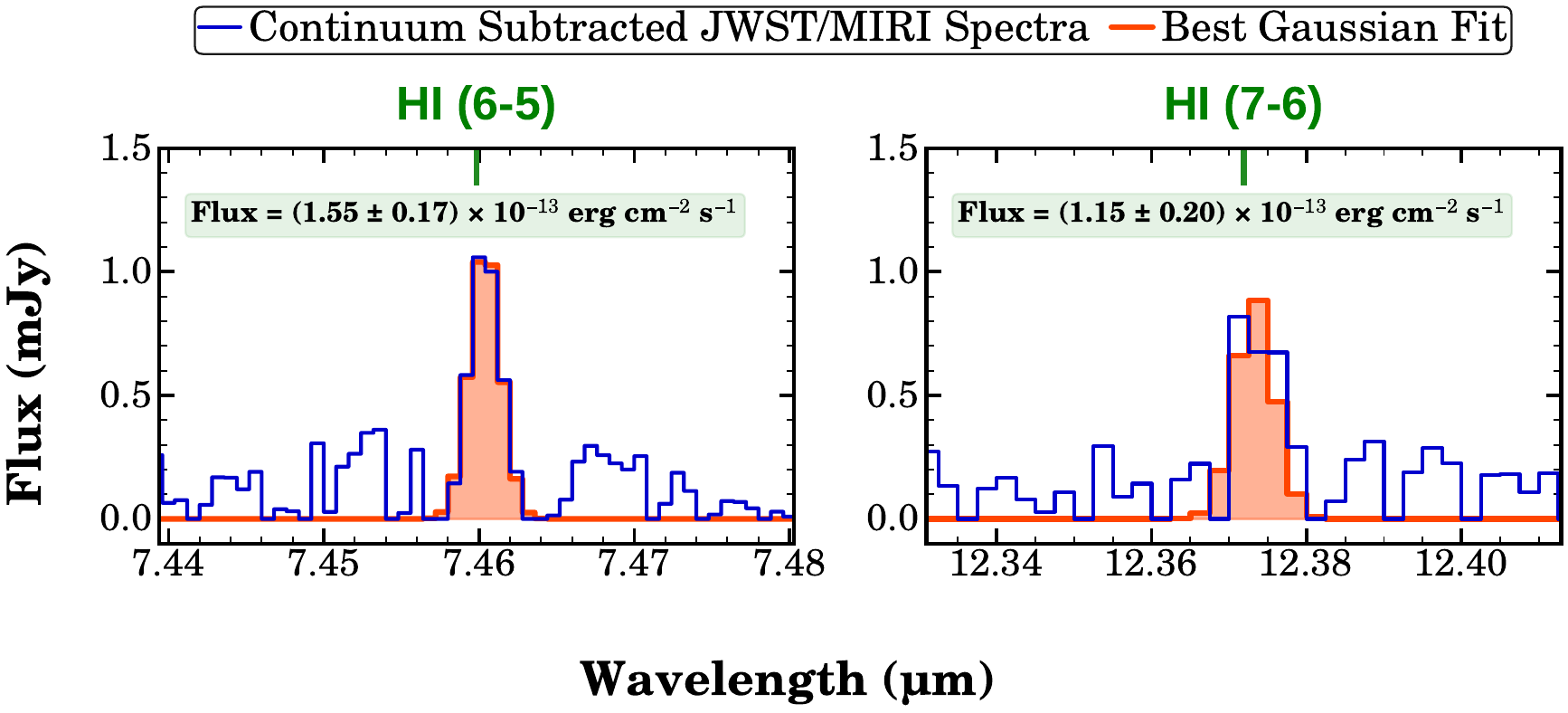}
\caption{Continuum-subtracted line profiles of the H\textsc{i} (6--5) and H\textsc{i} (7--6) transitions shown in blue. The red curves represent the best-fit Gaussian models, from which the integrated line fluxes were derived.}
\label{fig:HI_lines}
\end{figure}

In the MIRI-MRS spectrum of HV~Tau~C, we detect multiple H\,\textsc{i} recombination lines, including the Pfund~$\alpha$ (6--5) and Humphreys~$\alpha$ (7--6) transitions at 7.46~$\mu$m and 12.37~$\mu$m, respectively (see, Figure~\ref{fig:HI_lines}). These lines exhibit a higher S/N ratio, while a few higher-order transitions (i.e., 9--7, etc.) appear weaker and with a lower S/N ratio. All detected H\textsc{i} features are spatially unresolved, indicating that they originate close to the central star rather than from extended outflow regions, a trend that was noticed in \citet{Shridharan2026A&A...708A..22S} for a larger sample of Class~II sources.

We extracted the spectrum using an aperture radius equal to twice the PSF FWHM (as described in Section~\ref{sec:overall_1D_spectra}), appropriate for unresolved emission. A local continuum was subtracted around each line using \texttt{pybaseline}, and each line profile was fitted with a Gaussian function to measure its integrated flux $F_{\rm line}$ (following the procedure in Section~\ref{subsec:excitation_conditions_winds}). The observed line profiles and Gaussian fits are presented in Figure~\ref{fig:HI_lines}.

To investigate the effect of extinction, the measured $F_{\rm line}$ values were corrected using a set of five $A_V$ estimates derived in Section~\ref{sec:extinction_Av_text} (reported in Table~\ref{tab:Av_measurements_using_ices_and_H2}). These extinction values span the full plausible range for HV~Tau~C and probe different regions along the line of sight. At the low end, we adopted the value inferred from the 15~$\mu$m CO$_2$ ice feature ($A_V = 3.77 \pm 0.29$~mag), which traces cold outer material. A slightly larger value comes from the 6~$\mu$m H$_2$O ice band ($A_V = 4.31 \pm 0.22$~mag). To represent warmer gas, we included the extinction derived from the H$_2$ rotational diagram ($A_V = 12.8 \pm 2.5$~mag). In addition, two literature measurements sample more obscured conditions: near-IR H$_2$ line ratios give $A_V = 20.3 \pm 0.5$~mag, while earlier 3~$\mu$m H$_2$O ice measurements suggest even higher extinction ($A_V = 29.0$ and $44.1$~mag). Because these 3~$\mu$m values vary substantially and lack reported uncertainties, we adopt their mean ($A_V = 36.55$~mag) as an upper-limit scenario. Together, these $A_V$ values allow us to assess how the assumed extinction influences the inferred accretion rate.

The extinction-corrected line luminosities were calculated as
$L_{\rm line} = 4\pi d^{\,2} F_{\rm line}$,
adopting a distance of 138~pc for HV~Tau~C (Section~\ref{sec:introduction}). We did not subtract any H$_2$O contribution from the H\textsc{i} lines, as HV~Tau~C is an almost edge-on system and its JWST/MIRI-MRS spectrum shows very few molecular emission lines (e.g., CO, H$_2$O etc.). The accretion luminosity was then estimated using the latest empirical correlations between H\textsc{i} line luminosities and $L_{\rm acc}$ calibrated for T~Tauri stars \citep[see,][]{Shridharan2026A&A...708A..22S}. The relation takes the form

$$
\log\left(\frac{L_{\rm acc}}{L_{\odot}}\right) = a \times \left[\log\left(\frac{L_{\rm line}}{L_{\odot}}\right) - \mathrm{offset}\right] + b. 
$$

For the H\textsc{i} (6--5) Pfund~$\alpha$ transition, the best-fit parameters are $a = 1.08 \pm 0.14, ~ b = -0.70 \pm 0.08, ~ \text{offset} = -4.99 $. For the H\textsc{i} (7--6) Humphreys~$\alpha$ transition, the parameters are $a = 1.07 \pm 0.14, ~  b = -0.64 \pm 0.08, ~ \text{offset} = -5.32$.

Finally, the mass-accretion rate was computed using the standard relation
$$
\dot{M}_{\rm acc} = \frac{L_{\rm acc} R_\star}{G M_\star}
\left(1 - \frac{R_\star}{R_{\rm in}}\right)^{-1}, 
$$
following, e.g., \citet{Gullbring1998ApJ...492..323G, Muzerolle2003ApJ...592..266M, Franceschi2024A&A...687A..96F}. The stellar mass is taken to be $M_\star = 1.43\,M_\odot$, as derived in Section~\ref{sec:hv_tau_c_alma_characteristics}. Although HV~Tau~C is often classified as a K6-type source in the literature, its dynamical mass indicates that it is more consistent with an intermediate-mass pre-main-sequence star, likely of earlier spectral type \citep[e.g., F-type;][]{PecautMamajek2013ApJS..208....9P}.

Because no direct measurement of the stellar radius is available for HV~Tau~C, we adopt a representative value of $R_\star = 1.58\,R_\odot$ from the tabulated pre-main-sequence parameters of \citet{PecautMamajek2013ApJS..208....9P}, corresponding to a star of mass $M_\star \simeq 1.44\,M_\odot$. We emphasize that this choice is approximate and introduces an additional source of uncertainty in the derived accretion rate. We assume an inner truncation radius of $R_{\rm in} = 5~R_\star$, a typical value for magnetospheric accretion where the stellar magnetic field truncates the disk at several stellar radii \citep[see,][]{Rigliaco2015ApJ...801...31R}. This choice is commonly adopted in accretion studies of T~Tauri and intermediate-mass pre-main-sequence stars, and reflects the expected location at which disk material is channeled along magnetic field lines onto the stellar surface.

The derived accretion luminosities and accretion rates for various values of $A_V$ -- calculated using all three extinction laws -- are reported in Table~\ref{tab:accretion_rate_table}. These values are consistent with the low but steady accretion characteristic of evolved, edge-on Class~II disks. The results clearly show a dependence on the adopted extinction law, with derived accretion rates increasing for higher values of $A_V$. From Table~\ref{tab:accretion_rate_table}, we adopt an accretion-rate range of $\dot{M}_{\rm acc} \approx 2.4 \times 10^{-10}$--$1.7 \times 10^{-8}\,M_\odot\,{\rm yr^{-1}}$.

We further note that the inferred accretion rate depends on the adopted stellar mass approximately as $\dot{M}_{\rm acc} \propto M_\star^{-1}$, assuming a fixed stellar radius. Across the full ridge-to-edge stellar mass range derived in Section~\ref{subsec:dynamical_mass_using_keplerian_rotation} ($\sim$1.0--1.85~$M_\odot$), the resulting variation in $\dot{M}_{\rm acc}$ is at the level of $\sim$20--40\%. Accordingly, the adopted accretion-rate range would vary to approximately $(1.8$--$3.4)\times10^{-10}$ and $(1.3$--$2.4)\times10^{-8}\,M_\odot\,{\rm yr^{-1}}$, respectively, across the allowed stellar-mass range. This uncertainty remains smaller than the additional systematic uncertainties associated with extinction corrections, the assumed stellar radius, and the nearly edge-on geometry of the system. Therefore, the uncertainty in dynamical mass ($M_\star$) does not qualitatively affect our conclusions regarding the accretion properties of HV~Tau~C.

The relatively faint mid-IR H\,\textsc{i} lines, in contrast to the strong H$_2$ emission, suggest that the accretion rate in HV~Tau~C may be intrinsically low, episodic, or underestimated. However, several factors introduce significant systematic uncertainties in the accretion-rate estimates. First, HV~Tau~C is observed at a nearly edge-on inclination, implying that the central accretion region is likely heavily obscured. In such a geometry, the observed H\,\textsc{i} line emission may be dominated by scattered or reprocessed radiation rather than direct emission. Since empirical line--luminosity calibrations typically assume isotropic emission, a significant contribution from scattered light would imply that the effective emitting solid angle is smaller than $4\pi$, leading to an underestimate of the intrinsic accretion luminosity and hence $\dot{M}_{\rm acc}$. Second, the extinction toward the source remains poorly constrained, with estimates spanning a wide range depending on the adopted diagnostic (Section~\ref{sec:HI_accretion_measurement}). Although we explore this range explicitly, the corresponding spread in $A_V$ propagates into substantial uncertainty in the derived accretion luminosities and mass-accretion rates.

Additionally, a contribution from wind or jet emission to the observed mid-IR H\,\textsc{i} lines in Class~II sources is likely minimal, since these lines do not exhibit extended spatial morphology \citep{Shridharan2026A&A...708A..22S}. Taken together, these effects imply that the derived $\dot{M}_{\rm acc}$ values should be treated with caution and are likely lower limits rather than precise measurements.

\section{Discussion}\label{sec:discussion}

The combined JWST and ALMA observations of HV~Tau~C reveal clear evidence for a powerful, wide-angle molecular H$_2$ wind persisting into the Class~II stage of disk evolution. In this section, we present the observational arguments supporting a wind origin for the extended H$_2$ emission, examine its excitation and possible driving mechanisms, assess its impact on disk evolution, and discuss the implications for disk dispersal.

\subsection{\texorpdfstring{Evidence for a Molecular H$_2$ Wind in a Class II Disk}{Evidence for a Molecular H2 Wind in a Class II Disk}}

For the first time, enabled by the higher sensitivity and higher spatial resolution of JWST/MIRI-MRS, we detect pure-rotational H$_2$ lines in the mid-IR wavelength region as extended emission in HV~Tau~C. Previously, extended ro-vibrational H$_2$ emission from the $v$=1--0 S(1) at 2.12\,$\mu$m transition was detected in a few Class~II sources, including HV~Tau~C \citep[see, e.g.,][]{Beck2008ApJ...676..472B}. The detection of spatially extended pure-rotational H$_2$ emission naturally raises the question of its physical origin, namely whether it arises from the disk surface, the disk surroundings, or a disk-driven wind. To address this question, we examine the morphological, kinematical and dynamical properties of the H$_2$ emission. In this section, we emphasize four pieces of evidence demonstrating that the extended H$_2$ emission observed in HV Tau C arises in the outflowing gas driven by MHD disk winds.

First, the spatial structure of the H$_2$ emission does not resemble that expected from a compact disk surface. The spatial distribution of the H$_2$ emission -- encompassing all detected rotational transitions from S(1) to S(8) in the ground-vibrational state $v$=0--0, is significantly much more extended than the compact dust disk traced by ALMA at 887~$\mu$m, the cold $^{12}$CO~($J$=3--2) gas disk, and the scattered-light morphology that shows the extent of the upper disk layers at 1.15~$\mu$m with JWST/NIRCam, as discussed in Section~\ref{subsec:extended_H2_morphology}. As the excitation increases from the lower S(1) to the higher S(8) transitions, the H$_2$ emission becomes progressively narrower, forming a nested morphology evident in the continuum-subtracted linemaps (see Figure~\ref{fig:hv_tau_c_H2_v00_extended_linemaps}). However, this trend is not reflected in the derived semi-opening angles, which do not exhibit a systematic decrease with increasing transition level (Section~\ref{subsec:winds_semi_opening_angle}). This discrepancy likely arises from the limited spatial resolution, the lower pixel-by-pixel S/N ratio, and the resulting uncertainties in the semi-opening angle measurements. Taken together, this demonstrates that the H$_2$ emission extends far beyond the disk surface, indicating that the molecular gas detected is at elevated regions above the disk plane, consistent with a wide-angle molecular wind rather than compact disk emission. In addition, we examined the spatial distribution of other JWST/MIRI-MRS -- accessible UV tracers, including PAH emission, mid-IR OH lines, and the fine-structure lines [Ne\,{\sc ii}], [Ne\,{\sc iii}], and [S\,{\sc iii}], and compared them with the collimated jet traced by [Fe\,{\sc ii}] at 5.34~$\mu$m (see, Figure~\ref{fig:HV_Tau_C_comparison_of_H2_morphology_with_UV_tracers}). The H$_2$ emission traces a wider outflow component, particularly in the higher transitions, while the lower transition may also be affected by beam smearing as it extends both radially along the disk surface and vertically. The fine-structure lines lie along the jet but are broader than the [Fe\,{\sc ii}] jet, whereas the PAH ($\sim$11.3~$\mu$m) and OH emission ($\sim$10~$\mu$m) remain unresolved within the central $2\,\mathrm{PSF}$ FWHM region. Overall, this indicates that the extended H$_2$ emission shows a morphology distinct from the other line tracers and is likely associated with a wide-angle wind launched from the disk.

Second, the excitation analysis indicates gas temperatures significantly higher than those expected in a typical Class~II disk, where gas temperatures inferred from LTE line modeling are generally lower \citep[typically, a few 100~K, e.g.,][]{Grant2024A&A...689A..85G, Henning2024PASP..136e4302H, Arulanantham2025AJ....170...67A}. Fitting the rotational diagram (see Section~\ref{subsec:excitation_conditions_winds}) requires at least two temperature components: a warm component at $\sim$600~K and a hot component at $\sim$2000~K. The power-law fit to the rotational diagram further supports the presence of a broad temperature distribution (Fig.~\ref{fig:hv_tau_c_H2_rotational_diagrams_comparison}b). At distances of hundreds of au from the central star, such temperatures cannot be sustained by standard disk heating mechanisms alone. While addressing these temperatures, it is important to distinguish between the heating of the gas and the excitation of the H$_2$ molecules. UV heating likely contributes to the elevated gas temperatures, particularly closer to the star and the disk surface. However, at larger distances and higher disk heights, these temperatures are more naturally explained by internal shock heating within the outflow.

\begin{figure}[htbp!]
\centering
\includegraphics[width=\columnwidth]{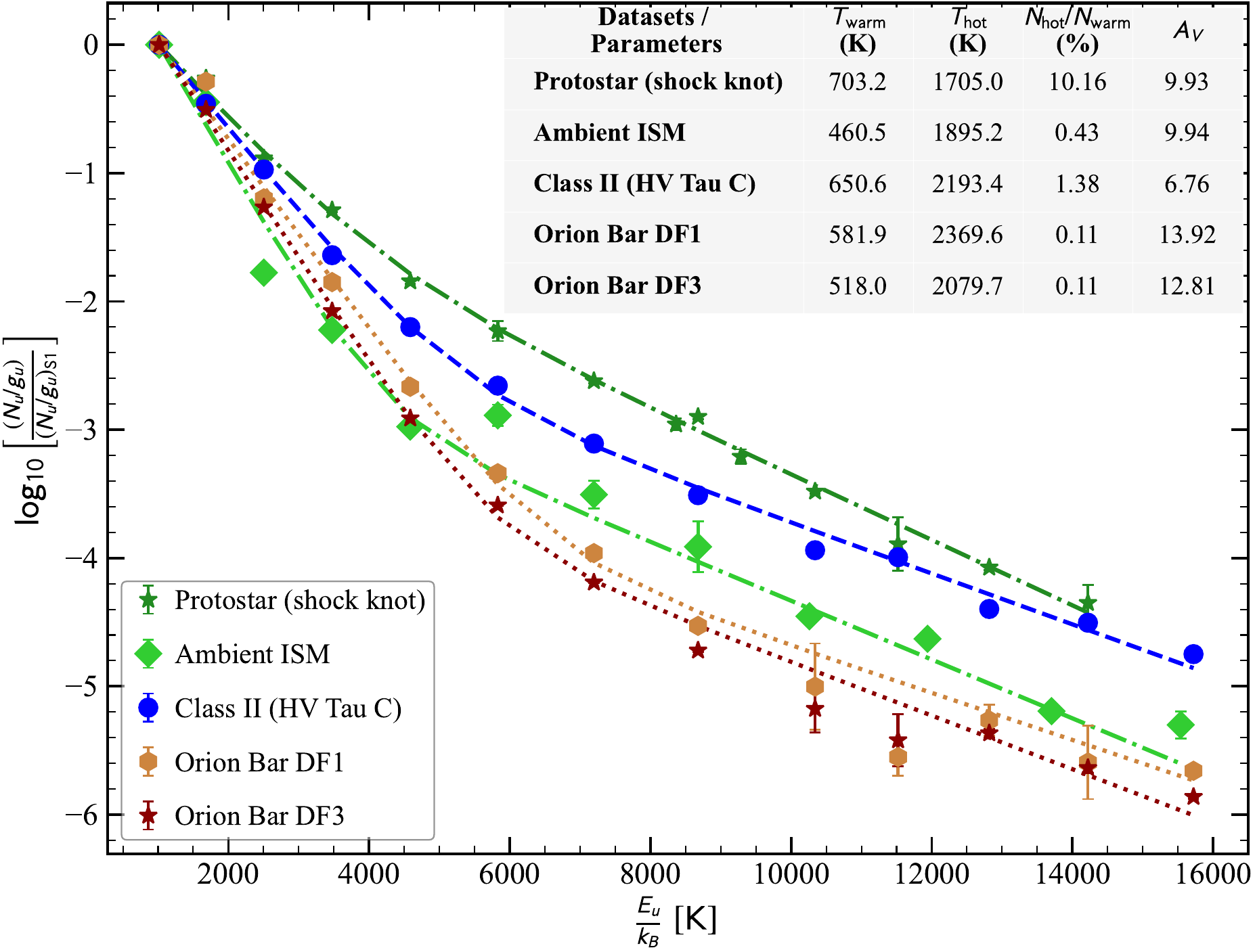}
\caption{Comparison of normalized two-component rotational diagrams for H$_2$ transitions in HV~Tau~C with those from different environments: (i) shock knot in the Class~0 protostar HOPS~370, (ii) ambient molecular gas surrounding the protostar IRAS~16253, and (iii) two PDR regions DF1 and DF3, in the Orion Bar (data taken from \citealt{H2OrionPDR2024A&A...687A..86V}). The normalized diagrams facilitate comparison of excitation conditions across different environments.}
\label{fig:Comparison_of_Rotational_diagrams_for_different_dataset}
\end{figure}

Third, while the gas may be heated by a combination of UV photons and shocks, the excitation mechanism of the extended H$_2$ emission is primarily collisional, rather than driven by radiative UV pumping (fluorescence). To demonstrate this, we compare the normalized H$_2$ rotational diagram of HV~Tau~C with those obtained for three representative environments: (i) Class~0 protostars (shock knot in HOPS~370), where H$_2$ emission is predominantly shock-excited in molecular outflows (Tyagi et al., in prep.); (ii) the ambient molecular gas surrounding protostars far beyond the envelope, where excitation is likely dominated by UV irradiation (Manoj et al., in prep.); and (iii) the Orion Bar photodissociation region (PDR), a benchmark UV-irradiated environment, including the Dissociation Fronts DF1 and DF3 positions from \citet{H2OrionPDR2024A&A...687A..86V}. To normalize the rotational diagram, all data points were divided by the value corresponding to the H$_2$ $v=0$--0 S(1) transition (see Figure~\ref{fig:Comparison_of_Rotational_diagrams_for_different_dataset}).

As shown in Figure~\ref{fig:Comparison_of_Rotational_diagrams_for_different_dataset}, the rotational diagram of HV~Tau~C closely follows the trend observed in protostellar shock regions and clearly deviates from the characteristic curvature seen in UV-dominated environments such as the Orion Bar and the ambient gas around protostars. In particular, the location of the ``knee'' (i.e., the slope change in the rotational diagram) occurs at relatively low excitation energies, corresponding to the S(3)--S(5) transitions ($E_{\rm up} \approx 4000$--6000~K), both in protostellar shock knots and in HV~Tau~C. In contrast, in UV-excited environments such as PDRs, this break appears at significantly higher excitation energies, typically between the S(6) and S(8) transitions ($E_{\rm up} \approx 6000$--9000~K).

Moreover, even when restricting the analysis to pure rotational transitions within the ground vibrational state ($v=0$--0; S(1)--S(8)), at least a two-temperature component fit is required to reproduce the H$_2$ excitation in YSOs, including protostellar shock knots and HV~Tau~C \citep[e.g.,][Tyagi et al., in prep.]{Narang2024ApJ...962L..16N, Schwarz2025ApJ...980..148S}. In contrast, PDR regions are well described by a single-temperature component for the $v=0$--0 lines, with the $v=1$--1 transitions forming a separate, hotter component \citep{H2OrionPDR2024A&A...687A..86V}. This difference likely distinguishes the rotational diagrams of collisionally-dominated environments from those of radiatively UV-pumped regions. We note that to definitively quantify the exact radiative contribution from UV pumping, observations of higher vibrational transitions (e.g., H$_2$ $v=2$--1, $v=3$--2 etc.) are necessary \citep[e.g.,][]{Black_vanDishoeck1987ApJ...322..412B, Assani2026arXiv260709407A}.

Despite this, the relative population of the hot H$_2$ component provides additional evidence for collisional excitation in the current dataset. In UV-dominated regions, such as the Orion Bar PDR regions (DF1 and DF3) and the ambient molecular gas surrounding protostars, the hot-to-warm column density ratio typically lies in the range $N_{\mathrm{hot}}/N_{\mathrm{warm}} \sim 0.1$--$0.2\%$ (see Figure~\ref{fig:Comparison_of_Rotational_diagrams_for_different_dataset}). In contrast, HV~Tau~C exhibits a significantly elevated ratio of $\sim 1$--$3\%$. This value overlaps with the lower end of the $N_{\mathrm{hot}}/N_{\mathrm{warm}} \sim 1$--$10\%$ range observed in strong protostellar shock knots, which suggests that the enhanced hot gas fraction in HV~Tau~C is difficult to explain by UV pumping alone and likely requires a significant contribution from collisional excitation driven by shock heating.

Furthermore, given the large spatial extent of the H$_2$ winds and the associated high column densities, the H$_2$ gas is expected to be highly self-shielded from UV. In such conditions, radiative UV pumping becomes inefficient at driving the extended emission, pointing strongly toward collisionally excited H$_2$ -- regardless of whether the underlying thermal energy comes from UV heating or shocks.

Fourth, the kinematics (as discussed in Section~\ref{subsec:winds_kinematics}) indicate that both the blue- and redshifted components trace material moving away with respect to the central source. PV cuts extracted along the wind axis reveal a strong velocity gradient from the lower to higher H$_2$ rotational transitions. The lower-$J$ lines, such as S(1) and S(2), show relatively small velocities of $\sim$8~km~s$^{-1}$ (inclination-corrected), tracing the outer regions of the flow. In contrast, the higher-$J$ transitions S(7) and S(8) exhibit substantially larger velocities of $\sim$40~km~s$^{-1}$ (inclination-corrected), corresponding to gas originating from deeper within the wind-launching region. This systematic increase in velocity with excitation is a natural signature of a nested wind morphology \citep[e.g.,][Tyagi et al. in prep.]{Delabrosse2024A&A...688A.173D, Narang2026ApJ..1000..184N}. The outward motion of material at such high velocities cannot be explained by gas merely residing on the disk surface or by a UV-irradiated photoevaporative layer alone. Instead, this behaviour is more naturally consistent with a stratified wind structure, such as an MHD disk wind \citep[e.g.,][]{Panoglou2012A&A...538A...2P, Bai2016ApJ...818..152B, Lesur2023ASPC..534..465L, Pascucci2023ASPC..534..567P}. In these theoretical models, the outflow is organized into nested streamlines: the higher-excitation lines preferentially trace inner, faster streamlines that are subjected to stronger irradiation and heating, whereas the lower-excitation lines trace the slower, cooler gas launched from larger disk radii. Even though the edge-on geometry blends the redshifted and blueshifted emission, the observed velocity gradient remains a clear signature of this structured outward flow.

Taken together, these results, combined with the morphological and kinematic evidence presented above, indicate that the extended H$_2$ emission in HV~Tau~C traces a wide-angle molecular wind launched from the disk surface. The excitation pattern is most consistent with shock-dominated heating, while a contribution from UV irradiation may also be present.

\subsection{Wind Strength and Persistence into the Class~II Stage}

\begin{figure}[htbp!]
\centering
\includegraphics[width=\columnwidth]{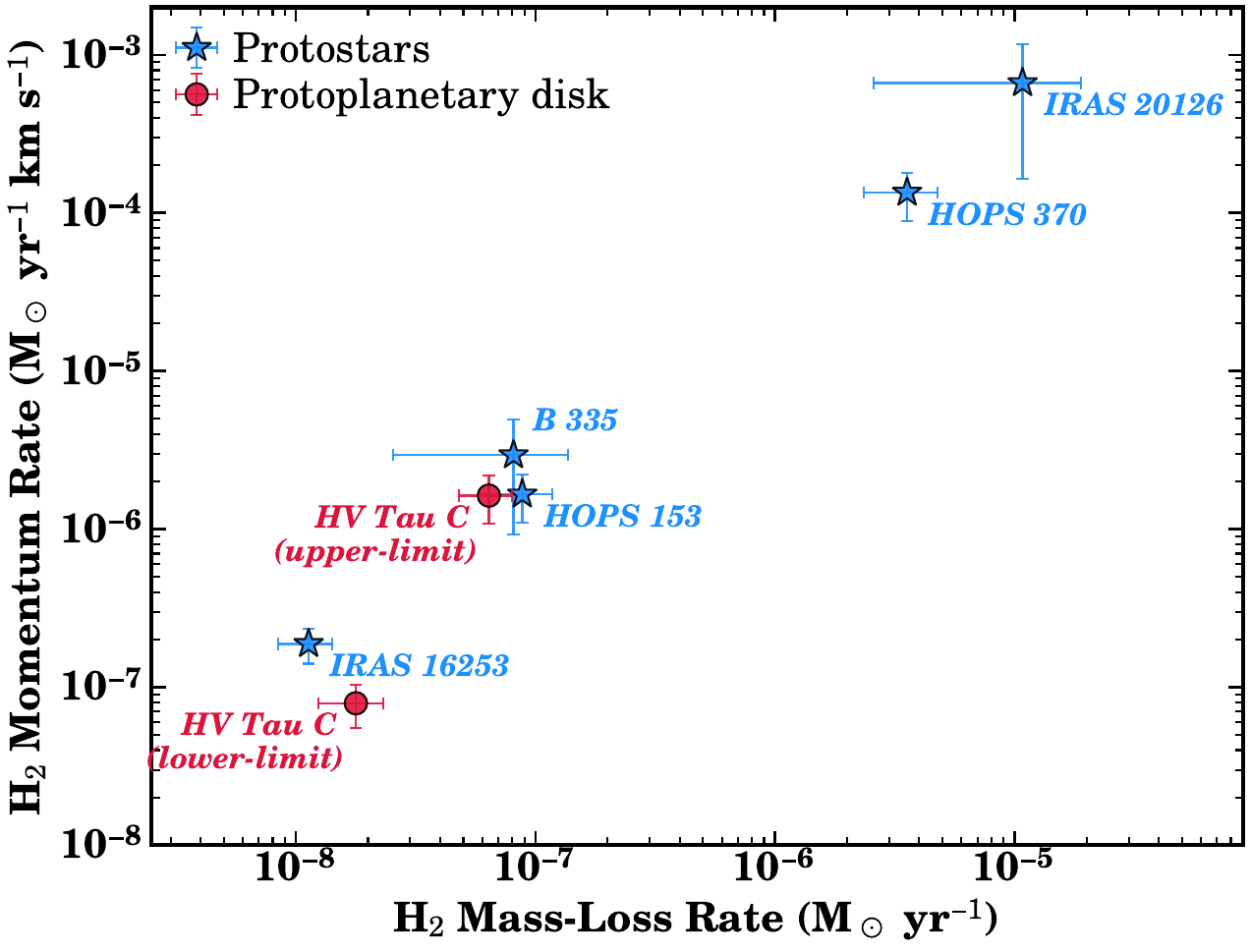}
\caption{Comparison of the dynamical quantities ($\dot{M}_{\rm loss}$ and $F_{\rm H_2}$) of HV~Tau~C (a Class~II source) with those of Class~0 protostars from the \textit{JWST} IPA program (Tyagi et al., in prep.). The plot compares the H$_2$ wind mass-loss rates and momentum rate, highlighting the relative strength of the molecular wind from HV~Tau~C in the context of earlier evolutionary stages. The upper and lower limits on the dynamical quantities of HV~Tau~C are described in Section~\ref{subsec:h2_wind_dynamics}.}
\label{fig:Comparison_of_dynamical_quantites_for_ClassO_and_ClassII_sources}
\end{figure}

A key result of this study is the unexpectedly strong molecular H$_2$ wind observed in HV~Tau~C. While the absolute dynamical quantities we derive -- such as the mass-loss rate and momentum rate of the H$_2$ flow -- depend on the evolutionary stage and the mass of the central object, we find that the relationship between mass loss and momentum is broadly consistent with that observed in Class~0 protostars (Tyagi et al., in prep.; see Figure~\ref{fig:Comparison_of_dynamical_quantites_for_ClassO_and_ClassII_sources}). This indicates that, although the overall wind properties evolve significantly from the embedded to the Class~II stage, the underlying coupling between mass-loss and momentum injection remains efficient. Our results therefore suggest that strong, wide-angled molecular winds can persist into the Class~II phase, where planet formation is actively underway \citep[see,][]{Narang2026ApJ..1004..188N}.

We note that the derived dynamical quantities are subject to systematic uncertainties stemming from the velocity structure of the flow. As detailed in Section~\ref{subsec:h2_wind_dynamics}, the assumption of a constant average wind speed ($\langle \mathrm{v} \rangle$) in an accelerating flow leads to an underestimated dynamical timescale, meaning the inferred $\dot{M}_{\mathrm{loss}}$ and $F_{\mathrm{H_2}}$ values are strictly upper limits. Despite these uncertainties, the relative placement of HV~Tau~C in the $\dot{M}_{\mathrm{loss}}$--$F_{\mathrm{H_2}}$ plane remains broadly consistent with the trend defined by Class~0 sources (see Figure~\ref{fig:Comparison_of_dynamical_quantites_for_ClassO_and_ClassII_sources}).

The dynamical mass of the HV~Tau~C protoplanetary disk is 1.43~$M_\odot$, placing it in the intermediate-mass regime and comparable to the Class~0 protostar HOPS~370, which has a similar stellar mass. However, the wind dynamical properties derived for HV~Tau~C are more similar to those measured in the lower-mass protostars B~335 and HOPS~153. This suggests that wind properties may not scale solely with stellar mass and may instead evolve with evolutionary stage. In particular, while molecular winds can persist into the Class~II phase, their dynamical impact may be reduced compared to the Class~0 stage for a given stellar mass source.

We emphasize, however, that currently there are only a few Class~II sources for which extended and strong H$_2$ winds have been detected and characterized (e.g., \citealt{Arulanantham2024ApJ...965L..13A, Schwarz2025ApJ...980..148S, Pascucci2025NatAs...9...81P, Bajaj2025AJ....169..296B, Narang2026ApJ..1004..188N, Somigliana2026arXiv260623794S}). This limited sample restricts our ability to draw broader conclusions about the evolution of molecular winds in protoplanetary disks. In addition, substructures such as ring-like features observed in some disks \citep[e.g., in FZ~Tau,][]{Pontoppidan2024ApJ...963..158P, Narang2026ApJ..1004..188N} may indicate episodic accretion, which could further complicate the connection between accretion and outflow activity.

\subsection{Mass ejection versus mass accretion}
\label{subsec:mass_ejection_vs_accretion}

The mass-loss rate in the molecular H$_2$ winds derived for HV~Tau~C lies in the range $\dot{M}_{\rm wind} \approx 2 \times 10^{-8}$ -- $5 \times 10^{-8}\,M_\odot\,{\rm yr^{-1}}$ (depending on the adopted dynamical timescale; see Figure~\ref{fig:Comparison_of_dynamical_quantites_for_ClassO_and_ClassII_sources}). As discussed in Section~\ref{sec:HI_accretion_measurement}, the accretion rate inferred from the mid-IR H\,{\sc i} diagnostics is highly uncertain owing to its strong dependence on the adopted $A_V$, the choice of extinction law, the nearly edge-on geometry of the source, and the uncertainty in the stellar mass. For HV~Tau~C, we estimate an accretion rate in the range $\dot{M}_{\rm acc} \approx 1.8 \times 10^{-10}$ -- $2.4 \times 10^{-8}\,M_\odot\,{\rm yr^{-1}}$, where the quoted range reflects plausible values of $A_V$ and different extinction laws appropriate for this nearly edge-on system. Using these values, the inferred wind-to-accretion ratio spans $\dot{M}_{\rm wind}/\dot{M}_{\rm acc} \approx 0.8$--280, with the lower value corresponding to the lowest assumed mass-loss rate and the highest assumed accretion rate, and the upper value corresponding to the highest assumed mass-loss rate and the lowest assumed accretion rate. Even under the most conservative assumption, the mass-loss rate is therefore comparable to the accretion rate, and may substantially exceed it. Such elevated wind-to-accretion ratios have also been reported in other evolved Class~II systems exhibiting molecular H$_2$ winds, including SY~Cha \citep{Schwarz2025ApJ...980..148S}, Tau042021 \citep{Arulanantham2024ApJ...965L..13A} and the JDISCS Class~II extended H$_2$ survey paper by \citet{Narang2026ApJ..1004..188N}.

The upper end of the inferred $\dot{M}_{\rm wind}/\dot{M}_{\rm acc}$ ratio is difficult to reconcile with steady, self-similar magneto-centrifugal MHD wind models, which typically predict $\dot{M}_{\rm wind} \sim 0.01$--$0.3\,\dot{M}_{\rm acc}$ depending on the wind-launching radius and magnetic-field geometry \citep[e.g.,][]{Pudritz2007prpl.conf..277P,Bai2016ApJ...818..152B}. However, the lower end of the observed range ($\sim$0.8--1) is more compatible with extended disk-wind models in which mass loading occurs over a broad radial extent. Several factors may therefore contribute to the apparently elevated wind-to-accretion ratio, and these possibilities are not mutually exclusive.

\begin{itemize}

\item \textit{Extended, multi-radius wind launching.}  
If mass loss occurs over a broad range of disk radii, including the outer disk, the integrated wind mass-loss rate can become substantial even if the local mass-loss rate per unit area remains modest \citep[e.g.,][]{Bai2016ApJ...818..152B, Lesur2021A&A...650A..35L, Pascucci2023ASPC..534..567P, Narang2026ApJ..1004..188N}. In standard collimated jet models, where the outflow is launched primarily from the innermost disk regions, the expected wind-to-accretion ratio is typically of order $\sim$0.1. By contrast, in extended disk-wind models with mass loading over a broader radial extent, the integrated $\dot{M}_{\rm wind}/\dot{M}_{\rm acc}$ ratio can become significantly larger and may approach values of $\sim$0.1--1, depending on the magnetic lever-arm parameter and the radial extent of the launching region (see Table~1 of \citealt{Pascucci2023ASPC..534..567P}). The ALMA-derived stellar mass ($\sim 1.43\,M_\odot$) and inferred disk properties are consistent with a disk capable of sustaining such extended launching.
\\
\item \textit{Non-steady behaviour (recent high-accretion episode).}  
The molecular H$_2$ flow exhibits a short dynamical timescale of a few $\times$ 10--100~yr, suggesting that the observed emission may trace the remnant of a recent episode of enhanced accretion and mass ejection. In such a scenario, the currently measured accretion rate may underestimate the accretion rate that originally powered the outflow. Moreover, the inferred mass-loss rate represents a quantity averaged over the dynamical timescale of the flow, whereas the accretion rate derived from the mid-IR H\,{\sc i} lines reflects the instantaneous accretion activity at the present epoch. Temporal variability in the accretion--ejection process may contribute to the observed discrepancy between the measured accretion and mass-loss rates.
\\
\item \textit{Uncertainties in extinction and accretion diagnostics.}  
The nearly edge-on geometry strongly attenuates UV and optical accretion tracers, potentially leading to a systematic underestimation of $\dot{M}_{\rm acc}$. The inferred accretion rate of HV~Tau~C is nearly two orders of magnitude lower than typical values for K6 pre-main-sequence stars in the Taurus star-forming region. If one instead adopts a median accretion rate of $\sim 10^{-8}\,M_\odot\,{\rm yr^{-1}}$, the implied extinction would exceed $A_V \sim 70$~mag. Moreover, for several of the adopted extinction laws, the derived $\dot{M}_{\rm acc}$ from the H\,{\sc i} 7--6 line is comparable to or larger than that obtained from the H\,{\sc i} 6--5 line when using the same $A_V$. Under a simple scattering-dominated interpretation (with a physically consistent extinction correction), one would instead expect $\dot{M}_{\rm acc}$(6--5) $\gtrsim$ $\dot{M}_{\rm acc}$(7--6), since scattering efficiency increases toward shorter wavelengths. The fact that this ordering is not always observed suggests that the adopted $A_V$ values may be underestimated, or that additional radiative-transfer effects are at play. This further highlights that the derivation of $\dot{M}_{\rm acc}$ carries significant uncertainties and that the quoted values should be regarded as tentative.
\\
\item \textit{H\,{\sc i} emission observed in scattered light.}  
As discussed earlier, the nearly edge-on disk geometry prevents a direct line of sight to the accretion hotspots on the stellar surface. The observed H\,{\sc i} recombination lines may therefore be dominated by scattered or reprocessed radiation rather than direct emission. In such a configuration, the detected H\,{\sc i} luminosity may represent only a fraction of the intrinsic accretion luminosity, corresponding to the portion of the radiation redirected into our line of sight by scattering off the disk surface. Consequently, the true accretion rate could be significantly higher than inferred, thereby reducing the apparent discrepancy between $\dot{M}_{\rm wind}$ and $\dot{M}_{\rm acc}$. The relatively faint mid-IR H\,\textsc{i} lines, compared to the strong extended H$_2$ emission, initially suggest that the accretion rate in HV~Tau~C may be intrinsically low or episodic. However, as established in Section~\ref{sec:extinction_Av_text}, the high degree of geometric obscuration and unconstrained extinction mean our derived $\dot{M}_{\rm acc}$ values must be treated as strict lower limits.

\end{itemize}

In summary, HV~Tau~C exhibits a large wind-to-accretion ratio, spanning $\dot{M}_{\rm wind}/\dot{M}_{\rm acc} \approx 0.8$--280 depending on the adopted accretion and mass-loss rates. The upper end of this range is difficult to reconcile with standard steady MHD wind models, whereas the lower end is more compatible with extended multi-radius disk-wind scenarios. Future observations of larger samples with similarly detailed H$_2$ diagnostics will be essential to determine whether such elevated wind-to-accretion ratios are common among evolved Class~II disks.

\section{Summary and Conclusions} \label{sec:conclusions}

We have carried out a multi-wavelength investigation of the nearly edge-on Class~II source HV~Tau~C, combining the unprecedented sensitivity of JWST/MIRI-MRS spectra, NIRCam imaging with the high angular resolution of ALMA observations. These data provide the first detailed characterization of an extended wide-angle molecular outflow traced by pure-rotational H$_2$ emission in the ground vibrational state from a Class~II protoplanetary disk, HV~Tau~C. Our main conclusions are summarized as follows:

\begin{enumerate}

    \item ALMA $^{12}$CO\,($J$=3--2) and 887~$\mu$m continuum observations reveal a compact dust disk embedded within a significantly more extended gaseous disk that exhibits a clear Keplerian velocity pattern. The gas disk extends to nearly three times the radius of the dust disk. A position--velocity analysis along the disk major axis yields a stellar mass of $1.43\,M_\odot$, identifying HV~Tau~C as an intermediate-mass T~Tauri star.
    
    \item The \textit{JWST}/MIRI-MRS spectrum shows numerous pure-rotational H$_2$ lines in the ground vibrational state, whose emission is spatially resolved and extends well beyond both the compact dust and CO gas disks, as well as the near-IR scattered light. The H$_2$ morphology is nested, with higher-excitation lines originating from inner regions of the outflow, indicating a stratified temperature and velocity structure.
    
    \item Several independent diagnostics indicate that the extended H$_2$ emission is consistent with a molecular disk wind rather than solely warm gas residing on the disk surface. The emission is significantly more extended than the compact disk components, its excitation requires at least two temperature components ($\sim$600~K and $\sim$2000~K), and the excitation pattern is consistent with a significant contribution from shock heating, although additional heating from UV irradiation or disk-wind processes cannot be excluded. PV diagram analysis along the flow axis reveals a systematic velocity gradient indicative of outward motion and inconsistent with purely Keplerian rotation.
    
    \item The inferred molecular wind mass-loss rate is of order $10^{-8},M_{\odot},\mathrm{yr^{-1}}$, comparable to values measured in some low-mass Class~0 protostars. This suggests that substantial molecular outflows may persist into the Class~II phase. The inferred mass-loss rate is comparable to or larger than the accretion rate derived from the mid-IR H,\textsc{i} recombination lines, although both estimates remain subject to significant systematic uncertainties, particularly the accretion rate. The mass-loss rate depends on the adopted dynamical timescale, while the accretion rate is highly uncertain owing to the nearly edge-on geometry resulting in scattering effects, and poorly constrained extinction.
    
    \item The observed properties are broadly consistent with an MHD-driven disk wind, potentially accompanied by shock heating and/or additional radiative heating processes within the flow. The presence of extended molecular H$_2$ emission in HV~Tau~C supports models in which disk winds contribute to angular momentum extraction and mass loss during the Class~II phase. Such winds may influence disk evolution, dispersal, and the physical conditions relevant for planet formation over a broad range of disk radii.
    
\end{enumerate}

This study highlights the unprecedented capability of \textit{JWST}/MIRI-MRS to trace extended H$_2$ emission in evolved protoplanetary disks. Systematic JWST surveys of Class~II sources spanning a range of inclinations, stellar masses, and accretion rates will be essential to establish how common such molecular winds are and to better quantify their role in disk evolution. HV~Tau~C provides evidence that extended molecular H$_2$ outflows can remain active into the Class~II stage, offering new insights into the dynamics and evolution of protoplanetary disks in the JWST era.

\begin{acknowledgements}
We thank the anonymous referee for the constructive comments and suggestions, which helped improve the manuscript. This work is based on observations mainly made with the NASA/ESA/CSA \textit{James Webb Space Telescope} (JWST). The data were obtained from the Mikulski Archive for Space Telescopes (MAST) at the Space Telescope Science Institute (STScI), which is operated by the Association of Universities for Research in Astronomy, Inc., under NASA contract NAS 5-03127. These observations are associated with program \#1282. V.P. and P.M. acknowledge support from the Department of Atomic Energy, Government of India, under Project Identification No.\ RTI 4002. B.S. acknowledges the Infosys Leading Edge travel support. A.C.G. acknowledges support from PRIN-MUR 2022 20228JPA3A, ``The path to star and planet formation in the JWST era (PATH)'', funded by NextGeneration EU; from INAF-GoG 2022, ``NIR-dark Accretion Outbursts in Massive Young stellar objects (NAOMY)''; and from Large Grant INAF-2024, ``Spectral Key Features of Young stellar objects: Wind-Accretion LinKs Explored in the InfraRed (SKYWALKER)''. Part of this research by M.N. was carried out at the Jet Propulsion Laboratory, California Institute of Technology, under a contract with the National Aeronautics and Space Administration (80NM0018D0004). G.P. gratefully acknowledges support from the Carlsberg Foundation, grant CF23-0481. The facilities used for this work include JWST/MIRI-MRS, JWST/NIRCam, and ALMA. The following software packages were used in this research: Astropy \citep{astropy:2013, astropy:2018, astropy:2022}, CARTA \citep{carta2021zndo...3377984C}, CASA \citep{CASA2022PASP..134k4501C}, Matplotlib \citep{Matplotlib2007CSE.....9...90H, matplotlib2024zndo..10669804C}, NumPy \citep{NumPyharris2020array}, SciPy \citep{SciPy2020SciPy-NMeth}, SpectralCube \citep{spectralcube2016ascl.soft09017R}, Astroquery \citep{astroquery2019AJ....157...98G}, and Pandas \citep{pandasreback2020pandas}.
\end{acknowledgements}

\bibliographystyle{aa}
\bibliography{references}

\clearpage

\onecolumn
\begin{appendix}

\section{\texorpdfstring{H$_2$ ($v=0$--0) Pure Rotational Transitions: Normalized Line Maps}{H2 (v=0--0) Pure Rotational Transitions: Normalized Line Maps}}
\label{appsec:v=0-0_transitions_nomrlaized_H2_linemaps}

In this appendix, we present normalized line-intensity maps (scaled to the range 0--1) for all continuum-subtracted pure-rotational H$_2$ transitions. These maps highlight the relatively faint emission visible in the S(3), S(4), and S(5) lines in the southwestern region of the wind structure.

\begin{figure*}[htbp!]
\centering
\includegraphics[width=\textwidth]{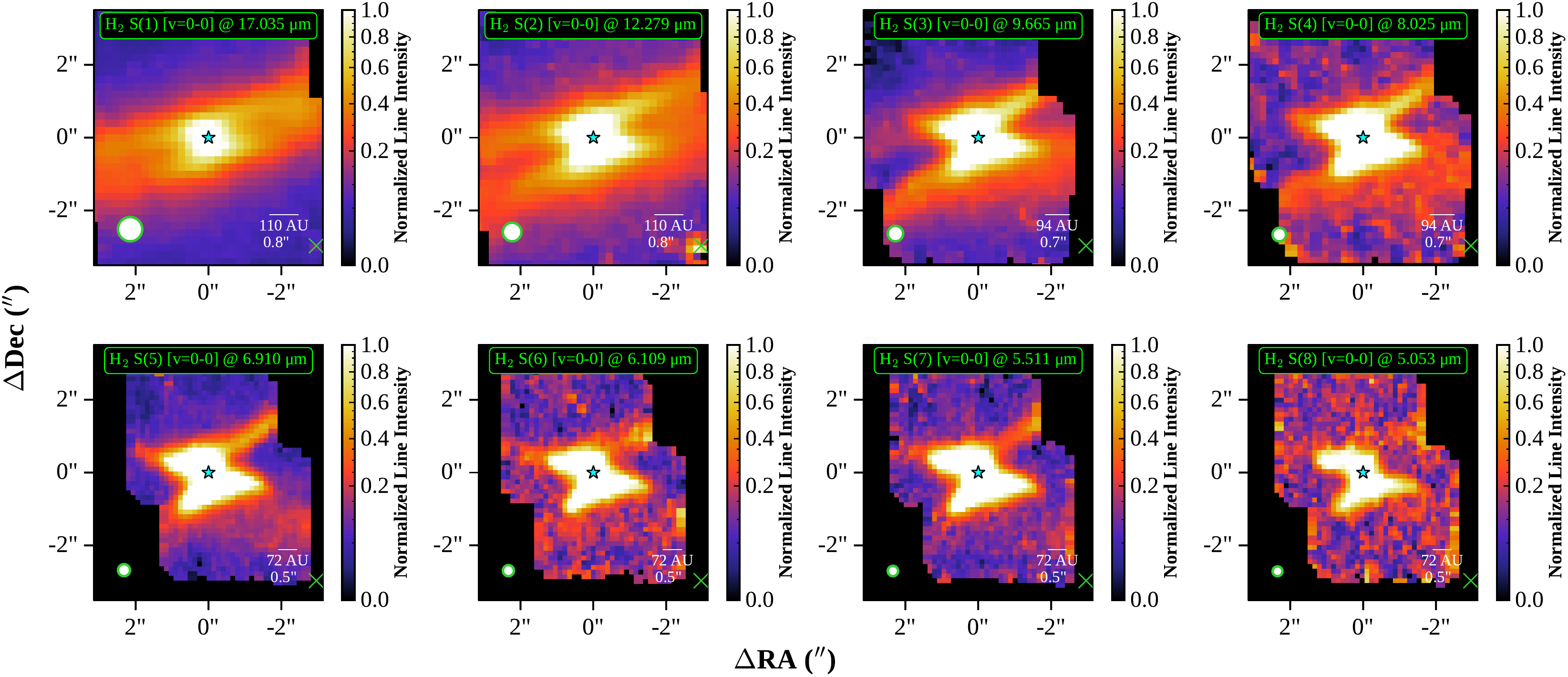}
\caption{Continuum-subtracted, normalized line-intensity maps of the H$_2$ $v=0$--0 pure-rotational transitions, showing faint H$_2$ emission in the southwestern region of the 2D field. The star symbol marks the JWST/MIRI-MRS 14.0~$\mu$m continuum position, and the cross marks the location of HV~Tau~AB (unresolved), located $\sim4\arcsec$ from the HV~Tau~C component.}
\label{appfig:hv_tau_c_H2_v_0_0_extended_lines_nomalized_maps}
\end{figure*}

\section{\texorpdfstring{H$_2$ ($v=1$--1) Pure Rotational Transitions Linemaps}{H2 (v=1--1) Pure Rotational Transitions Linemaps}} \label{appsec:v=1-1_transitions_H2_linemaps}

\begin{figure*}[htbp!]
\centering
\includegraphics[width=\textwidth]{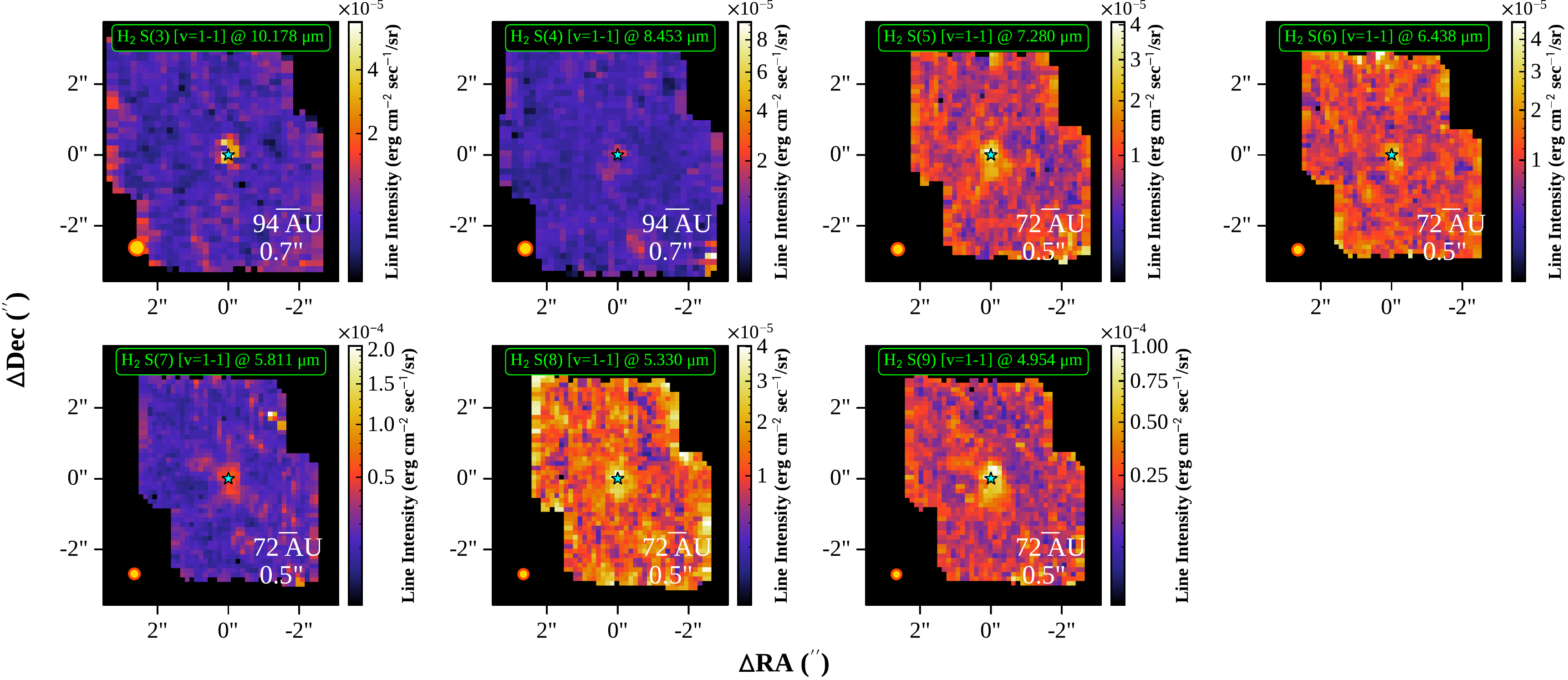}
\caption{Continuum-subtracted line intensity maps of the H$_2$ $v=1$--1 pure rotational transitions. The emission is compact and shows no evidence for extended structure beyond $\sim 2$ PSF FWHM. The star mark is the JWST/MIRI-MRS 14.0 $\mu$m continuum position.}
\label{appfig:hv_tau_c_H2_v_1_1non_extended_lines}
\end{figure*}

This appendix presents the continuum-subtracted line intensity maps of the pure rotational H$_2$ transitions in the first vibrational state ($v=1$--1), starting from the lowest transition detected within the MIRI-MRS wavelength coverage, S(3) at 10.178~$\mu$m, up to the highest transition detected, S(9) at 4.954~$\mu$m. These maps were generated using the same procedure described in Section~\ref{sec:extended_H2_emission}, allowing a direct comparison with the $v=0$--0 pure rotational linemaps discussed in the main text.

Inspection of the two-dimensional linemaps shows that the $v=1$--1 transitions do not exhibit the extended emission observed for the ground vibrational state. Instead, the emission remains compact and confined within roughly two PSF FWHM from the source position. This lack of spatially extended structure is likely due to the lower S/N ratio of the higher-excitation lines and the limited sensitivity of JWST/MIRI-MRS to intrinsically weak vibrationally excited H$_2$ transitions.

\section{\texorpdfstring{H$_2$ Wind Opening Angle Measurements: Edge Detection}{H2 Wind Opening Angle Measurements: Edge Detection}} \label{appsec:v=0-0_H2_lines_edge_detection_for_opening_angle_measurements}

This appendix shows the detected edge boundaries used in the semi-opening angle measurements for all eight pure rotational H$_2$ transitions S(1)--S(8) in the ground vibrational state $v=0-$0. These edges were extracted from the rotated and cropped 2D linemaps following the procedure described in Section~\ref{subsec:winds_semi_opening_angle}.

\begin{figure}[htbp!]
\centering
\includegraphics[width=\textwidth]{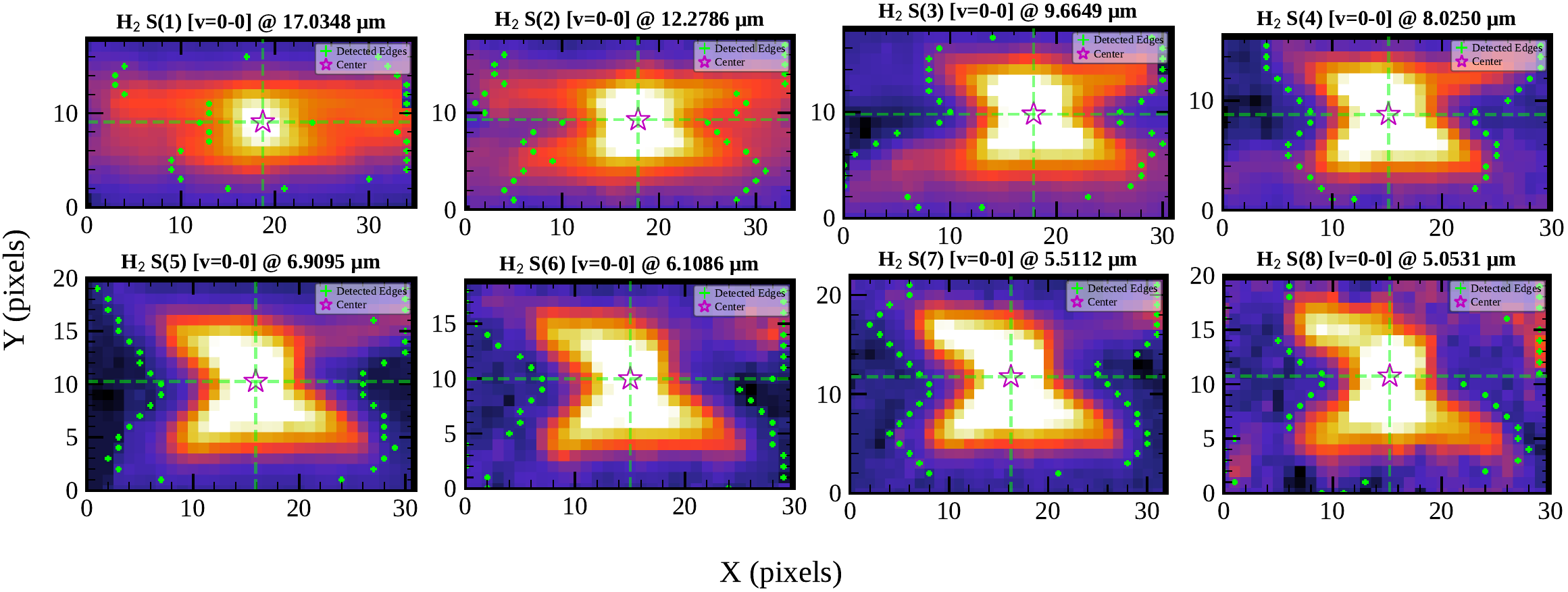}
\caption{Detected edges in the rotated linemaps of the eight H$_2$ $v=0$--0 pure rotational transitions used for the semi-opening angle measurements.}
\label{appfig:hv_tau_c_H2_v_0_0_lines_detected_edges}
\end{figure}

\section{Rotational Diagram Formalism}
\label{appendix:rotational_diagram_formalism}

Under the assumptions of optically thin emission and a Boltzmann distribution of level populations, the observed line fluxes $F_u$ are related to the upper-level column densities through

$$ 
\frac{4\pi F_u}{h c \nu \,\Omega\, g_u A_{ul}} = \frac{N_{\rm tot}}{Z(T)} \exp\!\left(-\frac{E_u}{k_{\rm B} T_{\rm rot}}\right),
$$

where $h$ is Planck's constant, $k_{\rm B}$ is Boltzmann's constant, $c$ is the speed of light, $\nu$ is the transition frequency, $\Omega$ is the solid angle subtended by the emitting region, $A_{ul}$ is the Einstein $A$ coefficient, and $g_u = 2J + 1$ is the statistical weight of the upper level. The quantities $N_{\rm tot}$ and $Q_{\rm rot}$ denote the total H$_2$ column density and the rotational partition function, respectively; $E_u$ is the upper-level energy, and $T_{\rm rot}$ is the rotational temperature.

For molecular hydrogen, the rotational partition function is well approximated by $ Z(T) = 0.0247\,T\, \left(1 - e^{-6000/T}\right)^{-1},$ as given in \citet{Herbst1996AJ....111.2403H} and adopted by \citet{H2OrionPDR2024A&A...687A..86V}.  
Throughout this analysis, we assume local thermodynamic equilibrium (LTE), which is justified because the critical densities of H$_2$ rotational transitions are typically lower than or comparable to those expected in warm molecular layers \citep[see, e.g.,][]{Franceschi2024A&A...687A..96F, Schwarz2025ApJ...980..148S}.

\subsection{Warm and Hot Gas Components: Two-Component Fit to the Rotational Diagram}
\label{appendix:two_component_model}

To model the observed curvature in the rotational diagram, we adopt a two-component LTE framework consisting of warm and hot gas populations. Rearranging the expression above and incorporating extinction as well as the ortho-to-para ratio (OPR), the model is written as

$$
y_{\rm model} = \ln \left[\sum_{i=1}^{2}\exp\!\left(-\frac{x}{T_i} + c_i\right)\right] - \tau_\lambda + \ln\!\left(\frac{{\rm OPR}}{{\rm OPR}_{\rm th}}\right),
$$

where
$$
x = \frac{E_u}{k_{\rm B}}, 
\qquad
y = \ln\!\left(\frac{4\pi F_u}{h c \nu_u g_u A_u}\right),
\qquad
c_i = \ln\!\left(\frac{N_{\mathrm{tot},i}}{Z(T_i)}\right),
\qquad
i=1~(\mathrm{warm}),~2~(\mathrm{hot}).
$$
Here, $c_i$ is the normalization determined by the total column density $N_{\mathrm{tot},i}$ and the rotational partition function $Z(T_i)$.

The extinction is related to the optical depth through $ A_\lambda = 1.086\,\tau_\lambda$. We adopt three independent extinction laws: \citetalias{HensleyDraine2023ApJ...948...55H}, \citetalias{Pontoppidan2024ApJ...963..158P}, and \citetalias{McClure2009ApJ...693L..81M}. From the fitted column densities, the total number of H$_2$ molecules within the emitting region is given by

$$
N_{\mathrm{H}_2} = N_{\mathrm{tot}} \times A, 
\qquad 
A = 3.00\arcsec \times 2.24\arcsec = 1.58 \times 10^{-10}\,\mathrm{sr}.
$$

The corresponding total molecular mass is
$$
M_{\mathrm{H}_2} = \mu_{\mathrm{H}_2}\,m_{\mathrm{H}}\,N_{\mathrm{H}_2}, 
\qquad 
\mu_{\mathrm{H}_2} = 2, 
\qquad 
m_{\mathrm{H}} = 1.67 \times 10^{-27}\,\mathrm{kg}.
$$

\subsection{Power-Law Temperature Distribution Model Fit to the Rotational Diagram}
\label{appendix:powerlaw_model}

To investigate whether the H$_2$ emission originates from gas spanning a continuous range of temperatures, we adopt a power-law temperature distribution. In this model, the differential column density of gas as a function of temperature is parameterized as

$$
\frac{dN}{dT} = a\,T^{-b}, \qquad T \in [T_{\min},T_{\max}],
$$

where $N$ is the H$_2$ column density (cm$^{-2}$), $a$ is a normalization constant, and $b>0$ is the power-law index describing the relative contribution of hot and cool gas (larger $b$ implies less hot gas).

\paragraph{Normalization of the temperature distribution:} The total column density of gas warmer than $T_{\min}$ is

$$ N_T \equiv N(>T_{\min}) = \int_{T_{\min}}^{T_{\max}} a\,T^{-b}\,dT . $$

Evaluating the integral gives

$$ 
b \neq 1:\quad N_T = a\,\frac{T_{\min}^{1-b}-T_{\max}^{1-b}}{b-1}, \quad \text{which yields,} \quad
a = N_T \frac{b-1}{T_{\min}^{1-b}-T_{\max}^{1-b}} = N_T \frac{1-b}{T_{\max}^{1-b}-T_{\min}^{1-b}}.
$$

$$
\text{For completeness, when $b=1:$} \quad N_T = a \ln\!\left(\frac{T_{\max}}{T_{\min}}\right),
\quad \text{which yields,} \quad a = \frac{N_T}{\ln(T_{\max}/T_{\min})}.
$$

\paragraph{Level populations in LTE: } For gas in local thermodynamic equilibrium (LTE), the fractional population of an upper level $u$ at temperature $T$ is
$
\frac{g_u\,e^{-E_u/(k_B T)}}{Z(T)},
$
where $g_u=(2J_u+1)g_{\rm nuc}$ is the statistical weight, $g_{\rm nuc}=1$ for para-H$_2$ and $3$ for ortho-H$_2$ (for OPR$=3$ in the high-temperature limit), $E_u$ is the upper level energy (commonly expressed as $E_u/k_B$ in Kelvin), $Z(T)$ is the rotational partition function.

The column density contribution from a temperature slice $[T,T+dT]$ is therefore

$$ 
dN_u = a\,T^{-b} \frac{g_u\,e^{-E_u/(k_B T)}}{Z(T)}\,dT.
$$

Integrating over the full temperature range gives

$$
N_u = a\,g_u \int_{T_{\min}}^{T_{\max}} \frac{e^{-E_u/(k_B T)}}{Z(T)} T^{-b} dT.
$$

The rotational partition function of H$_2$ can be approximated as (e.g., \citealt{Herbst1996AJ....111.2403H}),
$
Z(T) = 0.0247\,T\, \left(1 - e^{-6000/T}\right)^{-1},
$

where $T$ is in Kelvin and the factor $6000$~K approximates the vibrational temperature of H$_2$ ($\sim5980$~K).

Substituting Z(T) in above equation gives

$$
\frac{N_u}{g_u} = \frac{a}{0.0247} \int_{T_{\min}}^{T_{\max}} e^{-E_u/(k_B T)} \left(1-e^{-6000/T}\right) T^{-(1+b)} dT.
$$

\paragraph{Observed column densities: }Observed line fluxes $F_\ell$ (integrated over a beam) are converted to upper-level column densities via $ N_u^{\rm obs} = \frac{4\pi F_\ell}{A_{ul} h\nu \Omega_{\rm beam}},$ or for surface brightness $I_\ell$ (per steradian) $N_u^{\rm obs} = \frac{4\pi I_\ell}{A_{ul} h\nu}.$ Rotational diagrams are constructed by plotting $\log_{10}(N_u/g_u)$ versus $E_u/k_B$.

\paragraph{Including extinction: }Dust extinction attenuates the observed line flux. If $A_V$ is the visual extinction, the extinction at wavelength $\lambda$ is $ A_\lambda = \left(\frac{A_\lambda}{A_V}\right)A_V, $ where $(A_\lambda/A_V)$ is obtained from an adopted extinction curve (e.g., \citetalias{HensleyDraine2023ApJ...948...55H}, \citetalias{Pontoppidan2024ApJ...963..158P}, and \citetalias{McClure2009ApJ...693L..81M}. Extinction modifies the observed column density as
$$
\log_{10}\left(\frac{N_u}{g_u}\right)_{\rm obs} = \log_{10}\left(\frac{N_u}{g_u}\right)_{\rm intrinsic} - 0.4A_\lambda.
$$

\paragraph{Temperature integral: }Define the temperature integral
$$
I(E,b;T_{\min},T_{\max}) = \int_{T_{\min}}^{T_{\max}} e^{-E/T} (1-e^{-E_0/T}) T^{-(1+b)} dT,
$$
where $E\equiv E_u$ (in Kelvin) and $E_0=6000$~K.

Splitting the integral,

$$
I(E,b)=I_1-I_2, \quad \text{where,} \quad I_1 = \int_{T_{\min}}^{T_{\max}} e^{-E/T}T^{-(1+b)}dT \quad
\text{and} \quad I_2 = \int_{T_{\min}}^{T_{\max}} e^{-(E+E_0)/T}T^{-(1+b)}dT.
$$

Using the definition of Gamma integral function, integrals of the form $ \int u^{\,b-1} e^{-u} du $, can be written in terms of incomplete Gamma functions as $ \int_{u_1}^{u_2} u^{\,b-1} e^{-u} du = \Gamma(b,u_1) - \Gamma(b,u_2).$ Using the substitution $u=E/T$ gives,
$$
I_1 = E^{-b} \left[ \Gamma\!\left(b,\frac{E}{T_{\min}}\right) - \Gamma\!\left(b,\frac{E}{T_{\max}}\right) \right]
\quad \text{and} \quad
I_2 = (E+E_0)^{-b} \left[ \Gamma\!\left(b,\frac{E+E_0}{T_{\min}}\right) - \Gamma\!\left(b,\frac{E+E_0}{T_{\max}}\right) \right].
$$

Combining the two terms,
$$
I(E,b;T_{\min},T_{\max}) = E^{-b} \left[\Gamma\left(b,\frac{E}{T_{\min}}\right) - \Gamma\left(b,\frac{E}{T_{\max}}\right) \right]
-
(E+E_0)^{-b} \left[ \Gamma\left(b,\frac{E+E_0}{T_{\min}}\right) - \Gamma\left(b,\frac{E+E_0}{T_{\max}}\right)
\right].
$$
This relation is used in the evaluation of the temperature integral appearing in the power-law temperature distribution model.

\paragraph{Final fitting model: } Substituting the normalization constant, including extinction and OPR, yields the final model used to fit the rotational diagram:

$$
\log_{10}\!\left(\frac{N_u}{g_u}\right)_{\rm obs} = \log_{10}\!\left[ \frac{N_T (1-b)}{T_{\max}^{1-b}-T_{\min}^{1-b}} \cdot \frac{1}{0.0247} \cdot I(E_u,b;T_{\min},T_{\max}) \right] - 0.4\left(\frac{A_\lambda}{A_V}\right)A_V + \log_{10}\!\left(\frac{{\rm OPR}_i}{{\rm OPR}_{\rm th}}\right)
$$

For the fits presented in this work, we fix $T_{\max}=4000$~K. The free parameters of the model are therefore $b$, $T_{\min}$, $N_T$, $A_V$ and the average OPR.

\section{\texorpdfstring{Measuring Line Fluxes for all H$_2$ Transitions from rectangular aperture enclosing $v=0-$0 S(8) extended emission}{Measuring Line Fluxes for all H2 Transitions from rectangular aperture enclosing v=0-0 S(8) extended emission}} \label{appsec:line_flux_measurements_from_S8_rectangular_aperture}

\begin{figure*}[htbp]
\centering
\includegraphics[width=\textwidth]{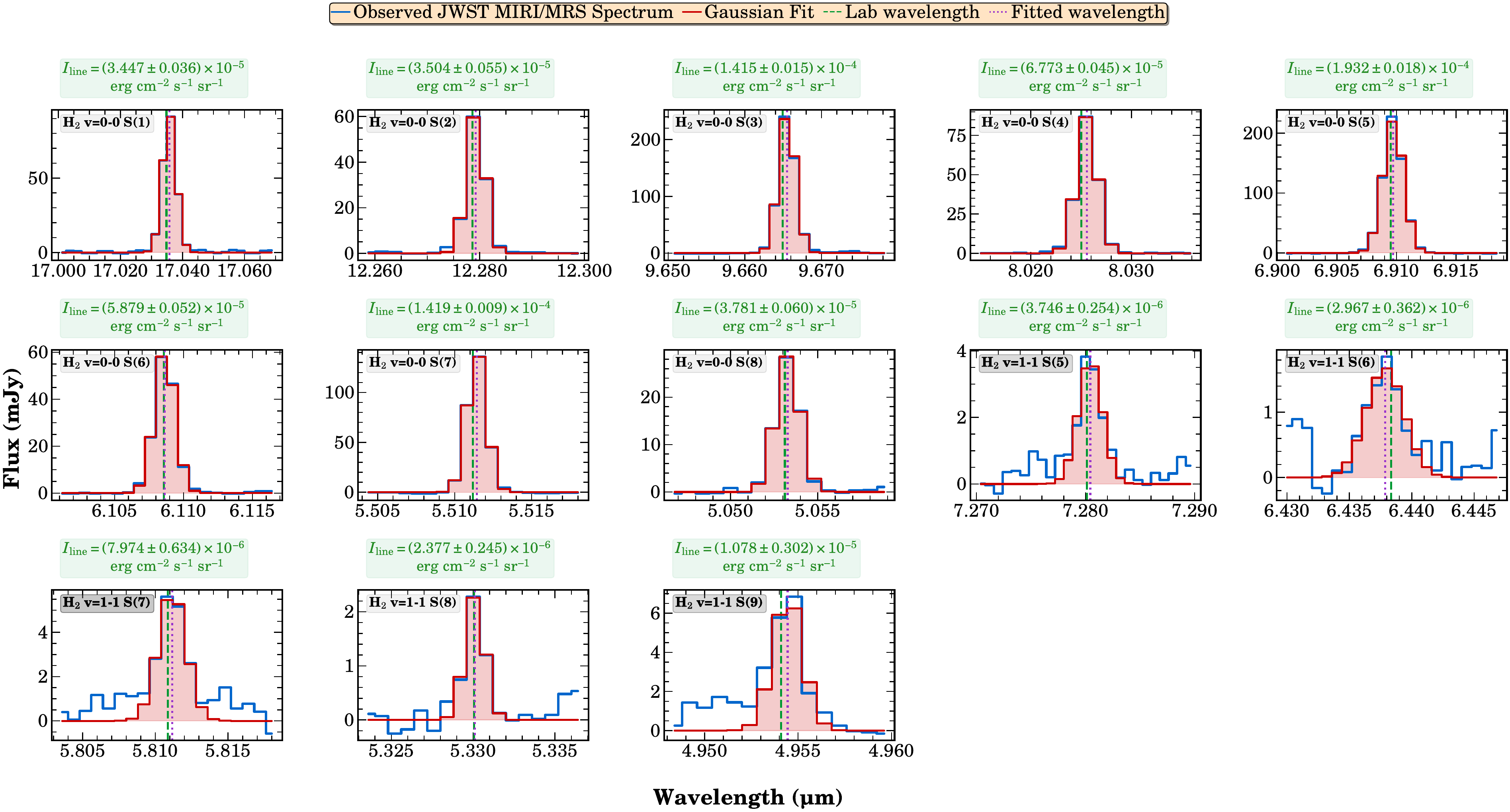}
\caption{Gaussian profile fits to the H$_2$ transitions used in the rotational diagrams. The blue curve shows the observed JWST/MIRI-MRS spectrum, the maroon curve is the best-fit Gaussian model, the green dashed line marks the lab wavelength, and the purple dotted line indicates the fitted centroid position.}
\label{appfig:hv_tau_c_all_H2_lines_gaussian_fit_profiles_for_rectangular_S8_aperture}
\end{figure*}

\begin{table}[htbp]
\caption{Measured H$_2$ line parameters from a rectangular aperture enclosing the H$_2$ $v=0$--0 S(8) transition.}
\label{apptab:hv_tau_c_all_H2_transitions_rotational_diagram_params_for_rectangular_S8_aperture}
\centering
\footnotesize
\begin{tabular}{cccccccc}
\hline\hline
Obs.\ $\lambda$ ($\mu$m) &
Transition &
$E_{\rm up}$ (K) &
$E_{\rm low}$ (K) &
$A$ (s$^{-1}$) &
$g_u$ &
$g_{\rm low}$ &
Line flux density \\
 & S($J$) & & & & & &
($I \pm \delta I$, erg cm$^{-2}$ s$^{-1}$ sr$^{-1}$) \\
\hline
\multicolumn{8}{c}{$v=0$--0 Transitions} \\
17.0362 & S(1) & 1015.08 & 170.48  & $4.758\times10^{-10}$ & 21 & 9  & $(3.447 \pm 0.036)\times10^{-5}$ \\
12.2788 & S(2) & 1681.64 & 509.86  & $2.753\times10^{-9}$  & 9  & 5  & $(3.504 \pm 0.055)\times10^{-5}$ \\
9.6652  & S(3) & 2503.74 & 1015.08 & $9.826\times10^{-9}$  & 33 & 21 & $(1.415 \pm 0.015)\times10^{-4}$ \\
8.0255  & S(4) & 3474.50 & 1681.64 & $2.641\times10^{-8}$  & 13 & 9  & $(6.773 \pm 0.045)\times10^{-5}$ \\
6.9096  & S(5) & 4586.06 & 2503.74 & $5.872\times10^{-8}$  & 45 & 33 & $(1.932 \pm 0.018)\times10^{-4}$ \\
6.1084  & S(6) & 5829.84 & 3474.50 & $1.140\times10^{-7}$   & 17 & 13 & $(5.879 \pm 0.052)\times10^{-5}$ \\
5.5116  & S(7) & 7196.71 & 4586.06 & $1.997\times10^{-7}$  & 57 & 45 & $(1.419 \pm 0.009)\times10^{-4}$ \\
5.0532  & S(8) & 8677.15 & 5829.84 & $3.229\times10^{-7}$  & 21 & 17 & $(3.781 \pm 0.060)\times10^{-5}$ \\
\multicolumn{8}{c}{$v=1$--1 Transitions} \\
7.2801  & S(5) & 10341.25 & 8364.94 & $5.407\times10^{-8}$  & 45 & 33 & $(3.746 \pm 0.254)\times10^{-6}$ \\
6.4383  & S(6) & 11521.12 & 9286.42 & $1.042\times10^{-7}$  & 17 & 13 & $(2.967 \pm 0.362)\times10^{-6}$ \\
5.8109  & S(7) & 12817.26 & 10341.25 & $1.810\times10^{-7}$  & 57 & 45 & $(7.974 \pm 0.634)\times10^{-6}$ \\
5.3300  & S(8) & 14220.49 & 11521.12 & $2.899\times10^{-7}$  & 21 & 17 & $(2.377 \pm 0.245)\times10^{-6}$ \\
4.9541  & S(9) & 15721.48 & 12817.26 & $4.346\times10^{-7}$  & 69 & 57 & $(1.078 \pm 0.302)\times10^{-5}$ \\
\hline
\end{tabular}
\tablefoot{Observed wavelengths, transitions, energy levels, Einstein coefficients, statistical weights, and measured line intensities for the eight $v=0$--0 and five $v=1$--1 H$_2$ transitions.}
\end{table}

Figure~\ref{appfig:hv_tau_c_all_H2_lines_gaussian_fit_profiles_for_rectangular_S8_aperture} shows the Gaussian profile fits for all H$_2$ lines used in constructing the rotational diagrams, including the eight pure rotational transitions S(1)-S(8) in the ground vibrational state ($v=0$--0) and five transitions in the first vibrational state ($v=1$--1). To measure the line fluxes, a spectral window of $\pm 8\Delta\lambda$ was extracted around the lab wavelength of each transition, where $\Delta\lambda$ corresponds to the spectral resolution of JWST/MIRI-MRS (mentioned earlier). The local continuum was removed using the \texttt{pybaseline} Python package, after which a single-component Gaussian profile was fitted to each line. The integrated flux of the fitted Gaussian is adopted as the line flux, and the associated uncertainty corresponds to the fit error. The resulting fluxes and uncertainties are listed in Table~\ref{apptab:hv_tau_c_all_H2_transitions_rotational_diagram_params_for_rectangular_S8_aperture}.

\section{\texorpdfstring{H$_2$ ($v=0$--0) Pure Rotational Transitions: Remaining PV Diagrams \& Radial Velocity Shift Maps}{H2 (v=0--0) Pure Rotational Transitions: Remaining PV Diagrams \& Radial Velocity Shift Maps}} \label{appsec:Remaining_PV_diagrams_and_Radial_Velocity_Shift_Maps}

\begin{figure*}[htbp!]
\centering
\includegraphics[width=\textwidth]{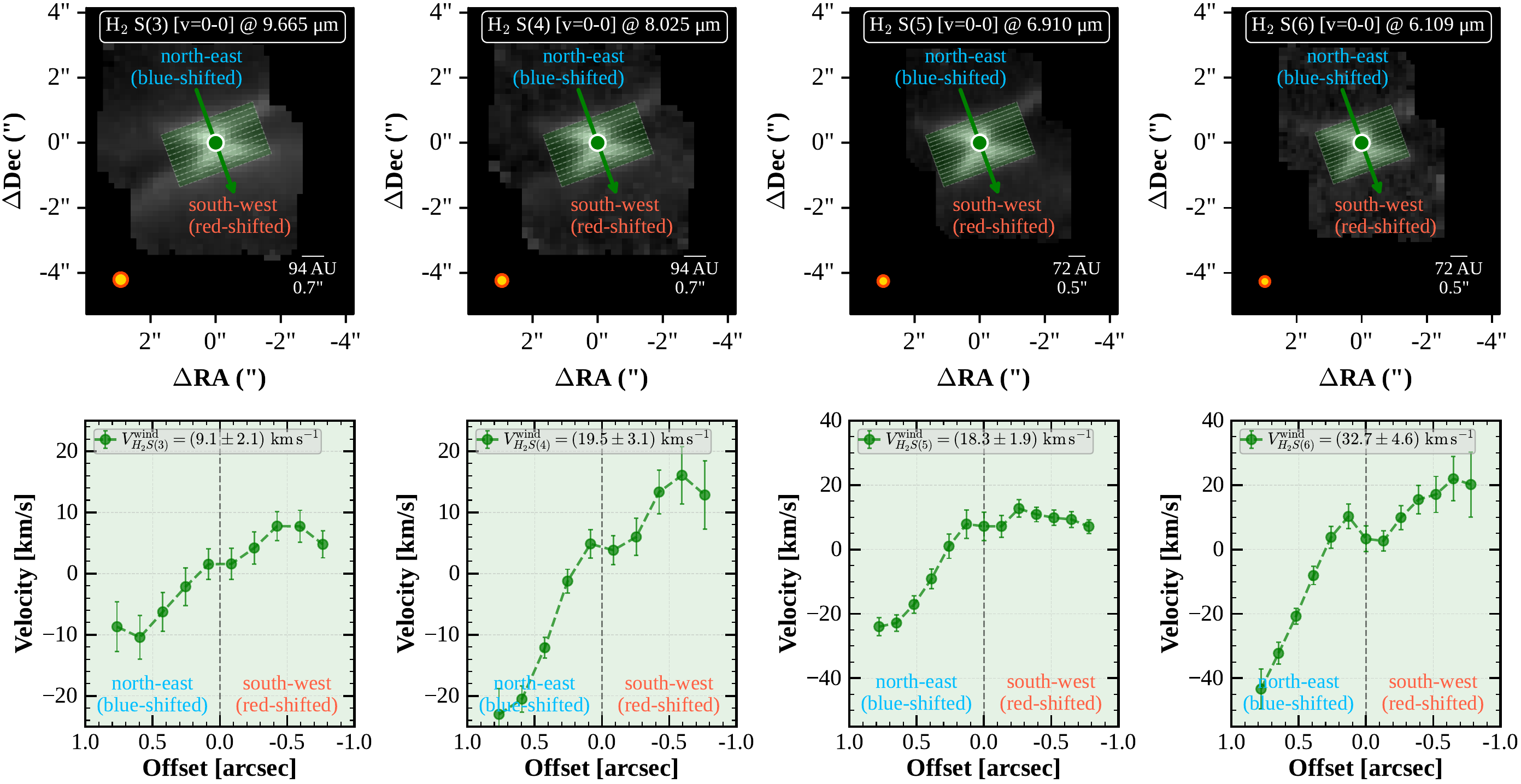}
\caption{PV diagrams for the H$_2$ $v=0$--0 pure rotational lines S(3)--S(6), extracted along the primary outflow axis. All velocities are corrected for inclination and systemic motion. The observed increase in maximum wind velocity from S(3) to S(6) indicates that higher rotational transitions trace progressively faster, more central regions of the wind. The green circle mark is the JWST/MIRI-MRS 14.0 $\mu$m continuum position.}
\label{appfig:hv_tau_c_H2_v=0-0_PV_diagrams_along_H2_outflow_axis_for_S3_S4_S5_S6_lines}
\end{figure*}

\begin{figure*}[htbp!]
\centering
\includegraphics[width=\textwidth]{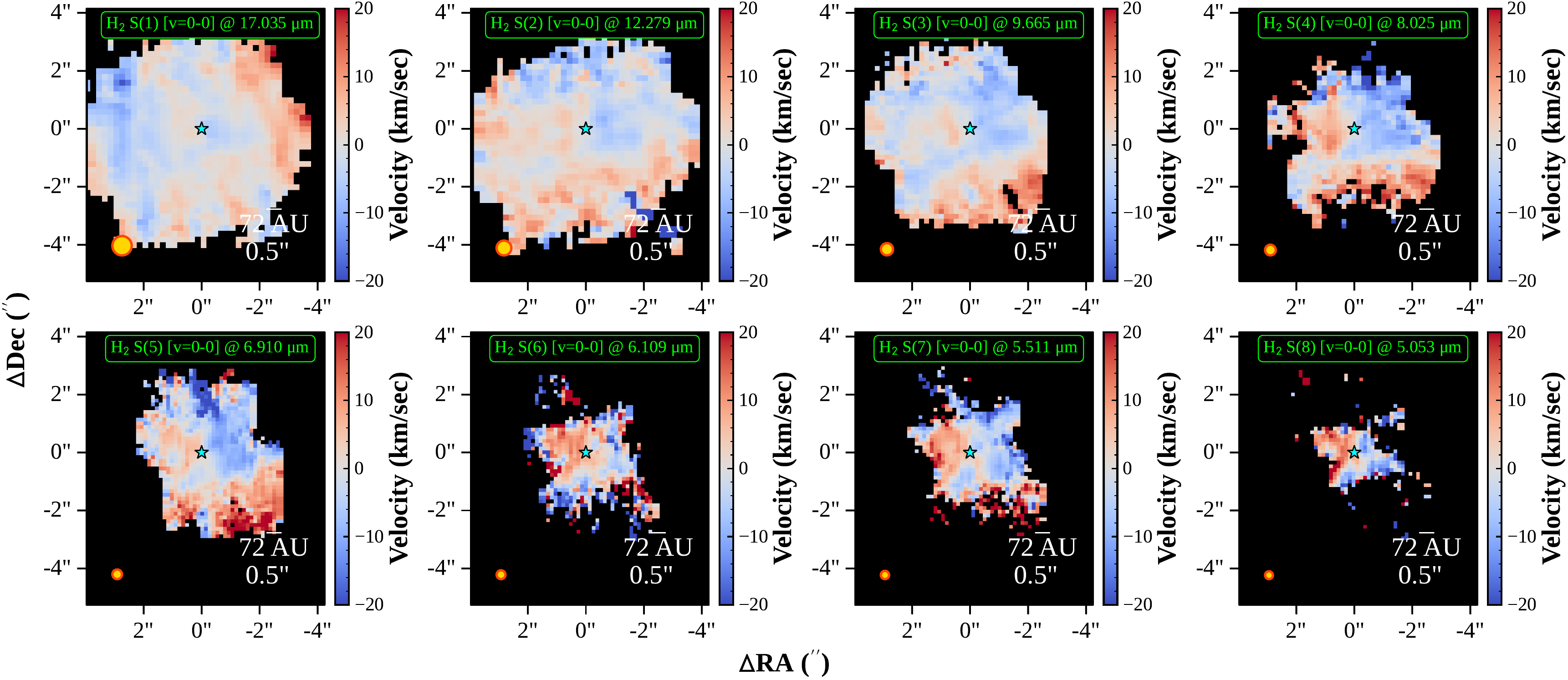}
\caption{Radial velocity-shift maps for the H$_2$ $v=0$--0 pure rotational lines S(1)--S(8). These maps show the line-of-sight wind velocity relative to the central star, derived from pixel-by-pixel Gaussian fitting and subtraction of the systemic velocity. The color scale indicates blueshifted and redshifted gas motion, revealing the velocity structure of the winds. The star mark is the JWST/MIRI-MRS 14.0 $\mu$m continuum position.}
\label{appfig:hv_tau_c_v=0-0_H2_lines_radial_velocity_shift_maps}
\end{figure*}

This appendix presents supplementary kinematic data for the pure rotational lines S(3)--S(8) in the ground vibrational state ($v=0$--0), complementing the analysis in Section~\ref{subsec:winds_kinematics}. Figure~\ref{appfig:hv_tau_c_H2_v=0-0_PV_diagrams_along_H2_outflow_axis_for_S3_S4_S5_S6_lines} displays PV diagrams for transitions S(3)--S(6), extracted along the H$_2$ outflow axis using the methodology described in Section~\ref{subsec:winds_kinematics}. All velocities have been corrected for inclination and systemic motion. A clear progression is evident as we move from lower (S(3)) to higher (S(6)) rotational transitions: the maximum observed wind velocity systematically increases. This trend indicates that higher-excitation transitions preferentially trace the inner, faster-moving regions of the molecular wind, providing a stratified view of the outflow kinematics.

Furthermore, we present radial velocity-shift maps for all eight pure rotational transitions S(1)--S(8) in Figure~\ref{appfig:hv_tau_c_v=0-0_H2_lines_radial_velocity_shift_maps}. These maps were constructed through the following procedure. For each H$_2$ line, we performed a pixel-by-pixel Gaussian fit to the line profile after continuum subtraction, resulting in integrated intensity maps (shown in Figure~\ref{fig:hv_tau_c_H2_v00_extended_linemaps}) and a corresponding map of the fitted mean wavelength, $\lambda_\text{fitted}$. The observed velocity at each pixel was then calculated using the standard Doppler formula:
$$
v_\text{pixel} = \frac{\lambda_\text{fitted} - \lambda_\text{lab}}{\lambda_\text{lab}} \cdot c,
$$
where $\lambda_\text{lab}$ is the laboratory rest wavelength (obtained from the HITRAN database) and $c$ is the speed of light. This yields a two-dimensional map of Doppler-shifted velocities. The systemic velocity of the source, $v_\text{sys}$, was determined by taking the mean velocity across the entire 2D map. Finally, the systemic velocity was subtracted to produce the final radial velocity-shift map, $v_\text{wind} = v_\text{pixel} - v_\text{sys}$, which shows the distribution of wind velocities relative to the central pre-main-sequence star. This process was repeated consistently for all eight transitions.

\section{HV Tau C JWST/NIRCam and JWST/MIRI Photometric Data Analysis}
\label{appsec:HV_Tau_C_JWST_NIRCam_and_MIRI_photometric_data_analysis}

\begin{figure*}[htbp!]
\centering
\includegraphics[width=\textwidth]{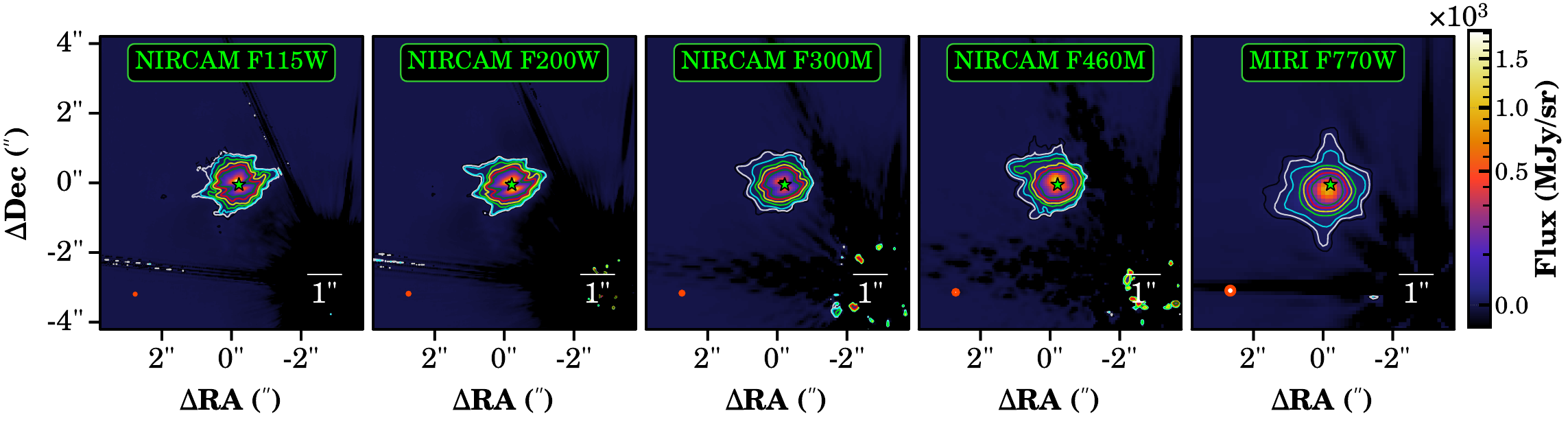}
\caption{JWST/NIRCam and JWST/MIRI images of HV~Tau~C in multiple filters: NIRCam F115W, F200W, F300M, F460M, and MIRI F770W. The star marker indicates the position of the 14.0~$\mu$m JWST continuum peak. Color contours represent emission at $5\sigma$ and higher significance levels, illustrating the scattered light structure across different wavelengths.}
\label{appfig:hv_tau_c_nircam_and_miri_photometric_images}
\end{figure*}

This appendix details the photometric analysis of HV~Tau~C using JWST/NIRCam and JWST/MIRI imaging data. We analyzed four NIRCam filters: F115W (wide band, 1.15~$\mu$m), F200W (wide band, 2.00~$\mu$m), F300M (medium band, 3.00~$\mu$m), and F460M (medium band, 4.60~$\mu$m), along with one MIRI filter: F770W (wide band, 7.70~$\mu$m). The corresponding images are presented in Figure~\ref{appfig:hv_tau_c_nircam_and_miri_photometric_images}. These photometric measurements were utilized in two primary analyses: the spectral energy distribution (SED) modeling discussed in Section~\ref{subsec:Av_from_ice}, and the investigation of extended H$_2$ morphology in Section~\ref{subsec:extended_H2_morphology}.

To obtain accurate photometry for HV~Tau~C, we first addressed the contamination from the nearby HV~Tau~AB binary system. Using the \texttt{stpsf} Python package (\url{https://stpsf.readthedocs.io/en/latest/}), we subtracted the point spread function (PSF) of the AB component from each image, effectively removing the scattered light that overlapped with the C component. Subsequently, we performed aperture photometry using a circular aperture centered on the HV~Tau~C continuum emission. The aperture size was carefully chosen to encompass the entire emission from the source while minimizing background contamination. The resulting flux measurements provide crucial multi-wavelength constraints for understanding the physical properties of this young stellar object.

\section{\texorpdfstring{Rotational Diagram Analysis for Various Rectangular Apertures in the H$_2$ Molecular Winds}{Rotational Diagram Analysis for Various Rectangular Apertures in the H2 Molecular Winds}}\label{Rotational_Diagram_Analysis_for_Various_Rectangular_Apertures_in_the_H2_Molecular_Winds}

\begin{figure*}[htbp!]
\centering

\begin{subfigure}[b]{0.24\textwidth}
    \centering
    \includegraphics[width=\linewidth]{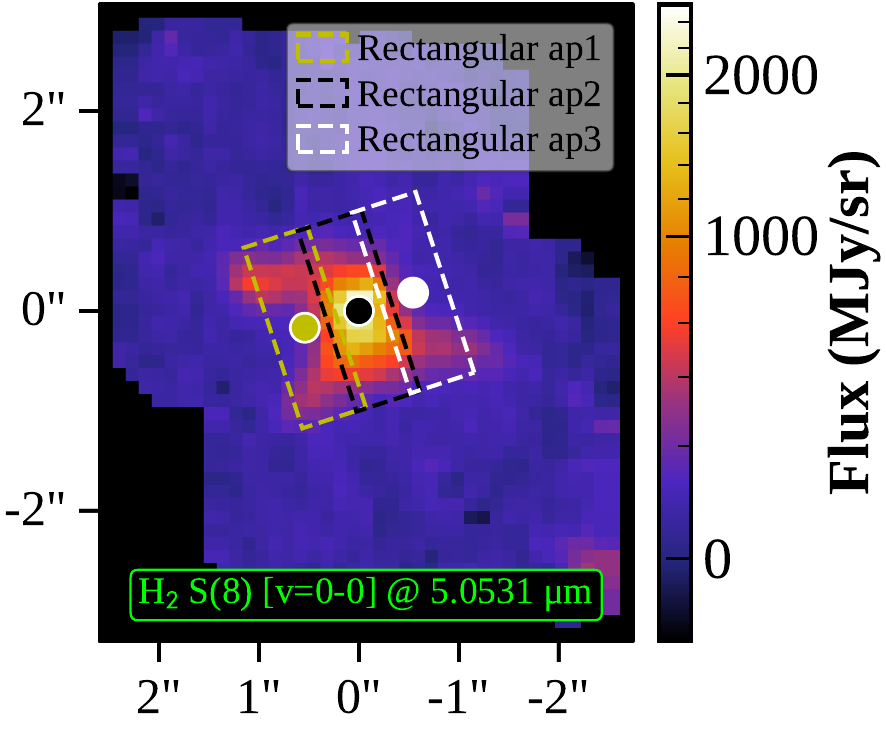}
    \caption{Datacube slice at the H$_2$ S(8) (v=0--0) transition with rectangular apertures AP1, AP2, and AP3 overplotted.}
    \label{appfig:ap123_slice}
\end{subfigure}
\hspace{0.08\textwidth} 
\begin{subfigure}[b]{0.31\textwidth}
    \centering
    \includegraphics[width=\linewidth]{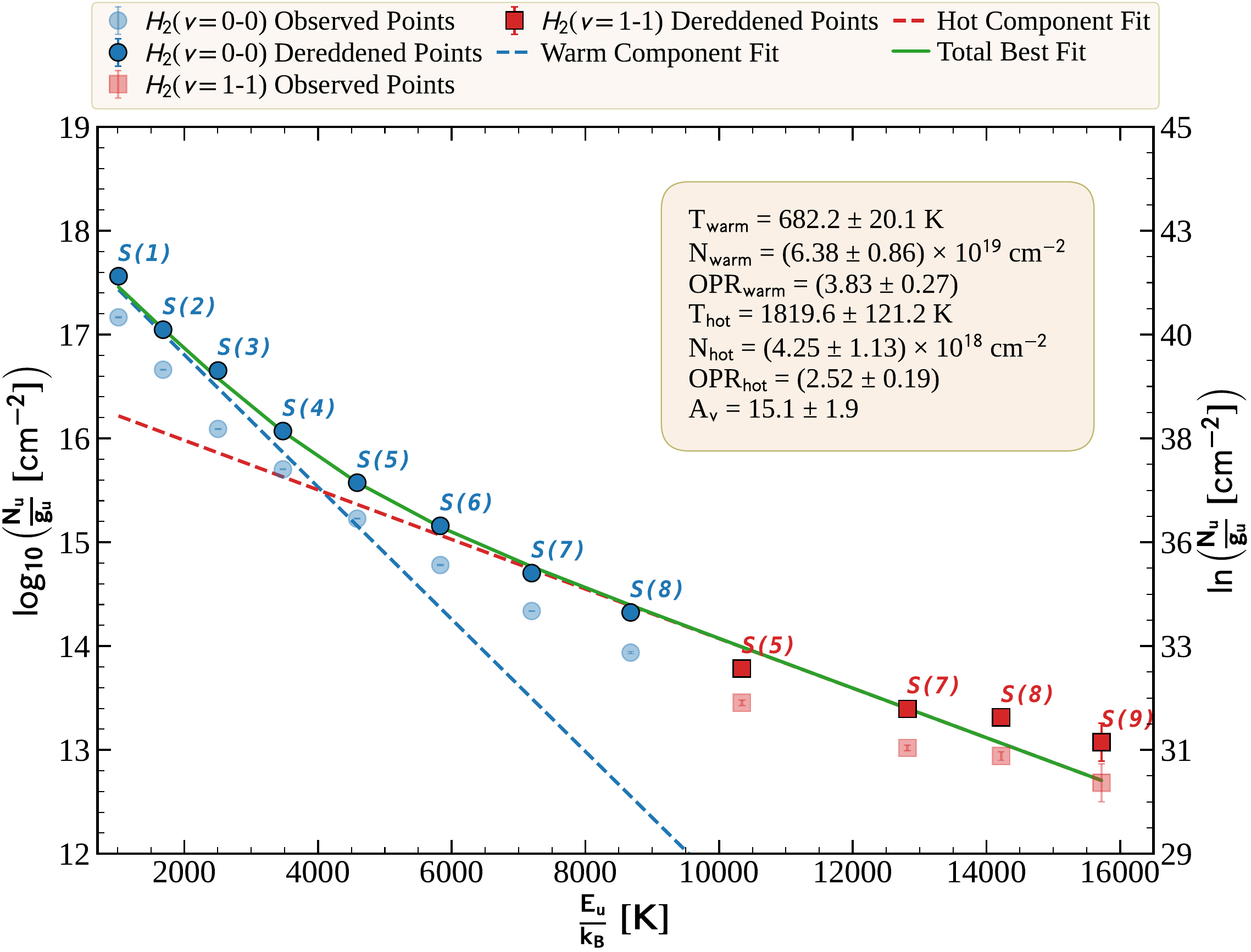}
    \caption{Two-component H$_2$ rotational diagram for AP1.}
    \label{appfig:ap1_rot}
\end{subfigure}

\vspace{0.05cm}

\begin{subfigure}[b]{0.31\textwidth}
    \centering
    \includegraphics[width=\linewidth]{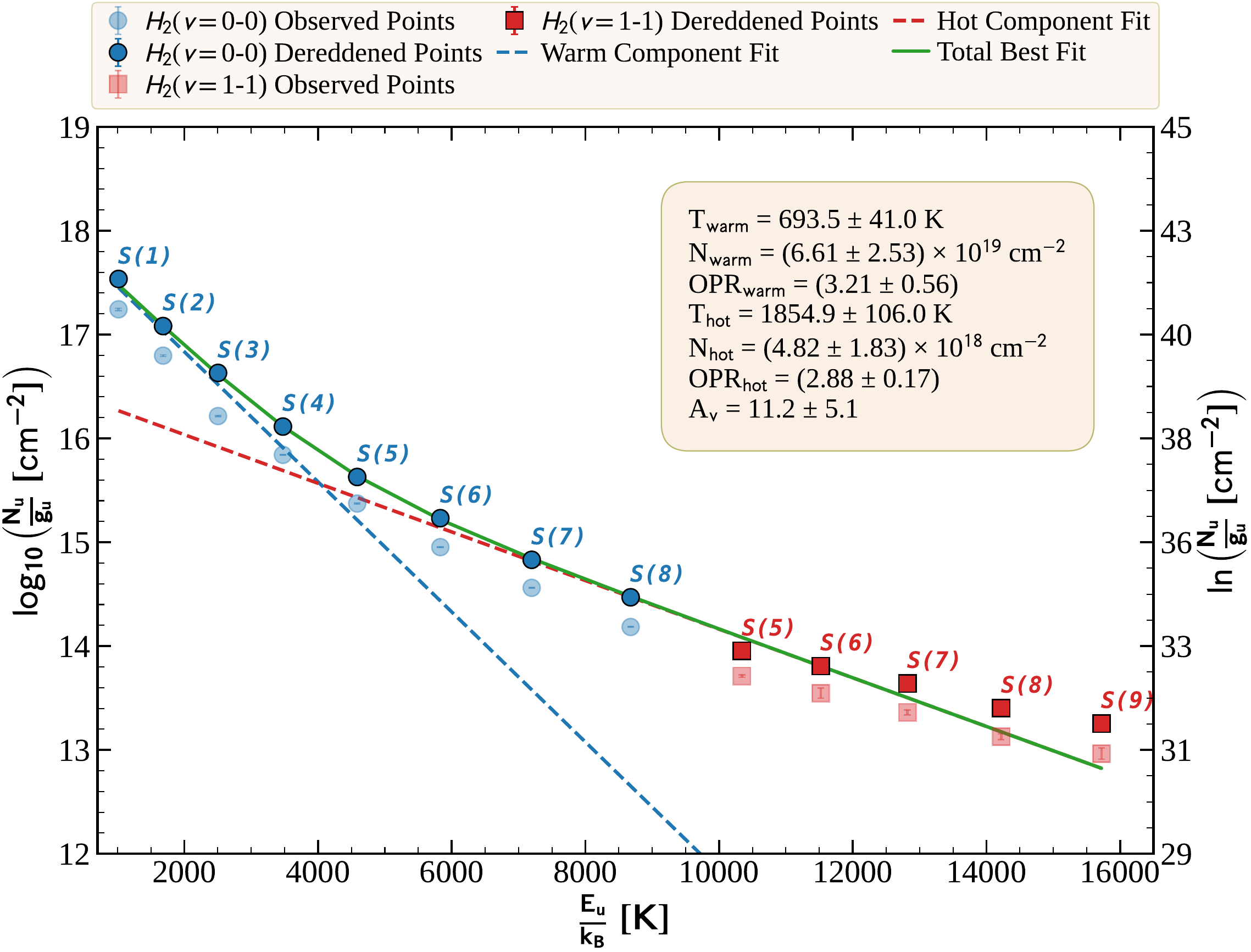}
    \caption{Two-component H$_2$ rotational diagram for AP2.}
    \label{appfig:ap2_rot}
\end{subfigure}
\hspace{0.02\textwidth}
\begin{subfigure}[b]{0.31\textwidth}
    \centering
    \includegraphics[width=\linewidth]{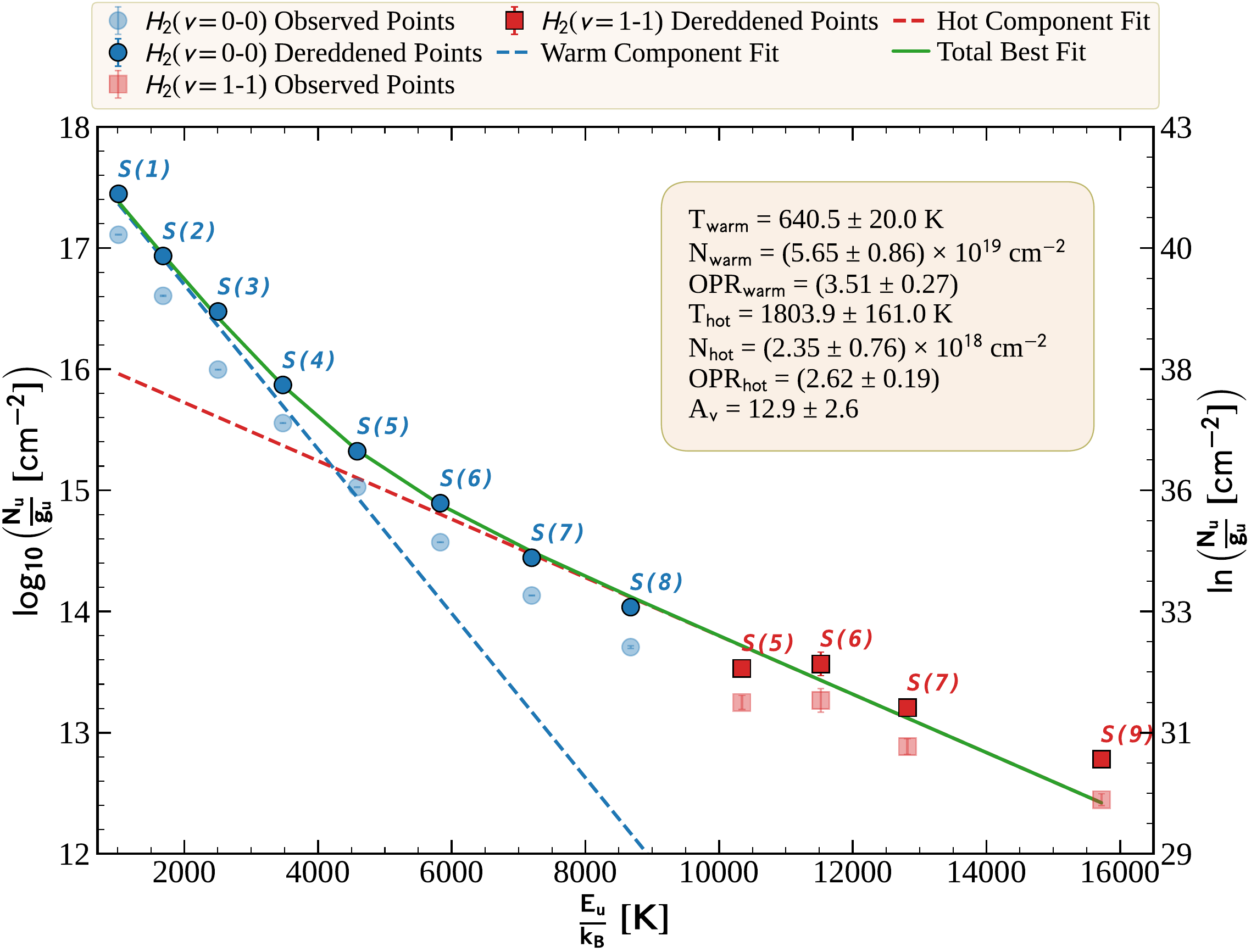}
    \caption{Two-component H$_2$ rotational diagram for AP3.}
    \label{appfig:ap3_rot}
\end{subfigure}
\caption{Spatial placement of rectangular apertures AP1, AP2, and AP3, and their corresponding two-component H$_2$ rotational diagrams. Each aperture has dimensions of $0.67\arcsec \times 1.9\arcsec$, consistent with the PSF FWHM at the longest H$_2$ transition wavelength (17.035~$\mu$m). The apertures are positioned sequentially along the disk axis to probe spatial variations in the excitation conditions.} 
\label{appfig:ap123}
\end{figure*}

\begin{figure*}[htbp!]
\centering

\begin{subfigure}[b]{0.24\textwidth}
    \centering
    \includegraphics[width=\linewidth]{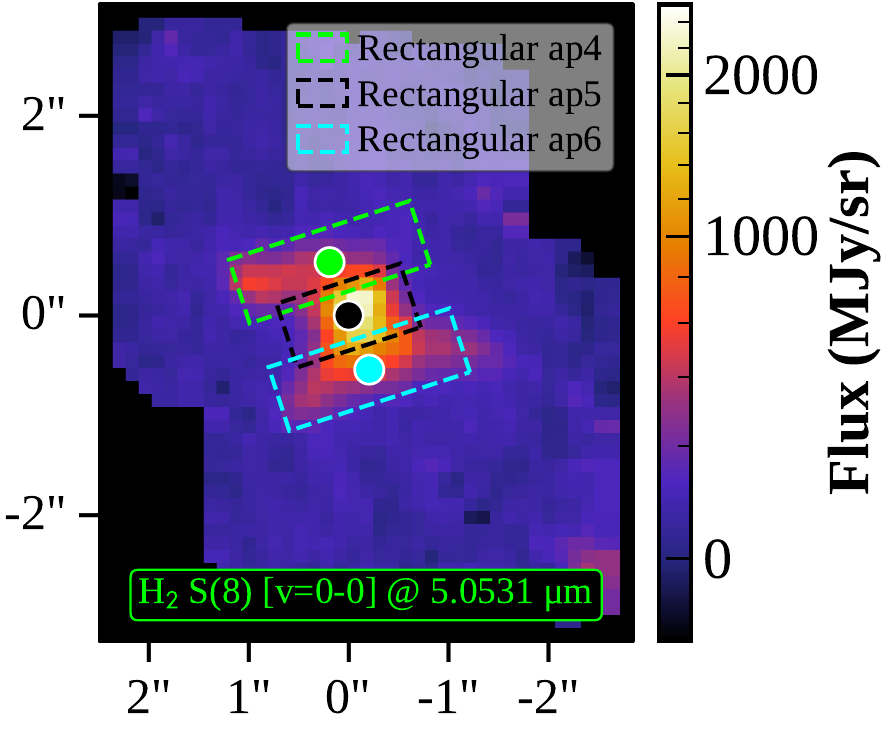}
    \caption{Datacube slice at the H$_2$ S(8) (v=0--0) transition with rectangular apertures AP4, AP5, and AP6 overplotted.}
    \label{appfig:ap456_slice}
\end{subfigure}
\hspace{0.08\textwidth} 
\begin{subfigure}[b]{0.31\textwidth}
    \centering
    \includegraphics[width=\linewidth]{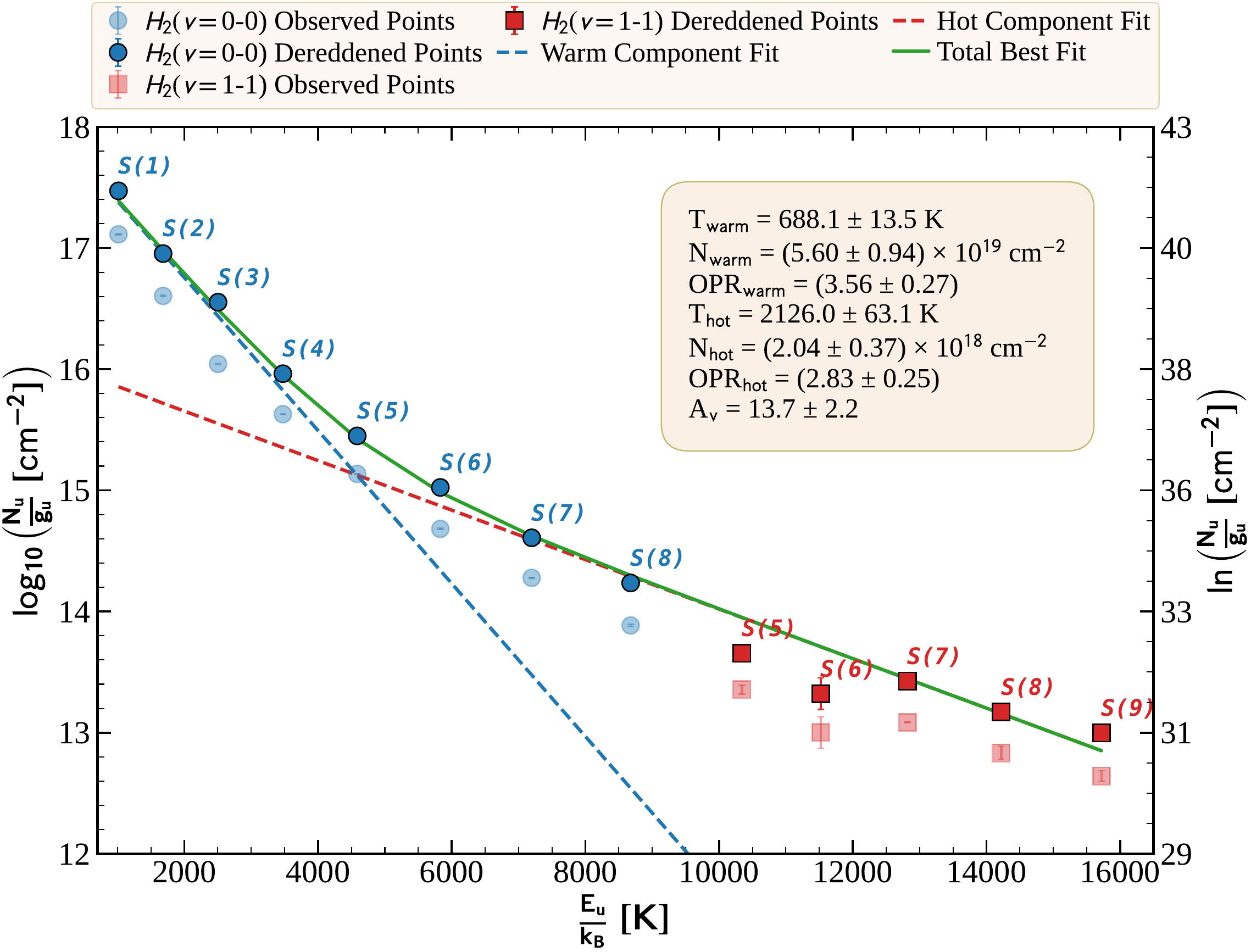}
    \caption{Two-component H$_2$ rotational diagram for rectangular aperture AP4.}
    \label{appfig:ap4_rot}
\end{subfigure}

\vspace{0.05cm}

\begin{subfigure}[b]{0.31\textwidth}
    \centering
    \includegraphics[width=\linewidth]{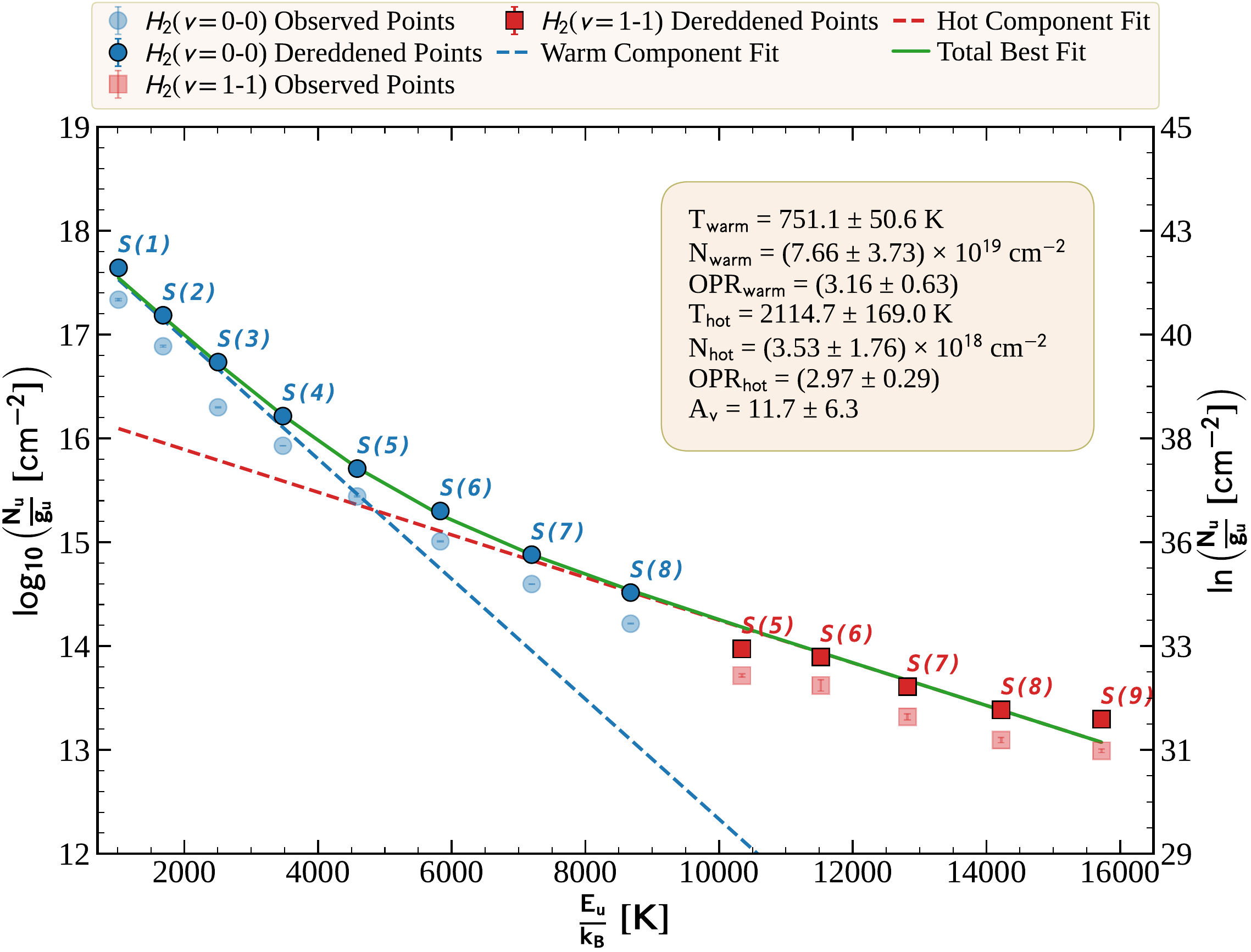}
    \caption{Two-component H$_2$ rotational diagram for rectangular aperture AP5.}
    \label{appfig:ap5_rot}
\end{subfigure}
\hspace{0.02\textwidth} 
\begin{subfigure}[b]{0.31\textwidth}
    \centering
    \includegraphics[width=\linewidth]{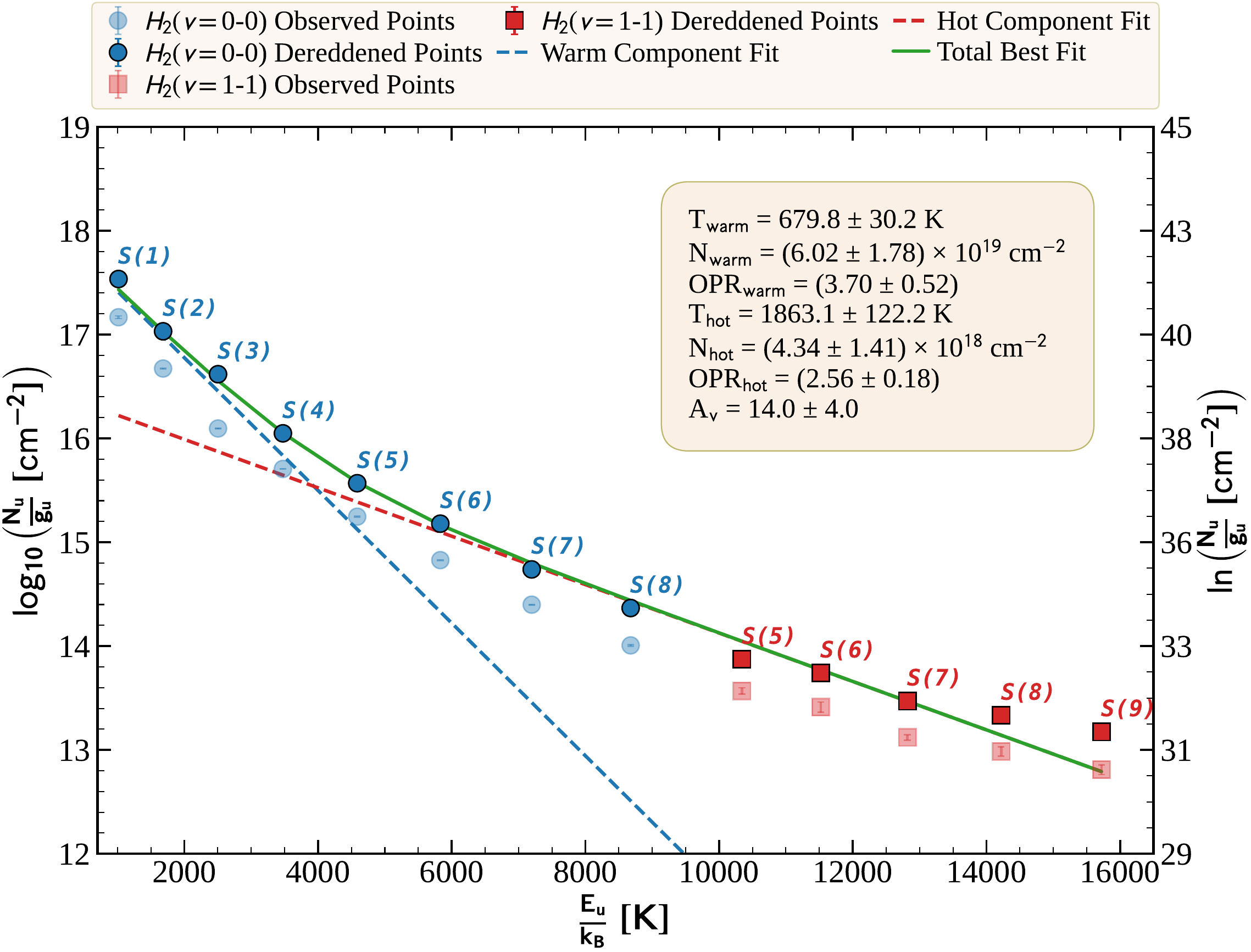}
    \caption{Two-component H$_2$ rotational diagram for rectangular aperture AP6.}
    \label{appfig:ap6_rot}
\end{subfigure}

\caption{Spatial placement of rectangular apertures AP4, AP5, and AP6, along with their corresponding two-component H$_2$ rotational diagrams. Apertures AP4 and AP6 have dimensions of $1.9\arcsec \times 0.67\arcsec$, while AP5 measures $1.3\arcsec \times 0.67\arcsec$. These apertures trace additional regions along the jet/wind axis to investigate spatial variations in molecular excitation.}
\label{appfig:ap456}
\end{figure*}

This appendix presents a detailed rotational diagram analysis applied to six distinct rectangular apertures positioned within the H$_2$ molecular wind outflows. The primary objectives are to investigate spatial variations in excitation conditions (as discussed in Section~\ref{subsec:excitation_conditions_winds}) and to all see the spatial variation of $A_V$ across different regions of the molecular winds (as addressed in Section~\ref{subsec:Av_from_ice}). We defined six rectangular apertures of consistent size, strategically positioned to sample different regions of the outflow. Apertures AP1, AP2, and AP3 are aligned along the disk plane (Figure~\ref{appfig:ap123_slice}), while apertures AP4, AP5, and AP6 are oriented along the wind/jet axis (Figure~\ref{appfig:ap456_slice}). For each aperture, we extracted spectra and subtracted the continuum using the \texttt{pybaseline} Python module. Gaussian profiles were then fitted to each H$_2$ transition to determine line intensities. All relevant line parameters (obtained from the HITRAN database) and the measured line fluxes are compiled in Table~\ref{apptab:H2_line_fluxes_for_various_rect_apertures}. 

Using these line intensities, we constructed rotational diagrams following the methodology outlined in Section~\ref{subsec:excitation_conditions_winds}. Each diagram was fitted with a two-temperature model characterized by seven free parameters: $T_\text{warm}$, $N_\text{warm}$, $\text{OPR}_\text{warm}$, $T_\text{hot}$, $N_\text{hot}$, $\text{OPR}_\text{hot}$, and $A_V$. To assess the sensitivity of our results to different dust properties, we employed three distinct extinction laws: \citetalias{HensleyDraine2023ApJ...948...55H}, \citetalias{Pontoppidan2024ApJ...963..158P}, and \citetalias{McClure2009ApJ...693L..81M}. Figures~\ref{appfig:ap1_rot}, \ref{appfig:ap2_rot}, and \ref{appfig:ap3_rot} display the rotational diagrams for apertures AP1--AP3 using the \citetalias{McClure2009ApJ...693L..81M} extinction law, while Figures~\ref{appfig:ap4_rot}, \ref{appfig:ap5_rot}, and \ref{appfig:ap6_rot} show the corresponding diagrams for apertures AP4--AP6. The best-fit parameters and their associated uncertainties derived from all three extinction laws are systematically presented in Table~\ref{tab:all_6_apertures_best_fit_params_from_H2_rotational_diagrams}.

\begin{sidewaystable*}[htbp]
\caption{Measured H$_2$ line fluxes for various rectangular apertures as shown in Figs.~\ref{appfig:ap123} and~\ref{appfig:ap456}.}
\label{apptab:H2_line_fluxes_for_various_rect_apertures}
\centering
\footnotesize
\resizebox{\textwidth}{!}{
\begin{tabular}{cccccccccccccc}
\hline\hline
$\lambda_{\rm lab}$ ($\mu$m) &
$v$ &
$J$ &
$E_{\rm up}$ (K) &
$E_{\rm low}$ (K) &
$A$ (s$^{-1}$) &
$g_{\rm up}$ &
$g_{\rm low}$ &
AP1 & AP2 & AP3 & AP4 & AP5 & AP6 \\
\hline
\multicolumn{14}{c}{0--0 Transitions} \\
17.0348 & 0--0 & S(1) & 1015.08 & 170.48 & $4.758\times10^{-10}$ & 21 & 9 &
4.279$\pm$0.050 & 5.131$\pm$0.132 & 3.772$\pm$0.021 & 3.805$\pm$0.035 & 6.355$\pm$0.177 & 4.306$\pm$0.127 \\
12.2786 & 0--0 & S(2) & 1681.64 & 509.86 & $2.753\times10^{-9}$  & 9  & 5 &
4.626$\pm$0.048 & 6.321$\pm$0.116 & 4.066$\pm$0.048 & 4.059$\pm$0.054 & 7.776$\pm$0.155 & 4.749$\pm$0.047 \\
9.6649  & 0--0 & S(3) & 2503.74 & 1015.08 & $9.826\times10^{-9}$  & 33 & 21 &
20.700$\pm$0.114 & 27.530$\pm$0.335 & 16.640$\pm$0.158 & 18.540$\pm$0.156 &
33.400$\pm$0.350 & 20.900$\pm$0.150 \\
8.0250  & 0--0 & S(4) & 3474.50 & 1681.64 & $2.641\times10^{-8}$  & 13 & 9 &
10.770$\pm$0.092 & 14.850$\pm$0.076 & 7.677$\pm$0.068 & 9.092$\pm$0.084 &
18.170$\pm$0.105 & 10.870$\pm$0.068 \\
6.9095  & 0--0 & S(5) & 4586.06 & 2503.74 & $5.872\times10^{-8}$  & 45 & 33 &
32.260$\pm$0.232 & 45.100$\pm$0.591 & 20.340$\pm$0.164 & 26.120$\pm$0.122 &
55.780$\pm$0.796 & 33.720$\pm$0.380 \\
6.1086  & 0--0 & S(6) & 5829.84 & 3474.50 & $1.140\times10^{-7}$   & 17 & 13 &
9.569$\pm$0.118 & 14.230$\pm$0.106 & 5.914$\pm$0.054 & 7.623$\pm$0.108 &
16.170$\pm$0.150 & 10.670$\pm$0.072 \\
5.5112  & 0--0 & S(7) & 7196.71 & 4586.06 & $1.997\times10^{-7}$  & 57 & 45 &
22.470$\pm$0.121 & 37.580$\pm$0.164 & 13.990$\pm$0.063 & 19.550$\pm$0.104 &
40.870$\pm$0.165 & 25.870$\pm$0.143 \\
5.0531  & 0--0 & S(8) & 8677.15 & 5829.84 & $3.229\times10^{-7}$  & 21 & 17 &
58.070$\pm$0.143 & 10.250$\pm$0.062 & 3.406$\pm$0.082 & 5.146$\pm$0.094 &
11.020$\pm$0.099 & 6.816$\pm$0.123 \\
\hline
\multicolumn{14}{c}{1--1 Transitions} \\
7.2801 & 1--1 & S(5) & 10341.25 & 8364.94 & $5.407\times10^{-8}$ & 45 & 33 &
0.476$\pm$0.029 & 0.860$\pm$0.023 & 0.297$\pm$0.039 & 0.378$\pm$0.031 &
0.871$\pm$0.032 & 0.618$\pm$0.044 \\
6.4383 & 1--1 & S(6) & 11521.12 & 9286.42 & $1.042\times10^{-7}$ & 17 & 13 &
-- & 0.484$\pm$0.054 & 0.254$\pm$0.057 & 0.138$\pm$0.041 &
0.575$\pm$0.073 & 0.356$\pm$0.040 \\
5.8109 & 1--1 & S(7) & 12817.26 & 10341.25 & $1.810\times10^{-7}$ & 57 & 45 &
0.928$\pm$0.058 & 2.043$\pm$0.111 & 0.681$\pm$0.099 & 1.081$\pm$0.011 &
1.851$\pm$0.125 & 1.172$\pm$0.068 \\
5.3300 & 1--1 & S(8) & 14220.49 & 11521.12 & $2.899\times10^{-7}$ & 21 & 17 &
0.498$\pm$0.046 & 0.766$\pm$0.048 & -- & 0.389$\pm$0.048 &
0.711$\pm$0.041 & 0.553$\pm$0.057 \\
4.9541 & 1--1 & S(9) & 15721.48 & 12817.26 & $4.346\times10^{-7}$ & 69 & 57 &
1.463$\pm$0.615 & 2.789$\pm$0.344 & 0.846$\pm$0.094 & 1.328$\pm$0.133 &
2.972$\pm$0.125 & 1.955$\pm$0.021 \\
\hline
\end{tabular}
}
\tablefoot{Fluxes are given in units of $\times ~10^{-5}$~erg cm$^{-2}$ sec$^{-1}$ sr$^{-1}$ and are measured within the rectangular apertures AP1--AP6 shown in Figs.~\ref{appfig:ap123} and~\ref{appfig:ap456}.}
\end{sidewaystable*}

\end{appendix}

\end{document}